\documentclass[11pt,a4paper]{article}

\usepackage[utf8]{inputenc}
\usepackage[T1]{fontenc}
\usepackage[english]{babel}
\usepackage{a4wide}
\usepackage{indentfirst}
\usepackage{amsmath}
\usepackage{amssymb}
\usepackage{graphicx}
\usepackage{subfigure}
\usepackage{color}

\newcommand{\valos}{\mathbb{R}}

\newcommand{\eps}{\varepsilon}

\newcommand{\ordo}{\mathcal{O}}

\newcommand{\ket}[1]{{\left|#1\right\rangle}}
\newcommand{\bra}[1]{{\left\langle #1\right|}}

\newcommand{\vev}[1]{\left\langle #1 \right\rangle}

\newcommand{\sih}{\operatorname{sh} }
\newcommand{\coh}{\operatorname{ch} }

\usepackage{ifpdf}

\ifpdf
\usepackage{epstopdf}
\usepackage[pdftex,ps2pdf,dvips,colorlinks,urlcolor=blue,citecolor=blue,linkcolor=blue]{hyperref}
\else
\usepackage[hypertex,ps2pdf,dvips,colorlinks,urlcolor=blue,citecolor=blue,linkcolor=blue]{hyperref}
\fi
\begin{document}

\title{Form factor expansion for thermal correlators}

\author{B. Pozsgay\\
Institute for Theoretical Physics, Universiteit van Amsterdam\\
Science Park 904, Postbus 94485, 1090 GL Amsterdam, The Netherlands\\
{ }\\
G. Takács\\
HAS Theoretical Physics Research Group\\
1117 Budapest, Pázmány Péter sétány 1/A, Hungary}

\date{21st August 2010}
\maketitle

\begin{abstract}
We consider finite temperature correlation functions in massive
integrable Quantum Field Theory.
Using a regularization by putting the system in finite volume, we
develop a novel approach (based on multi-dimensional residues) to 
the form factor expansion for thermal
correlators. The first few terms are obtained explicitly in theories
with diagonal scattering. We also discuss the validity of the
LeClair-Mussardo proposal.
\end{abstract}

\numberwithin{equation}{section}

\section{Introduction}

One of the central tasks in many-body quantum physics is the
calculation of correlation functions. They yield considerable amount
of information about the dynamics of the system and their Fourier modes can be
measured, for example with elastic neutron scattering
experiments. In addition to the correlations in the ground state it is
also important to calculate correlation functions at finite
temperatures, in which case the system is populated by a number of
excited states. 

In this paper we consider finite temperature correlation functions in
1+1 dimensional integrable models. A common property of these theories
is that their Hamiltonian can be 
diagonalized analytically using the Bethe Ansatz \cite{XXX,korepinBook}. Moreover, there are powerful methods
available to obtain correlation functions, the most general being
the so-called form factor expansion. The idea is to obtain the matrix
elements of local operators (form factors) in the eigenstate
basis of the Hamiltonian, and then to sum up the spectral series.
The difference as compared to usual approaches (like perturbation theory) is
that in integrable models both the spectrum and the form factors can be
calculated exactly. This presents a unique opportunity to study
strongly correlated quantum systems in situations where conventional
methods break down. Interest in integrable
models has recently been renewed, in large part due to the developments in recent years 
that made it possible to realize certain  
models with the help of optical and magnetic traps
\cite{low-D-trapped,2003PhRvL..91y0402M,2003cond.mat.12003L,vanDruten-YangYang}.

The ideas of the form factor expansion are quite general, however the
methods to obtain the form factors and to sum up the spectral series
can be different. One framework is provided by the Algebraic Bethe
Ansatz (ABA) \cite{korepinBook,Faddeev-ABA-intro,korepin-izergin},
which was applied successfully to a number of models, 
most prominently the Heisenberg spin chains
\cite{Korepin-Izergin-XXZ,ABA-XXZ-corr,2002hep.th....1045K} and the 1D
Bose 
gas \cite{Lieb-Liniger,korepin-slavnov}. Correlation functions are obtained typically in the form of
integral series \cite{korepin-LL1,iz-kor-resh} or multiple integral
formulas \cite{ABA-XXZ-corr,2002hep.th....1045K}, or one can resort to
numerical summation schemes \cite{Caux-Abacus-XXZ,Caux-Calabrese1}.
 Moreover, there are
methods to handle the finite temperature situation, either through 
generalizations of the basic techniques \cite{korepinBook} or developing an alternative
description using the so-called Quantum Transfer Matrix 
\cite{2004LNP...645..349K,2004JPhA...37.7625G,2007JSMTE..08...30S}.

Integrable Quantum Field Theory provides a different framework to
obtain form factors and correlation functions. In these theories the basic object
is the factorized S-matrix \cite{zam-zam,Mussardo:1992uc}, and the relation to a microscopic
description is rather indirect. It can be shown, that the form factors
satisfy a certain set of equations (the form factor bootstrap
equations) which follow from general
field theoretical arguments supplemented with the special analytic properties of the
S-matrix \cite{Karowski:1978vz,smirnov_ff,zam_Lee_Yang,Delfino:1996nf}. 
The idea is to provide a general solution to these equations,
and then to identify those solutions which correspond to a given local
operator \cite{Delfino:2008ia}. The resulting form factor functions can then be used to construct
correlation functions. An important feature is that the form factors
are calculated in the infinite volume asymptotic state basis. This is
to be contrasted with the situation in ABA, where one starts with a
finite system and the infinite volume limit is only performed
afterwards. Interesting connections between the form factor bootstrap
and the ABA were pointed out recently in \cite{nonrelFF}.

The problem of zero-temperature correlations in integrable QFT is
well understood. Although an analytic summation of the spectral series
is in general not possible (except in some simple models
\cite{Korepin:1998vp,Korepin:1998rj,Oota:1998tr}), the series
has very good 
convergence properties in massive models and can be evaluated
numerically to a desired precision \cite{zam_Lee_Yang,ising_ff1}. On
the other hand, the problem of finite temperature correlation functions is less
understood and it has been subject to an active research in the last
ten years \cite{Leclair:1996bf,Leclair:1999ys,Saleur:1999hq,
Lukyanov:2000jp,Delfino:2001sz,Mussardo:2001iv,Konik:2001gf,Essler:2004ht,Altshuler:2005ty}.  
The idea is to use the zero-temperature form factors in the spectral
series, however, the thermal average is ill-defined in infinite volume. The problem
is related to the appearance of disconnected 
terms in the expansion, which lead to formally divergent expressions. Following Balog 
 it can be shown that the divergent parts cancel with contributions from the partition 
function \cite{Balog:1992gf}, however it is a highly non-trivial task
to obtain the finite left-over pieces.

There have been attempts to write down a regularized version of the
spectral series. In particular, LeClair and Mussardo proposed an
expansion for the one-point and two-point 
functions in terms of form factors dressed by appropriate occupation
number factors containing the pseudo-energy function from the thermodynamical
Bethe Ansatz \cite{Leclair:1999ys}. Their proposal for the two-point function
was questioned by Saleur \cite{Saleur:1999hq} (see also \cite{CastroAlvaredo:2002ud}); on the
other hand, he also gave a proof of the LeClair-Mussardo formula for one-point
functions provided the operator considered is the density of some
local conserved charge.
However, it was demonstrated in the case of 
one-point functions that the results obtained by naive regularization are 
ambiguous; this is shown in particular by the difference \cite{Mussardo:2001iv} 
between the formulae proposed by LeClair and Mussardo and by Delfino \cite{Delfino:2001sz} 

This motivated the present authors to develop a regularization method based 
on finite volume form factors \cite{Pozsgay:2007kn,Pozsgay:2007gx} which was applied to 
one-point functions giving a confirmation of the LeClair-Mussardo formula. The central 
idea is to use a finite volume setting to regularize the divergences, and to compute the 
physical quantities in finite volume. At the end of the calculation, the volume is taken 
to infinity. If one computes only quantities meaningful in infinite volume, the divergences 
cancel and the end result is well-defined. Because finite volume is not an ad hoc, but a 
physical regulator (note that physically realizable systems are always of finite size) one 
is virtually guaranteed to obtain the correct result provided the calculations are performed 
correctly. The existence of a mass gap $m$ is essential in this
approach: the Boltzmann-factor $e^{-m/T}$ provides a natural small
parameter for the finite temperature expansion. The result is
an integral series, where the $N$th term represents
$N$-particle processes over the Fock-vacuum. The contributions with a
low number of particles can be interpreted as disconnected terms of
matrix elements calculated in a thermal state with a large number of particles
\cite{Altshuler:2005ty,pozsi-LM}. In this sense the approach is
similar to the one used in the seminal works in ABA
\cite{korepin-LL1,iz-kor-resh}. However, a distinctive feature is that
the calculations do not use any information about the form factors
other than their singularity properties.

The finite volume form factor approach was extended to boundary operators as
well \cite{Kormos:2007qx}, which was used to compute finite temperature one-point functions 
of boundary operators \cite{Takacs:2008ec}. Another application of the bulk finite volume 
form factors is the construction of one-point functions of bulk operators on a finite interval
\cite{Kormos:2010ae}. One of the present authors have also used this formalism in 
\cite{Takacs:2009fu} to construct the 
form factor perturbation expansion in non-integrable field theories (originally proposed by 
Delfino et al. \cite{Delfino:1996xp}) beyond the lowest order.
The finite volume regularization method was also exploited in a recent work by 
Essler and Konik \cite{Essler:2007jp,Essler:2009zz}. They computed the first nontrivial 
contribution to the dynamical spin-spin correlation function of a spin chain using the 
O(3) sigma model. However, the methods they use do not have any obvious extension to higher 
order, albeit it is reasonably clear that the finite volume regularization must work.

In this paper we develop a systematic method to compute the finite temperature form factor 
expansion to arbitrary orders, relying on an application of the residue theorem for multiple 
complex variables. We also demonstrate that the same method can be applied to computing the 
form factor perturbation theory contributions, and the zero-temperature three-point function.

The outline of the paper is as follows: section \ref{section2} summarizes the necessary background on form 
factor bootstrap and finite volume form factors. Also, it sets up the framework for the 
expansion of the thermal two-point function, and discusses the scope
and validity of the approach. In section \ref{section3} we present the calculation
of the (zero-temperature) 3-point functions as a warm-up example.
The expansion for the finite temperature 2-point function is derived
in section \ref{section4}, except for some technical details that are relegated to  
appendices. The discussion of the LeClair-Mussardo proposal in light
of our results is presented in subsection \ref{LM-discussion}. We present an
application of our method to form factor perturbation
theory in section \ref{section5}, and conclude in section \ref{section6}. 

\section{Form factor expansion for the thermal two-point function}

\label{section2}

\subsection{The form factor bootstrap}

Here we give a very brief summary of the equations of the form factor
bootstrap, in order to set up notations and to provide background
for later arguments; the interested reader is referred to Smirnov's
review \cite{Smirnov:1992vz} for more details. For the sake of simplicity 
let us suppose 
that the spectrum of the model consists of a single particle mass $m$. 
The energy and the momentum of an on-shell particle is parametrized 
by the rapidity variable as $E=m\cosh \theta$ and $p=m\sinh \theta$.
Because of integrability, multi-particle scattering amplitudes factorize
into the product of pairwise two-particle scatterings, which is described 
by a pure phase, which we denote by $S\left(\theta\right)$
where $\theta$ is the relative rapidity of the incoming particles. 
Incoming and outgoing asymptotic states can
be distinguished by ordering of the rapidities:
\begin{equation}
|\theta_{1},\dots,\theta_{n}\rangle=\begin{cases}
|\theta_{1},\dots,\theta_{n}\rangle^{in} & :\;\theta_{1}>\theta_{2}>\dots>\theta_{n}\\
|\theta_{1},\dots,\theta_{n}\rangle^{out} & :\;\theta_{1}<\theta_{2}<\dots<\theta_{n}\end{cases}
\end{equation}
and states which only differ in the order of rapidities are related
by
\begin{equation}
|\theta_{1},\dots,\theta_{k},\theta_{k+1},\dots,\theta_{n}\rangle=
S(\theta_{k}-\theta_{k+1})
|\theta_{1},\dots,\theta_{k+1},\theta_{k},\dots,\theta_{n}\rangle
\end{equation}
from which the $S$ matrix of any multi-particle scattering process
can be obtained. The normalization of these states is specified by
the following inner product for the one-particle states:\begin{equation}
\langle\theta^{'}|\theta\rangle=2\pi\delta(\theta^{'}-\theta)\end{equation}
The form factors of a local operator $\mathcal{O}(t,x)$ are defined
as
\begin{equation}
F_{mn}^{\mathcal{O}}(\theta_{1}^{'},\dots,\theta_{m}^{'}|\theta_{1},\dots,\theta_{n})_=
\langle\theta_{1}^{'},\dots,\theta_{m}^{'}\vert\mathcal{O}(0,0)\vert\theta_{1},\dots,\theta_{n}\rangle
\label{eq:genff}
\end{equation}
With the help of the crossing relations
\begin{eqnarray}
\nonumber
 &  & F_{mn}^{\mathcal{O}}(\theta_{1}^{'},\dots,\theta_{m}^{'}|\theta_{1},\dots,\theta_{n})= F_{m-1n+1}^{\mathcal{O}}(\theta_{1}^{'},\dots,\theta_{m-1}^{'}|\theta_{m}^{'}+i\pi,\theta_{1},\dots,\theta_{n})\\
 &  & +\sum_{k=1}^{n}2\pi\delta(\theta_{m}^{'}-\theta_{k})\prod_{l=1}^{k-1}S(\theta_{l}-\theta_{k}) F_{m-1n-1}^{\mathcal{O}}(\theta_{1}^{'},\dots,\theta_{m-1}^{'}
|\theta_{1},\dots,\theta_{k-1},\theta_{k+1}\dots,\theta_{n})
\label{eq:ffcrossing}
\end{eqnarray}
all form factors can be expressed in terms of the elementary form
factors
\begin{equation}
F_{n}^{\mathcal{O}}(\theta_{1},\dots,\theta_{n})=\langle0\vert\mathcal{O}(0,0)\vert\theta_{1},\dots,\theta_{n}\rangle
\end{equation}
which satisfy the following equations: 

I. Lorentz transformation: 

\begin{equation}
F_{n}^{\mathcal{O}}(\theta_{1}+\Lambda,\theta_{2}+\Lambda,\dots,\theta_{n}+\Lambda)
=\exp\left(s_\mathcal{O}\Lambda\right)F_{n}^{\mathcal{O}}(\theta_{1},\theta_2,\dots,\theta_{n})\label{eq:shiftaxiom}\end{equation}
where $s_\mathcal{O}$ denotes the Lorentz spin of the operator $\mathcal{O}$.

II. Exchange:

\begin{eqnarray}
 &  & F_{n}^{\mathcal{O}}(\theta_{1},\dots,\theta_{k},\theta_{k+1},\dots,\theta_{n})=\nonumber \\
 &  & \qquad S(\theta_{k}-\theta_{k+1})F_{n}^{\mathcal{O}}(\theta_{1},\dots,\theta_{k+1},\theta_{k},\dots,\theta_{n})
\label{eq:exchangeaxiom}
\end{eqnarray}

III. Cyclic permutation: \begin{equation}
F_{n}^{\mathcal{O}}(\theta_{1}+2i\pi,\theta_{2},\dots,\theta_{n})
=F_{n}^{\mathcal{O}}(\theta_{2},\dots,\theta_{n},\theta_{1})\label{eq:cyclicaxiom}\end{equation}

IV. Kinematical singularity\begin{equation}
-i\mathop{\textrm{Res}}_{\theta=\theta^{'}}F_{n+2}^{\mathcal{O}}(\theta+i\pi,\theta^{'},\theta_{1},\dots,\theta_{n})
=\left(1-\prod_{k=1}^{n}S(\theta'-\theta_{k})\right)F_{n}^{\mathcal{O}}(\theta_{1},\dots,\theta_{n})
\label{eq:kinematicalaxiom}\end{equation}

There is also a further equation related to bound states which we do not need in the sequel. These equations 
are supplemented by the assumption of maximum analyticity (i.e. that the form factors are meromorphic functions 
which only have the singularities prescribed by the equations) and possible further conditions
expressing properties of the particular operator whose form factors are sought.

\subsection{The thermal two-point function}

Let us take a field theory at finite temperature $T$, which can be formulated in a periodic Euclidean time 
\begin{equation}
t\equiv t+R\quad\mathrm{where}\quad R=1/T
\end{equation}
Our aim is to determine the following correlation function
\begin{equation}
\langle\mathcal{O}_{1}(x,t)\mathcal{O}_{2}(0)\rangle^{R}=
\frac{\mathrm{Tr}\left(\mathrm{e}^{-RH}\mathcal{O}_{1}(x,t)\mathcal{O}_{2}(0)\right)}
{\mathrm{Tr}\left(\mathrm{e}^{-RH}\right)}
\label{finiteTcorr}\end{equation}
Naively, one can proceed by inserting two complete sets of states to obtain the spectral representation
\begin{equation}
\langle\mathcal{O}_{1}(x,t)\mathcal{O}_{2}(0)\rangle^{R}=
\frac
{\displaystyle\sum_{m,n}\mathrm{e}^{-(R-t)E_{n}}\mathrm{e}^{-tE_{m}}\mathrm{e}^{ix(P_{n}-P_{m})}
\langle n|\mathcal{O}_1(0)|m\rangle\langle m|\mathcal{O}_2(0)|n\rangle}
{\displaystyle\sum_{n}\mathrm{e}^{-RE_{n}}}
\end{equation}
where $E_m$ and $E_n$ denote the total energies, while $P_m$ and $P_n$ denote the total momenta of the states inserted. Using asymptotic completeness, the sets of  states $\{|n\rangle\}$ and $\{|m\rangle\}$ can be chosen as the basis of 
asymptotic in (or out) states. In this case, the matrix elements appearing in the above formula are just 
the form factors of the operators $\mathcal{O}_1$ and $\mathcal{O}_2$.

However, the above expression is ill-defined because according to the crossing relation (\ref{eq:ffcrossing}) any term in which there is at least one particle with the same rapidities in the states $|n\rangle$ and $|m\rangle$ both matrix elements contains a $\delta$ function term, and so the expression contains squares of $\delta$ functions. Similar divergences occur in the partition function in the denominator. A standard combinatorial consideration of disconnected terms shows that the divergent parts cancel out 
between the numerator and denominator \cite{Balog:1992gf}. The issue is therefore to compute the finite remainder.

The simplest idea is to put the system in a finite spatial volume $L$ with periodic boundary conditions
\begin{equation}
x\equiv x+L
\end{equation}
in which case the expression becomes
\begin{equation}
\langle\mathcal{O}_{1}(x,t)\mathcal{O}_{2}(0)\rangle_{L}^{R}=
\frac{\text{Tr}_{L}\left(\mathrm{e}^{-RH_{L}}\mathcal{O}_{1}(x,t)\mathcal{O}_{2}(0)\right)}{\text{Tr}_{L}\left(\mathrm{e}^{-RH_{L}}\right)}\end{equation}
where $\mathrm{Tr}_{L}$ denotes the trace over the finite-volume states, $H_{L}$ is the Hamiltonian in volume $L$. This expression can be expanded inserting two complete sets of states
\begin{equation}
\text{Tr}_{L}\left(\mathrm{e}^{-RH_{L}}\mathcal{O}_{1}(x,t)\mathcal{O}_{2}(0)\right)=\sum_{m,n}\mathrm{e}^{-RE_{n}(L)}\langle n|\mathcal{O}_1(x,t)|m\rangle_{L}\langle m|\mathcal{O}_2(0)|n\rangle_{L}\label{eq:2ptexp}\end{equation}
where the matrix elements of local operators are also taken in the finite volume system. 

\subsection{Form factors in finite volume}\label{finvolFF}

The next ingredient we need is the description of form factors in finite volume. Previously this was achieved using semi-classical techniques \cite{Smirnov:1998kv,Mussardo:2003ji}, but that is not suitable for our purposes here. We need a formalism that gives the exact quantum form factors to all orders in $L^{-1}$ (i.e. up to corrections that decay exponentially with the volume). The relevant results were derived by us in \cite{Pozsgay:2007kn,Pozsgay:2007gx}, Following our conventions in those papers, the finite volume multi-particle states can be denoted
\begin{equation}
\vert\{ I_{1},\dots,I_{n}\}\rangle_{L}
\end{equation}
where the $I_{k}$ are momentum quantum numbers. We can order the momentum quantum numbers 
in a monotonically decreasing sequence: $I_{1}\geq\dots\geq I_{n}$, which is just a
matter of convention. The corresponding energy levels are determined
by the Bethe-Yang equations
\begin{equation}
\mathrm{e}^{imL\sinh\tilde{\theta}_{k}}\prod_{l\neq k}S(\tilde{\theta}_{k}-\tilde{\theta}_{l})=1
\end{equation}
We define the two-particle phase shift $\delta(\theta)$ by the relation
\begin{equation}
S(\theta)=-\mathrm{e}^{i\delta(\theta)} 
\label{deltadef}\end{equation}
where the $-$ sign ensures that (due to the generic feature $S(0)=-1$ and the bootstrap relation 
$S(\theta)S(-\theta)=1$)\footnote{$S(0)=-1$ is valid in any known integrable model except for the free boson, which we do not consider 
here (as the issues of this paper can be solved trivially), while the other relation is a consequence of the unitarity and Hermitian 
analyticity of the $S$ matrix.}
the phase-shift can be defined as a continuous and odd function of $\theta$. In addition we introduce the following notation
\begin{equation}
\varphi(\theta)=\frac{\partial\delta(\theta)}{\partial\theta} 
\label{phidef}\end{equation}
for the derivative of the phase shift. Using these definitions we can write
\begin{equation}
Q_{k}(\tilde{\theta}_{1},\dots,\tilde{\theta}_{n})=mL\sinh\tilde{\theta}_{k}+\sum_{l\neq k}\delta(\tilde{\theta}_{k}-\tilde{\theta}_{l})=2\pi I_{k}\quad,\quad k=1,\dots,n
\label{eq:betheyang}
\end{equation}
where the quantum numbers $I_k$ take integer/half-integer values for odd/even numbers of particles respectively. 
Eqns. \eqref{eq:betheyang} must be solved with respect to the particle rapidities $\tilde{\theta}_{k}$, where
the energy (relative to the finite volume vacuum state) can be computed as
\begin{equation}
\sum_{k=1}^{n}m\cosh\tilde{\theta}_{k}
\end{equation}
up to corrections which decay exponentially with $L$. 
The density of $n$-particle states in rapidity space can be calculated as 
\begin{equation}
\rho(\theta_{1},\dots,\theta_{n})=\det\mathcal{J}^{(n)}\qquad,\qquad\mathcal{J}_{kl}^{(n)}=\frac{\partial Q_{k}(\theta_{1},\dots,\theta_{n})}{\partial\theta_{l}}\quad,\quad k,l=1,\dots,n\label{eq:byjacobian}
\end{equation}
The finite
volume behaviour of local matrix elements can be given as \cite{Pozsgay:2007kn}
\begin{eqnarray}
\langle\{ I_{1}',\dots,I_{m}'\}\vert\mathcal{O}(0,0)\vert\{ I_{1},\dots,I_{n}\}\rangle_{L}= \frac{F_{m+n}^{\mathcal{O}}(\tilde{\theta}_{m}'+i\pi,\dots,\tilde{\theta}_{1}'+i\pi,\tilde{\theta}_{1},\dots,\tilde{\theta}_{n})}{\sqrt{\rho(\tilde{\theta}_{1},\dots,\tilde{\theta}_{n})\rho(\tilde{\theta}_{1}',\dots,\tilde{\theta}_{m}')}}
+O(\mathrm{e}^{-\mu L})\label{eq:genffrelation}
\end{eqnarray}
where $\tilde{\theta}_{k}$ ($\tilde{\theta}_{k}'$) are the solutions
of the Bethe-Yang equations (\ref{eq:betheyang}) corresponding to
the state with the specified quantum numbers $I_{1},\dots,I_{n}$
($I_{1}',\dots,I_{n}'$) at the given volume $L$. The above relation
is valid provided there are no disconnected terms i.e. the left and
the right states do not contain particles with the same rapidity, i.e. 
the sets $\left\{ \tilde{\theta}_{1},\dots,\tilde{\theta}_{n}\right\} $
and $\left\{ \tilde{\theta}_{1}',\dots,\tilde{\theta}_{m}'\right\} $
are disjoint. The lower limit on the exponent $\mu$ is independent of the operator and the 
states considered (it is related to the bound state pole structure of the infinite volume 
scattering theory).

It is easy to see that in the presence
of nontrivial scattering there are only two cases when exact equality
of (at least some of) the rapidities can occur \cite{Pozsgay:2007gx}:

\begin{enumerate}
\item The two states are identical, i.e. $n=m$ and 
\begin{equation}
\{ I_{1}',\dots,I_{m}'\}  =  \{ I_{1},\dots,I_{n}\}
\end{equation}
in which case the corresponding diagonal matrix element
can be written as a sum over all bipartite divisions of the set of
the $n$ particles involved (including the trivial ones when $A$
is the empty set or the complete set $\{1,\dots,n\}$)
\begin{equation}
\langle\{ I_{1}\dots I_{n}\}|\mathcal{O}|\{ I_{1}\dots I_{n}\}\rangle_{L} 
= 
\frac{\sum_{A\subset\{1,2,\dots n\}}\mathcal{F}(A)_{L}\rho(\{1,\dots,n\}\setminus A)_{L}}
{\rho(\{1,\dots,n\})_{L}}
+O(\mathrm{e}^{-\mu L})
\label{diagff}
\end{equation}
where $|A|$ denotes the cardinal number (number of elements) of the
set $A$ 
\begin{equation}
\rho(\{ k_{1},\dots,k_{r}\})_{L}=\rho(\tilde{\theta}_{k_{1}},\dots,\tilde{\theta}_{k_{r}})\end{equation}
is the $r$-particle Bethe-Yang Jacobi determinant (\ref{eq:byjacobian})
involving only the $r$-element subset $1\leq k_{1}<\dots<k_{r}\leq n$
of the $n$ particles, and
\begin{eqnarray}
\mathcal{F}(\{ k_{1},\dots,k_{r}\})_{L} & = & F_{2r}^{s}(\tilde{\theta}_{k_{1}},\dots,\tilde{\theta}_{k_{r}})
\nonumber\\
F_{2l}^{s}(\theta_{1},\dots,\theta_{l}) & = & \lim_{\epsilon\rightarrow0}F_{2l}^{\mathcal{O}}(\theta_{l}+i\pi+\epsilon,\dots,\theta_{1}+i\pi+\epsilon,\theta_{1},\dots,\theta_{l})\end{eqnarray}
is the so-called symmetric evaluation of diagonal multi-particle matrix
elements.
\item Both states are parity symmetric states in the spin zero sector, i.e.
\begin{eqnarray}
\{ I_{1},\dots,I_{n}\} & \equiv & \{-I_{n},\dots,-I_{1}\}\nonumber\\
\{ I_{1}',\dots,I'_{m}\} & \equiv & \{-I'_{m},\dots,-I'_{1}\}\end{eqnarray}
Furthermore, both states must contain one (or possibly more, in a theory with more than
one species) particle of quantum number $0$, whose rapidity is then
exactly $0$ for any value of the volume $L$ due to the symmetric
assignment of quantum numbers. Writing $m=2k+1$ and $n=2l+1$ and defining
\begin{eqnarray}
 &  & \mathcal{F}_{k,l}(\theta_{1}',\dots,\theta_{k}'|\theta_{1},\dots,\theta_{l})=\nonumber \\
 &  & \lim_{\epsilon\rightarrow0}F_{2k+2l+2}(i\pi+\theta_{1}'+\epsilon,\dots,i\pi+\theta_{k}'+\epsilon,i\pi-\theta_{k}'+\epsilon,\dots,i\pi-\theta_{1}'+\epsilon,\nonumber \\
 &  & i\pi+\epsilon,0,\theta_{1},\dots,\theta_{l},-\theta_{l},\dots,-\theta_{1})\label{eq:oddoddlimitdef}
\end{eqnarray}
the formula for the finite-volume matrix element takes the form
\begin{eqnarray}
&  & 
\langle\{ I_{1}',\dots,I_{k}',0,-I_{k}',\dots,-I_{1}'\}|\mathcal{O}|\{ I_{1},\dots,I_{l},0,-I_{l},\dots,-I_{1}\}\rangle_{L}
\label{eq:oddoddlyrule}\\
&=&
\left(
\rho_{2k+1}(\tilde{\theta}_{1}',\dots,\tilde{\theta}_{k}',0,-\tilde{\theta}_{k}',\dots,-\tilde{\theta}_{1}')
\rho_{2l+1}(\tilde{\theta}_{1},\dots,\tilde{\theta}_{l},0,-\tilde{\theta}_{l},\dots,-\tilde{\theta}_{1})
\right)^{-1/2}
\nonumber \\
&  & \times
\Big[\mathcal{F}_{k,l}(\tilde{\theta}_{1}',\dots,\tilde{\theta}_{k}'|\tilde{\theta}_{1},\dots,\tilde{\theta}_{l})
\nonumber \\
&  &+mL\, F_{2k+2l}(i\pi+\tilde{\theta}_{1}',\dots,i\pi+\tilde{\theta}_{k}', 
i\pi-\tilde{\theta}_{k}',\dots,i\pi-\tilde{\theta}_{1}',\tilde{\theta}_{1},\dots,\tilde{\theta}_{l},-\tilde{\theta}_{l},\dots,-\tilde{\theta}_{1})
\Big]
\nonumber \\
&  &+O(\mathrm{e}^{-\mu L})\nonumber 
\end{eqnarray}
 
\end{enumerate}

\subsection{The form factor expansion using finite volume regularization}\label{subsec:2ptorganization}

Using the finite volume description introduced in subsection \ref{finvolFF} we can write 
\begin{equation}
    \langle\mathcal{O}_{1}(x,t)\mathcal{O}_{2}(0)\rangle_{L}^{R}=
\frac{1}{Z} \sum_{N,M} C_{NM}
\label{Imn_def}
\end{equation}
where
\begin{eqnarray}
\nonumber
C_{NM}&=&\sum_{I_1\dots I_N}\sum_{J_1\dots J_M}
\bra{\{I_1\dots I_N\}}\mathcal{O}_1(0)\ket{\{J_1\dots J_M\}}_L \times\\
&&\bra{\{J_1\dots J_M\}}\mathcal{O}_2(0)\ket{\{I_1\dots I_N\}}_L
e^{i(P_1-P_2)x}e^{-E_1(R-t)}e^{-E_2t}
\label{cnm}
\end{eqnarray}
and $E_{1,2}$ and $P_{1,2}$ are the total energies and momenta of
the multi-particle states $\ket{\{I_1\dots I_N\}}_L$ and $\ket{\{J_1\dots J_M\}}$. The task is to calculate 
the sum in finite volume and then take the limit $L\rightarrow\infty$.

It is easy to see, that the terms $N=0,M=0\dots\infty$ and $N=0\dots\infty,M=0$ add up to
zero-temperature correlation functions
\begin{equation}
\lim_{L\to \infty}  \Big(\sum_{M=0}^\infty C_{0M}\Big)=
  \langle\mathcal{O}_{1}(x,t)\mathcal{O}_{2}(0)\rangle
\quad\quad
\lim_{L\to \infty}   \Big(\sum_{N=0}^\infty C_{N0}\Big)=
  \langle\mathcal{O}_{1}(x,R-t)\mathcal{O}_{2}(0)\rangle
\label{eq:zerotemppart}\end{equation}
In particular
\begin{equation}
  \lim_{L\to\infty} C_{0M}=
\frac{1}{M!}\int \frac{d\theta_1}{2\pi}\dots \frac{d\theta_M}{2\pi}
F^{\mathcal{O}_1}(\theta_1,\dots,\theta_M)
F^{\mathcal{O}_2}(\theta_M,\dots,\theta_1)
e^{-im(\sum_j \sinh\theta_j)x -m(\sum_j \cosh\theta_j)t}
\label{Cdefinition}\end{equation}
and similarly for $C_{N0}$.

Let us introduce two auxiliary variables $u$ and $v$ to
keep track of the orders
of $e^{-mt}$ and $e^{-m(R-t)}$
(at the end both will be set to $1$). Then \eqref{Imn_def}
takes the form 
\begin{equation}
\label{complete1}
\langle\mathcal{O}_{1}(x,t)\mathcal{O}_{2}(0)\rangle_{L}^{R} =\frac{1}{Z}\sum_{N,M} u^N v^M C_{NM}
\end{equation}
We define a similar expansion for the partition function
\begin{equation}
  Z=\sum_N (uv)^N Z_N
\end{equation}
with $Z_N$ denoting the $N$-particle contribution to the partition function. The first few 
terms are given by
\begin{equation}
 Z_0=1\qquad  Z_1=\sum_I e^{-E_IR}
\qquad 
  Z_2=\sum_{I\ne J} e^{-(E_I+E_J)R}  
\end{equation}
The inverse of the partition function is expanded as
\begin{equation}
  Z^{-1}=\sum_N (uv)^N \bar{Z}_N
\end{equation}
where
\begin{equation}
  \bar{Z}_0=1\qquad \bar{Z}_1=-Z_1\qquad \bar{Z}_2=Z_1^2-Z_2
\end{equation}
Putting this together we can rewrite the expansion as 
\begin{equation}
\label{complete}
\langle\mathcal{O}_{1}(x,t)\mathcal{O}_{2}(0)\rangle_{L}^{R} = \sum u^N v^N \tilde{D}_{NM}
\end{equation}
with
\begin{equation}
\label{dnm}
  \tilde{D}_{NM}=\sum_l  C_{N-l,M-l}\bar{Z}_l
\end{equation}
The first few nontrivial terms are given by
\begin{equation}
\begin{split}
  \tilde{D}_{1M}&=C_{1M}-Z_1C_{0,M-1}\\
\tilde D_{2M}&=C_{2M}-Z_1 C_{1,M-1}+(Z_1^2-Z_2)C_{0,M-2}
\end{split}
\label{firstfewDtildes}\end{equation}
In this way we produce a double series expansions in powers of the variables 
$e^{-mt}$ and $e^{-m(R-t)}$. Since these variables are independent, each quantity 
$\tilde D_{NM}$ must have a well-defined $L\to\infty$ limit which we denote as
\begin{equation}
\label{dnm-uj}
    D_{NM}=\lim_{L\to\infty} \tilde D_{NM}
\end{equation}
and we obtain that 
\begin{equation}
  \langle\mathcal{O}_{1}(x,t)\mathcal{O}_{2}(0)\rangle^{R}=\lim_{L\to\infty}
\langle\mathcal{O}_{1}(x,t)\mathcal{O}_{2}(0)\rangle_{L}^{R}=
\sum_{N,M} D_{NM}
\label{eq:2ptwithD}\end{equation}
The reordering of the series in (\ref{complete}) using the coefficients $\tilde D_{NM}$ is also an integral part of Essler 
and Konik's calculation in \cite{Essler:2009zz}; we used a similar reordering for the expansion of the one-point function in powers of $e^{-mR}$ \cite{Pozsgay:2007gx}.

Note that individual terms contributing in (\ref{dnm}) to $\tilde D_{NM}$ 
may contain divergent pieces which scale with positive powers of $L$. Similarly to the considerations for 
the one-point function in \cite{Pozsgay:2007gx}, it turns out that the $N$-particle terms 
which are most singular in the large-$L$ limit carry a factor of $(mL e^{-mR})^N$. 
Since it is also necessary 
that the exponential corrections to the finite volume form factors in eqns. (\ref{eq:genffrelation}) and
(\ref{diagff}) are small, the finite-volume expansion is valid in the domain 
\begin{equation}
  1\ll mL \ll e^{mR}
\label{validitydomain}\end{equation}
For the limit (\ref{dnm-uj}) to exist the positive powers of $L$ must drop out, 
therefore one can understand the $L\to\infty$ limit of the series as an analytic continuation to very large values 
of $L$ outside the domain (\ref{validitydomain}). Eventually, the condition that the coefficients $\tilde D_{nm}$ 
must have a finite large volume limit can be used as a nontrivial check to verify our calculations.

It is evident from \eqref{cnm} that the quantities $D_{NM}$ with $N>M$
can be obtained from those with $N<M$ after a trivial exchange of $t$
and $R-t$. Therefore we will only consider the case $N\le M$.

\section{Warm-up example: the zero-temperature three-point function}

\label{section3}

Before tackling the central issue of the paper, we consider a simpler
problem which allows us to introduce the central ideas without too
many complications. Let us consider the three-point function 
\begin{equation}
\langle0|\mathcal{O}_{1}(t_{1},x_{1})\mathcal{O}_{2}(t_{2},x_{2})\mathcal{O}_{3}(0)|0\rangle\end{equation}
in the Euclidean theory with non-compact time direction (i.e. $T=0$).
Suppose that $t_{1}>t_{2}$ and to shorten the formulae we also omit
the dependence on $x_{1}$ and $x_{2}$ (it can be reintroduced easily).
The spectral decomposition takes the form
\begin{eqnarray}
&&\langle0|\mathcal{O}_{1}(t_{1},x_{1})\mathcal{O}_{2}(t_{2},x_{2})\mathcal{O}(0,0)|0\rangle\nonumber\\
&&=\sum_{m,n}\langle0|\mathcal{O}_{1}(0)|m\rangle\langle m|\mathcal{O}_{2}(0)|n\rangle\langle n|\mathcal{O}_{3}(0)|0\rangle\mathrm{e}^{-E_{m}(t_{1}-t_{2})}\mathrm{e}^{-E_{n}t_{2}}
\end{eqnarray}
What makes this example simpler is that the disconnected terms appear
linearly since they only enter from the $\mathcal{O}_{2}$ matrix
element. Following the example of the three point function, we can
introduce a finite volume regularization
\begin{equation}
\langle0|\mathcal{O}_{1}(t_{1},x_{1})\mathcal{O}_{2}(t_{2},x_{2})\mathcal{O}(0,0)|0\rangle_{L}=\sum_{N,M}C_{NM}^{(3)}\end{equation}
where\begin{eqnarray}
C_{NM}^{(3)} & = & \sum_{I_{1}\dots I_{N}}\sum_{J_{1}\dots J_{M}}\left\langle 0\left|\mathcal{O}_{1}(0)\right|\left\{ I_{1}\dots I_{N}\right\} \right\rangle _{L}\left\langle \left\{ I_{1}\dots I_{N}\right\} \left|\mathcal{O}_{2}(0)\right|\left\{ J_{1}\dots J_{M}\right\} \right\rangle _{L}\\
 & \times & \left\langle \left\{ J_{1}\dots J_{M}\right\} \left|\mathcal{O}_{3}(0)\right|\right\rangle _{L}\mathrm{e}^{-E_{1}(L)(t_{1}-t_{2})}\mathrm{e}^{-E_{2}(L)t_{2}}\end{eqnarray}
There is no denominator $Z$ to supply counter terms for the $L$
dependence, therefore each of these expressions must have a finite
limit as $L\rightarrow\infty$:\begin{equation}
D_{NM}^{(3)}=\lim_{L\to\infty}C_{NM}^{(3)}\end{equation}
and\begin{equation}
\langle0|\mathcal{O}_{1}(t_{1},x_{1})\mathcal{O}_{2}(t_{2},x_{2})\mathcal{O}(0,0)|0\rangle=\sum_{N,M}D_{NM}^{(3)}\end{equation}
Eventually, it is trivial to write down some terms of the expansion:\begin{eqnarray}
D_{0M}^{(3)} & = & \left\langle \mathcal{O}_{1}\right\rangle \frac{1}{M!}\int\frac{d\theta_{1}}{2\pi}\dots\int\frac{d\theta_{M}}{2\pi}F_{M}^{\mathcal{O}_{2}}(\theta_{1},\dots,\theta_{M})F_{M}^{\mathcal{O}_{3}}(\theta_{M}+i\pi,\dots,\theta_{1}+i\pi)
\nonumber\\
&\times&\exp\left(-mt_{2}\sum_{i=1}^{M}\cosh\theta_{i}\right)\nonumber\\
D_{N0}^{(3)} & = & \frac{1}{N!}\int\frac{d\theta_{1}}{2\pi}\dots\int\frac{d\theta_{N}}{2\pi}F_{N}^{\mathcal{O}_{1}}(\theta_{1},\dots,\theta_{N})F_{N}^{\mathcal{O}_{2}}(\theta_{N}+i\pi,\dots,\theta_{1}+i\pi)\nonumber\\
&\times&\exp\left(-m(t_{1}-t_2)\sum_{i=1}^{N}\cosh\theta_{i}\right)\left\langle \mathcal{O}_{3}\right\rangle \end{eqnarray}
and also\begin{equation}
D_{11}^{(3)}=\int\frac{d\theta_{1}}{2\pi}\int\frac{d\theta_{1}'}{2\pi}F_{1}^{\mathcal{O}_{1}}F_{2}^{\mathcal{O}_{2}}(\theta_{1}+i\pi,\theta_{1}')F_{1}^{\mathcal{O}_{3}}\mathrm{e}^{-m(t_{2}-t_{1})\cosh\theta_{1}}\mathrm{e}^{-mt_{1}\cosh\theta_{1}'}\end{equation}
since the two-particle form factor has no kinematical singularities.

\subsection{The contribution $D_{12}^{(3)}$}

The first nontrivial contribution is $D_{12}^{(3)}$, for which the finite
volume expression is \begin{equation}
C_{12}^{(3)}=\sum_{I_{1}}\sum_{J_{1}<J_{2}}\frac{F_{1}^{\mathcal{O}_{1}}F_{3}^{\mathcal{O}_{2}}(\theta_{1}+i\pi,\theta_{1}',\theta_{2}')F_{2}^{\mathcal{O}_{3}}(\theta_{2}',\theta_{1}')}{\rho_{1}(\theta_{1})\rho_{2}(\theta_{1}',\theta_{2}')}\mathrm{e}^{-m(t_{1}-t_{2})\cosh\theta_{1}}\mathrm{e}^{-mt_{2}(\cosh\theta_{1}'+\cosh\theta_{2}')}\end{equation}
where the rapidities satisfy the appropriate Bethe-Yang quantization
relations. We can make a choice whether to perform first the one-particle
or two-particle summation. The latter proceeds by an application of the multi-dimensional 
residue theorem \eqref{multidimresidue} and illustrates one of the central ideas that make the expansion of
the thermal correlator feasible.

\subsubsection{Summing over one-particle states first}

We can substitute the sum over $I_{1}$\begin{equation}
\sum_{I_{1}}\rightarrow\sum_{I_{1}}\oint_{C_{I_{1}}}\frac{d\theta_{1}}{2\pi}\frac{\rho_{1}(\theta_{1})}{\mathrm{e}^{iQ_{1}(\theta_{1})}-1}\end{equation}
where\begin{equation}
Q_{1}(\theta_{1})=mL\sinh\theta_{1}\qquad\rho_{1}(\theta_{1})=Q'(\theta_{1})\end{equation}
and $C_{I_{1}}$ are small closed curves surrounding the solution
of\begin{equation}
Q_{1}(\theta_{1})=2\pi I_{1}\end{equation}
in the complex $\theta_{1}$ plane. Now we open these circles and
join them to obtain the contour \begin{equation}
C=C_{+}+C_{-}\end{equation}
the $C_{+}$ running from $\infty+i\epsilon$ to $-\infty+i\epsilon$
(i.e. backward in $\Re e\,\theta_{1}$) while $C_{-}$ running from
$-\infty-i\epsilon$ to $+\infty-i\epsilon$ (forward in $\Re e\,\theta_{1}$).
However, by this operation we also include the contribution of two
poles at $\theta_{1}=\theta_{1}'$ and $\theta_{1}=\theta_{2}'$ which
must be subtracted. Using (\ref{eq:kinematicalaxiom}), the singularity of the integrand at
$\theta_{1}\sim\theta_{1}'$ can be written as
\begin{equation}
\frac{1}{\theta_{1}-\theta_{1}'}\frac{iF_{1}^{\mathcal{O}_{1}}(1-S(\theta_{1}'-\theta_{2}'))F_{1}^{\mathcal{O}_{2}}F_{2}^{\mathcal{O}_{3}}(\theta_{2}',\theta_{1}')}{\rho_{2}(\theta_{1}',\theta_{2}')\left(\mathrm{e}^{iQ_{1}(\theta_{1}')}-1\right)}\mathrm{e}^{-mt_{1}\cosh\theta_{1}'}\mathrm{e}^{-mt_{2}\cosh\theta_{2}'}\label{eq:3ptc12sing1}
\end{equation}
The quantization relation of the two-particle state can also be written
as \begin{equation}
\mathrm{e}^{imL\sinh\theta_{1}'}S(\theta_{1}'-\theta_{2}')=1\end{equation}
and so we can rewrite (\ref{eq:3ptc12sing1}) as \begin{equation}
\frac{1}{\theta_{1}-\theta_{1}'}\frac{iF_{1}^{\mathcal{O}_{1}}F_{1}^{\mathcal{O}_{2}}F_{2}^{\mathcal{O}_{3}}(\theta_{1}',\theta_{2}')}{\rho_{2}(\theta_{1}',\theta_{2}')}\mathrm{e}^{-mt_{1}\cosh\theta_{1}'}\mathrm{e}^{-mt_{2}\cosh\theta_{2}'}\end{equation}
The contribution of this singularity can be evaluated as 
\begin{eqnarray}
&&\oint_{\theta_{1}'}\frac{d\theta_{1}}{2\pi}\frac{1}{\theta_{1}-\theta_{1}'}\frac{iF_{1}^{\mathcal{O}_{1}}F_{1}^{\mathcal{O}_{2}}F_{2}^{\mathcal{O}_{3}}(\theta_{1}',\theta_{2}')}{\rho_{2}(\theta_{1}',\theta_{2}')}\mathrm{e}^{-mt_{1}\cosh\theta_{1}'}\mathrm{e}^{-mt_{2}\cosh\theta_{2}'}=
\nonumber\\
&&-\frac{F_{1}^{\mathcal{O}_{1}}F_{1}^{\mathcal{O}_{2}}F_{2}^{\mathcal{O}_{3}}(\theta_{1}',\theta_{2}')}{\rho_{2}(\theta_{1}',\theta_{2}')}\mathrm{e}^{-mt_{1}\cosh\theta_{1}'}\mathrm{e}^{-mt_{2}\cosh\theta_{2}'}
\end{eqnarray}
The contribution of the $\theta_{1}=\theta_{2}'$ pole can be obtained
in a similar way. These must be subtracted from the $\theta_{1}$
integral and therefore we obtain\begin{eqnarray}
C_{12}^{(3)} & = & \sum_{J_{1},J_{2}}\Bigg[\oint_{C}\frac{d\theta_{1}}{2\pi}\frac{F_{1}^{\mathcal{O}_{1}}F_{3}^{\mathcal{O}_{2}}(\theta_{1}+i\pi,\theta_{1}',\theta_{2}')F_{2}^{\mathcal{O}_{3}}(\theta_{2}',\theta_{1}')}{\rho_{2}(\theta_{1}',\theta_{2}')\left(\mathrm{e}^{iQ_{1}(\theta_{1})}-1\right)}\mathrm{e}^{-m(t_{1}-t_{2})\cosh\theta_{1}}\mathrm{e}^{-mt_{2}(\cosh\theta_{1}'+\cosh\theta_{2}')}\nonumber\\
 &  & +\frac{F_{1}^{\mathcal{O}_{1}}F_{1}^{\mathcal{O}_{2}}F_{2}^{\mathcal{O}_{3}}(\theta_{1}',\theta_{2}')}{\rho_{2}(\theta_{1}',\theta_{2}')}\mathrm{e}^{-mt_{1}\cosh\theta_{1}'}\mathrm{e}^{-mt_{2}\cosh\theta_{2}'}
\nonumber\\
 &  &
+\frac{F_{1}^{\mathcal{O}_{1}}F_{1}^{\mathcal{O}_{2}}F_{2}^{\mathcal{O}_{3}}(\theta_{2}',\theta_{1}')}{\rho_{2}(\theta_{1}',\theta_{2}')}\mathrm{e}^{-mt_{1}\cosh\theta_{2}'}\mathrm{e}^{-mt_{2}\cosh\theta_{1}'}\Bigg]
\end{eqnarray}
Taking the large $L$ limit, we can substitute the discrete sum with
an integral\begin{equation}
\sum_{J_{1}<J_{2}}\rightarrow\frac{1}{2}\iint\frac{d\theta_{1}'}{2\pi}\frac{d\theta_{2}'}{2\pi}\rho_{2}(\theta_{1}',\theta_{2}')\end{equation}
and so we obtain\begin{eqnarray}
&&C_{12}^{(3)}=\frac{1}{2}\iint\frac{d\theta_{1}'}{2\pi}\frac{d\theta_{2}'}{2\pi}\Bigg[
\nonumber\\ &&-\int_{-\infty}^{\infty}\frac{d\theta_{1}}{2\pi}\frac{F_{1}^{\mathcal{O}_{1}}F_{3}^{\mathcal{O}_{2}}(\theta_{1}+i(\pi+\epsilon),\theta_{1}',\theta_{2}')F_{2}^{\mathcal{O}_{3}}(\theta_{2}',\theta_{1}')}{\mathrm{e}^{iQ_{1}(\theta_{1}+i\epsilon)}-1}
\mathrm{e}^{-m(t_{1}-t_{2})\cosh(\theta_{1}+i\epsilon)-mt_{2}(\cosh\theta_{1}'+\cosh\theta_{2}')}
\nonumber\\
 && + \int_{-\infty}^{\infty}\frac{d\theta_{1}}{2\pi}\frac{F_{1}^{\mathcal{O}_{1}}F_{3}^{\mathcal{O}_{2}}(\theta_{1}+i(\pi-\epsilon),\theta_{1}',\theta_{2}')F_{2}^{\mathcal{O}_{3}}(\theta_{2}',\theta_{1}')}{\mathrm{e}^{iQ_{1}(\theta_{1}-i\epsilon)}-1}
\mathrm{e}^{-m(t_{1}-t_{2})\cosh(\theta_{1}-i\epsilon)-mt_{2}(\cosh\theta_{1}'+\cosh\theta_{2}')}
\nonumber\\
 && + F_{1}^{\mathcal{O}_{1}}F_{1}^{\mathcal{O}_{2}}F_{2}^{\mathcal{O}_{3}}(\theta_{1}',\theta_{2}')\mathrm{e}^{-mt_{1}\cosh\theta_{1}'-mt_{2}\cosh\theta_{2}'}
\nonumber\\
 && +  F_{1}^{\mathcal{O}_{1}}F_{1}^{\mathcal{O}_{2}}F_{2}^{\mathcal{O}_{3}}(\theta_{2}',\theta_{1}')\mathrm{e}^{-mt_{1}\cosh\theta_{2}'-mt_{2}\cosh\theta_{1}'}\Bigg]\end{eqnarray}
Note that
\begin{equation}
iQ_{1}(\theta_{1}\pm i\epsilon)=\mp mL\cosh\theta_{1}\sin\epsilon+imL\sinh\theta_{1}\cos\epsilon
\end{equation}
and therefore the $L\rightarrow\infty$ limit yields
\begin{eqnarray}
&&D_{12}^{(3)}  =  \frac{1}{2}\iint\frac{d\theta_{1}'}{2\pi}\frac{d\theta_{2}'}{2\pi}\Bigg[
 \int_{-\infty}^{\infty}\frac{d\theta_{1}}{2\pi}F_{1}^{\mathcal{O}_{1}}F_{3}^{\mathcal{O}_{2}}(\theta_{1}+i(\pi+\epsilon),\theta_{1}',\theta_{2}')F_{2}^{\mathcal{O}_{3}}(\theta_{2}',\theta_{1}')
\nonumber\\
 &&\times
 \mathrm{e}^{-m(t_{1}-t_{2})\cosh(\theta_{1}+i\epsilon)-mt_{2}(\cosh\theta_{1}'+\cosh\theta_{2}')}
\nonumber\\
 && + F_{1}^{\mathcal{O}_{1}}F_{1}^{\mathcal{O}_{2}}F_{2}^{\mathcal{O}_{3}}(\theta_{1}',\theta_{2}')\mathrm{e}^{-mt_{1}\cosh\theta_{1}'-mt_{2}\cosh\theta_{2}'}
 \nonumber\\
 && + F_{1}^{\mathcal{O}_{1}}F_{1}^{\mathcal{O}_{2}}F_{2}^{\mathcal{O}_{3}}(\theta_{2}',\theta_{1}')\mathrm{e}^{-mt_{1}\cosh\theta_{2}'-mt_{2}\cosh\theta_{1}'}\Bigg]\label{eq:c12res1}\end{eqnarray}

\subsubsection{Evaluating the two-particle summation first}

Using the multi-dimensional residue theorem (\ref{multidimresidue}) we can represent the two-particle
sum as\begin{equation}
\sum_{J_{1}>J_{2}}\frac{1}{\rho_{2}(\theta_{1}',\theta_{2}')}\rightarrow\sum_{J_{1}>J_{2}}\oint\oint_{C_{J_{1}J_{2}}}\frac{d\theta_{1}'}{2\pi}\frac{d\theta_{2}'}{2\pi}\frac{1}{\left(\mathrm{e}^{iQ_{1}(\theta_{1}',\theta_{2}')}+1\right)\left(\mathrm{e}^{iQ_{2}(\theta_{1}',\theta_{2}')}+1\right)}\end{equation}
where $C_{J_{1}J_{2}}$ is a multi-contour (a direct product of two
curves in the variables $\theta_{1}'$ and $\theta_{2}'$) surrounding
the solution of\begin{eqnarray}
Q_{1}(\theta_{1}',\theta_{2}') & = & mL\sinh\theta_{1}'+\delta(\theta_{1}'-\theta_{2}')=2\pi J_{1}\nonumber \\
Q_{2}(\theta_{1}',\theta_{2}') & = & mL\sinh\theta_{2}'+\delta(\theta_{2}'-\theta_{1}')=2\pi J_{2}\label{eq:3ptQ1Q2}\end{eqnarray}
where due to the definition\[
S=-\mathrm{e}^{i\delta}\]
$J_{1}$ and $J_{2}$ take half-integer values. Since the form factors
vanish when any two of their arguments coincide, we can extend the
sum by adding the diagonal\[
\sum_{J_{1}>J_{2}}\rightarrow\frac{1}{2}\sum_{J_{1},J_{2}}\]
and so\begin{equation}
C_{12}^{(3)}=\sum_{I_{1}}\frac{\tilde{C}_{12}(\theta_{1})}{\rho_{1}(\theta_{1})}\end{equation}
where \begin{eqnarray}
\tilde{C}_{12}(\theta_{1})&=&\frac{1}{2}\sum_{J_{1},J_{2}}\oint\oint_{C_{J_{1}J_{2}}}\frac{d\theta_{1}'}{2\pi}\frac{d\theta_{2}'}{2\pi}\frac{F_{1}^{\mathcal{O}_{1}}F_{3}^{\mathcal{O}_{2}}(\theta_{1}+i\pi,\theta_{1}',\theta_{2}')F_{2}^{\mathcal{O}_{3}}(\theta_{2}',\theta_{1}')}{\left(\mathrm{e}^{iQ_{1}(\theta_{1}',\theta_{2}')}+1\right)\left(\mathrm{e}^{iQ_{2}(\theta_{1}',\theta_{2}')}+1\right)}
\nonumber\\
&&\times\mathrm{e}^{-m(t_{1}-t_{2})\cosh\theta_{1}}\mathrm{e}^{-mt_{2}(\cosh\theta_{1}'+\cosh\theta_{2}')}
\end{eqnarray}
Now we open the contours to surround the whole of the real $\theta_{1}'$
and $\theta_{2}'$ axes (but close enough so as to avoid all singularities
of the $S$ matrix). Just as before it is necessary to subtract the
contributions of any singularities encountered in the process. There
are two classes of such singularities:
\begin{itemize}
\item $\theta_{1}'=\theta_{1}$ and $\mathrm{e}^{iQ_{2}}+1=0$
\item $\theta_{2}'=\theta_{1}$ and $\mathrm{e}^{iQ_{1}}+1=0$
\end{itemize}
There are no triple singularities because $\theta_{1}$ satisfies\begin{equation}
mL\sinh\theta_{1}=2\pi I_{1}\qquad I_{1}\in\mathbb{Z}\end{equation}
and it is impossible for the three Bethe-Yang conditions $Q_{1}$,
$Q_{2}$ and $Q_{3}$ to be satisfied simultaneously. As before, it
is enough to evaluate the first case; the second can be obtained by
swapping $\theta_{1}'$ and $\theta_{2}'$. In the large $L$ limit\begin{equation}
\mathrm{e}^{iQ_{1,2}(\theta_{1}'\pm i\epsilon_1,\theta_{2}'\pm i\epsilon_2)}\rightarrow\begin{cases}
0 & +\mbox{ sign}\\
\infty & -\mbox{ sign}\end{cases}\end{equation}
therefore we obtain
\begin{eqnarray}
&&\tilde{C}_{12}(\theta_{1})  =  
\nonumber\\
&&\frac{1}{2}\iint_{C_{++}}\frac{d\theta_{1}'}{2\pi}\frac{d\theta_{2}'}{2\pi}
F_{1}^{\mathcal{O}_{1}}F_{3}^{\mathcal{O}_{2}}(\theta_{1}+i\pi,\theta_{1}',\theta_{2}')
F_{2}^{\mathcal{O}_{3}}(\theta_{2}',\theta_{1}')
\mathrm{e}^{-m(t_{1}-t_{2})\cosh\theta_{1}-mt_{2}(\cosh(\theta_{1}')+\cosh(\theta_{2}'))}\nonumber \\
 &  & -\frac{1}{2}\Bigg\{\sum_{J_{2}\in\mathbb{Z}+1/2}i^{2}
\mathop{\mbox{Res}}_{{\theta_{1}'=\theta_{1}\atop Q_{2}=J_{2}}}
\Bigg[
\frac{F_{1}^{\mathcal{O}_{1}}F_{3}^{\mathcal{O}_{2}}(\theta_{1}+i\pi,\theta_{1}',\theta_{2}')
F_{2}^{\mathcal{O}_{3}}(\theta_{2}',\theta_{1}')}
{\left(\mathrm{e}^{iQ_{1}(\theta_{1}',\theta_{2}')}+1\right)
\left(\mathrm{e}^{iQ_{2}(\theta_{1}',\theta_{2}')}+1\right)}
\nonumber \\
&  & \times
\mathrm{e}^{-m(t_{1}-t_{2})\cosh\theta_{1}-mt_{2}(\cosh\theta_{1}'+\cosh\theta_{2}')}
\Bigg]+\left(\theta_{1}'\leftrightarrow\theta_{2}'\right)\Bigg\}\end{eqnarray}
where 
$C_{++}$ denotes the part of the two-particle multi-contour that survives in the $L\to\infty$ limit; 
it corresponds to an integration parallel to the real axes in $\theta_{1}'$ and $\theta_{2}'$ with a shift in the positive imaginary direction, i.e.
\begin{equation}
\iint_{C_{++}}\frac{d\theta_{1}'}{2\pi}\frac{d\theta_{2}'}{2\pi}f(\theta_{1}',\theta_{2}')=
\int_{\mathbb{R}}\frac{d\theta_{1}'}{2\pi}\int_{\mathbb{R}}\frac{d\theta_{2}'}{2\pi}
f(\theta_{1}'+i\epsilon_1,\theta_{2}'+i\epsilon_2)
\label{Cppcontourdef}
\end{equation}
Recalling (\ref{eq:3ptQ1Q2}) we get
\begin{eqnarray}
 &  & \mathop{\mbox{Res}}_{{\theta_{1}'=\theta_{1}\atop Q_{3}=I_{3}}}
\frac{F_{3}^{\mathcal{O}_{2}}(\theta_{1}+i\pi,\theta_{1}',\theta_{2}')
F_{2}^{\mathcal{O}_{3}}(\theta_{2}',\theta_{1}')\mathrm{e}^{-m(t_{1}-t_{2})\cosh\theta_{1}-mt_{2}(\cosh\theta_{1}'+\cosh\theta_{2}')}}
{\left(\mathrm{e}^{iQ_{1}(\theta_{1}',\theta_{2}')}+1\right)\left(\mathrm{e}^{iQ_{2}(\theta_{1}',\theta_{2}')}+1\right)}\nonumber \\
 & = & \frac{-i\left(1-S(\theta_{1}-\theta_{2}')\right)F_{1}^{\mathcal{O}_{2}}F_{2}^{\mathcal{O}_{3}}(\theta_{2}',\theta_{1})\mathrm{e}^{-mt_{1}\cosh\theta_{1}-mt_{2}\cosh\theta_{2}'}}{\left(1-S(\theta_{1}-\theta_{2}')\right)(-i)\left(mL\cosh\theta_{2}'+\varphi(\theta_{2}'-\theta_{1})\right)}\end{eqnarray}
Substituting these into the expression for $C_{12}^{(3)}$ and taking $L\rightarrow\infty$
we obtain
\begin{eqnarray}
D_{12}^{(3)} & = & \frac{1}{2}\iint_{C_{++}}\frac{d\theta_{1}'}{2\pi}\frac{d\theta_{2}'}{2\pi}
\Bigg[\int_{-\infty}^{\infty}\frac{d\theta_{1}}{2\pi}F_{1}^{\mathcal{O}_{3}}
F_{3}^{\mathcal{O}_{2}}(\theta_{1}+i\pi,\theta_{1}',\theta_{2}')
F_{2}^{\mathcal{O}_{1}}(\theta_{2}',\theta_{1}')\\
 & \times & \mathrm{e}^{-m(t_{1}-t_{2})\cosh\theta_{1}}\mathrm{e}^{-mt_{1}(\cosh(\theta_{1}')+\cosh(\theta_{2}'))}\nonumber \\
 & + & F_{1}^{\mathcal{O}_{1}}F_{1}^{\mathcal{O}_{2}}F_{2}^{\mathcal{O}_{3}}(\theta_{1}',\theta_{2}')\mathrm{e}^{-mt_{2}\cosh\theta_{1}'}\mathrm{e}^{-mt_{1}\cosh\theta_{2}'}
\nonumber\\
 & + & F_{1}^{\mathcal{O}_{1}}F_{1}^{\mathcal{O}_{2}}F_{2}^{\mathcal{O}_{3}}(\theta_{2}',\theta_{1}')\mathrm{e}^{-mt_{2}\cosh\theta_{2}'}\mathrm{e}^{-mt_{1}\cosh\theta_{1}'}\Bigg]\nonumber \end{eqnarray}
After substituting $\theta_{1,2}'\rightarrow-\theta_{1,2}'$ and using
\[
F_{2}^{\mathcal{O}_{3}}(-\theta_{1}',-\theta_{2}')=F_{2}^{\mathcal{O}_{1}}(\theta_{2},'\theta_{1}')\]
we can make a combined shift of the three contours in the triple integral term to obtain 
\begin{eqnarray}
D_{12}^{(3)} & = & \frac{1}{2}\iint\frac{d\theta_{1}'}{2\pi}\frac{d\theta_{2}'}{2\pi}\Bigg[\int_{-\infty}^{\infty}\frac{d\theta_{1}}{2\pi}F_{1}^{\mathcal{O}_{1}}F_{3}^{\mathcal{O}_{2}}(\theta_{1}+i(\pi-\epsilon),\theta_{1}',\theta_{2}')F_{2}^{\mathcal{O}_{3}}(\theta_{2}',\theta_{1}')\label{eq:c12res2}\\
 & \times & \mathrm{e}^{-m(t_{1}-t_{2})\cosh(\theta_{1}-i\epsilon)}\mathrm{e}^{-mt_{1}(\cosh\theta_{1}'+\cosh\theta_{2}')}\nonumber \\
 & + & F_{1}^{\mathcal{O}_{1}}F_{1}^{\mathcal{O}_{2}}F_{2}^{\mathcal{O}_{3}}(\theta_{1}',\theta_{2}')\mathrm{e}^{-mt_{2}\cosh\theta_{1}'}\mathrm{e}^{-mt_{1}\cosh\theta_{2}'}
\nonumber\\
 & + & F_{1}^{\mathcal{O}_{1}}F_{1}^{\mathcal{O}_{2}}F_{2}^{\mathcal{O}_{3}}(\theta_{2}',\theta_{1}')\mathrm{e}^{-mt_{2}\cosh\theta_{2}'}\mathrm{e}^{-mt_{1}\cosh\theta_{1}'}\Bigg]\nonumber \end{eqnarray}
It can easily be shown that this expression agrees with (\ref{eq:c12res1});
the difference due to the $\epsilon\rightarrow-\epsilon$ change drops
out: \begin{eqnarray}
 &  & \int_{-\infty}^{\infty}\frac{d\theta_{1}}{2\pi}F_{1}^{\mathcal{O}_{1}}F_{3}^{\mathcal{O}_{2}}(\theta_{1}+i(\pi+\epsilon),\theta_{1}',\theta_{2}')F_{2}^{\mathcal{O}_{3}}(\theta_{2}',\theta_{1}')\mathrm{e}^{-m(t_{2}-t_{1})\cosh(\theta_{1}+i\epsilon)}\mathrm{e}^{-mt_{1}(\cosh\theta_{1}'+\cosh\theta_{2}')}-\nonumber \\
 & - & \int_{-\infty}^{\infty}\frac{d\theta_{1}}{2\pi}F_{1}^{\mathcal{O}_{1}}F_{3}^{\mathcal{O}_{2}}(\theta_{1}+i(\pi-\epsilon),\theta_{1}',\theta_{2}')F_{2}^{\mathcal{O}_{3}}(\theta_{2}',\theta_{1}')\mathrm{e}^{-m(t_{2}-t_{1})\cosh(\theta_{1}-i\epsilon)}\mathrm{e}^{-mt_{1}(\cosh\theta_{1}'+\cosh\theta_{2}')}\nonumber \\
 & = & i\left(\mathop{\mbox{Res}}_{\theta_{1}=\theta_{1}'}+\mathop{\mbox{Res}}_{\theta_{1}=\theta_{2}'}\right)F_{1}^{\mathcal{O}_{1}}F_{3}^{\mathcal{O}_{2}}(\theta_{1}+i\pi,\theta_{1}',\theta_{2}')F_{2}^{\mathcal{O}_{3}}(\theta_{2}',\theta_{1}')\mathrm{e}^{-m(t_{2}-t_{1})\cosh\theta_{1}}\mathrm{e}^{-mt_{1}(\cosh\theta_{1}'+\cosh\theta_{2}')}\nonumber \\
 & = & -\left(F_{1}^{\mathcal{O}_{1}}F_{1}^{\mathcal{O}_{2}}\left(F_{2}^{\mathcal{O}_{1}}(\theta_{2}',\theta_{1}')-F_{2}^{\mathcal{O}_{1}}(\theta_{1}',\theta_{2}')\right)\mathrm{e}^{-mt_{2}\cosh\theta_{1}'}\mathrm{e}^{-mt_{1}\cosh\theta_{2}'}-\left(\theta_{1}'\leftrightarrow\theta_{2}'\right)\right)\end{eqnarray}
and the integral of the last expression over $\theta_{1,2}'$ vanishes
by symmetry.

\subsection{$D_{22}^{(3)}$}

The new aspect in this case is that the contribution must be split
into two parts: one with the two two-particle states being different
and the {}``diagonal'' when these states are the same. Using the
finite volume form factor formulae (\ref{eq:genffrelation}) and (\ref{diagff}) we can write\begin{eqnarray}
C_{22}^{(3)} & = & \sum_{\{I_{1},I_{2}\}\neq\{J_{1},J_{2}\}}
\frac{F_{2}^{\mathcal{O}_{1}}(\theta_{1},\theta_{2})
F_{4}^{\mathcal{O}_{2}}(\theta_{2}+i\pi,\theta_{1}+i\pi,\theta_{3},\theta_{4})
F_{2}^{\mathcal{O}_{3}}(\theta_{2}',\theta_{1}')}
{\rho_{2}(\theta_{1},\theta_{2})\rho_{2}(\theta_{1}',\theta_{2}')}\nonumber \\
 & \times & 
\mathrm{e}^{-m(t_{1}-t_{2})(\cosh\theta_{1}+\cosh\theta_{2})}
\mathrm{e}^{-mt_{2}(\cosh\theta_{1}'+\cosh\theta_{2}')}\nonumber \\
 & + & \sum_{\{I_{1},I_{2}\}}
\frac{
F_{4s}^{\mathcal{O}_{2}}(\theta_{1},\theta_{2})+F_{2s}^{\mathcal{O}_{2}}\left(\rho_{1}(\theta_{1})
+\rho_{1}(\theta_{2})\right)+\left\langle \mathcal{O}_{2}\right\rangle \rho_{2}(\theta_{1},\theta_{2})}
{\rho_{2}(\theta_{1},\theta_{2})^{2}}\nonumber \\
 & \times & F_{2}^{\mathcal{O}_{1}}(\theta_{1},\theta_{2})F_{2}^{\mathcal{O}_{3}}(\theta_{2},\theta_{1})\mathrm{e}^{-mt_{1}(\cosh\theta_{1}+\cosh\theta_{2})}\end{eqnarray}
The diagonal part can be written as 
\begin{eqnarray}
C_{22}^{(3)\mathrm{diag}} & = & 
\sum_{\{I_{1},I_{2}\}}
\frac
{F_{4,s}^{\mathcal{O}_{2}}(\theta_{1},\theta_{2})+F_{2,s}^{\mathcal{O}_{2}}\left(\rho_{1}(\theta_{1})+\rho_{1}(\theta_{2})\right)+\left\langle \mathcal{O}_{2}\right\rangle \rho_{2}(\theta_{1},\theta_{2})}
{\rho_{2}(\theta_{1},\theta_{2})^{2}}\nonumber \\
 & \times & F_{2}^{\mathcal{O}_{1}}(\theta_{1},\theta_{2})F_{2}^{\mathcal{O}_{3}}(\theta_{2},\theta_{1})
\mathrm{e}^{-mt_{1}(\cosh\theta_{1}+\cosh\theta_{2})}\nonumber \\
 & {\longrightarrow\atop L\rightarrow\infty} & \frac{1}{2}\iint\frac{d\theta_{1}}{2\pi}\frac{d\theta_{2}}{2\pi}F_{2}^{\mathcal{O}_{1}}(\theta_{1},\theta_{2})\left\langle \mathcal{O}_{2}\right\rangle F_{2}^{\mathcal{O}_{3}}(\theta_{2},\theta_{1})\mathrm{e}^{-mt_{1}(\cosh\theta_{1}+\cosh\theta_{2})}\end{eqnarray}
For the non-diagonal part we need to evaluate
\begin{equation}
\tilde{C}(\theta_{1}',\theta_{2}')  =  \frac{1}{2}\sum_{I_{1},I_{2}}
\frac{F_{4}^{\mathcal{O}_{2}}(\theta_{2}+i\pi,\theta_{1}+i\pi,\theta_{1}',\theta_{2}')}
{\rho_{2}(\theta_{1},\theta_{2})}K_{t_{1},t_{2}}(\theta_{1},\theta_{2},\theta_{1}',\theta_{2}')
\end{equation}
where
\begin{equation}
K_{t_{1},t_{2}}(\theta_{1},\theta_{2},\theta_{1}',\theta_{2}')=
F_{2}^{\mathcal{O}_{3}}(\theta_{1},\theta_{2})
F_{2}^{\mathcal{O}_{1}}(\theta_{1}',\theta_{2}')
\mathrm{e}^{-m(t_{1}-t_{2})(\cosh\theta_{1}+\cosh\theta_{2})}
\mathrm{e}^{-mt_{2}(\cosh\theta_{1}'+\cosh\theta_{2}')} 
\end{equation}
(again we extended the $I_{1}<I_{2}$ summation by symmetry and included
the diagonal $I_{1}=I_{2}$ where the form factors vanish). Using
the residue trick it can be represented as \begin{equation}
\frac{1}{2}\sum_{I_{1},I_{2}}\oint\oint_{C_{I_{1}I_{2}}}\frac{d\theta_{1}}{2\pi}\frac{d\theta_{2}}{2\pi}\frac{F_{4}^{\mathcal{O}_{2}}(\theta_{2}+i\pi,\theta_{1}+i\pi,\theta_{1}',\theta_{2}')}{\left(\mathrm{e}^{iQ_{1}(\theta_{1},\theta_{2})}+1\right)\left(\mathrm{e}^{iQ_{2}(\theta_{1},\theta_{2})}+1\right)}K_{t_{1},t_{2}}(\theta_{1},\theta_{2},\theta_{1}',\theta_{2}')\end{equation}
where \begin{eqnarray}
Q_{1}(\theta_{1},\theta_{2}) & = & mL\sinh\theta_{1}+\delta(\theta_{1}-\theta_{2})\nonumber \\
Q_{2}(\theta_{1},\theta_{2}) & = & mL\sinh\theta_{2}+\delta(\theta_{2}-\theta_{1})\end{eqnarray}
To open the contour we need to find the singularities that do not
result as solutions of $Q_{1,2}=2\pi I_{1,2}$. There are the following
possibilities:
\begin{itemize}
\item $Q_{2}=2\pi I_{2}$ and $\theta_{1}=\theta_{1}'$ or $\theta_{1}=\theta_{2}'$
\item $Q_{1}=2\pi I_{1}$ and $\theta_{2}=\theta_{1}'$ or $\theta_{2}=\theta_{2}'$
\item $\theta_{1}=\theta_{1}'$ and $\theta_{2}=\theta_{2}'$ or $\theta_{2}=\theta_{1}'$
and $\theta_{1}=\theta_{2}'$. Albeit the form factor is regular at
this point, the denominator has a double zero due to the quantization
condition satisfied by $\theta_{1}'$ and $\theta_{2}'$:\begin{eqnarray}
mL\sinh\theta_{1}'+\delta(\theta_{1}'-\theta_{2}') & = & 2\pi J_{1}\nonumber \\
mL\sinh\theta_{2}'+\delta(\theta_{2}'-\theta_{1}') & = & 2\pi J_{2}\label{eq:3ptprimedby}\end{eqnarray}
 These were excluded and their contribution calculated in the diagonal
part $C_{22}^\mathrm{diag}$.
\end{itemize}
As an example we consider the contribution from the singularity $Q_{2}=2\pi I_{2}$
and $\theta_{1}=\theta_{1}'$. Using (\ref{eq:3ptprimedby}) we can
evaluate\begin{equation}
\mathrm{e}^{iQ_{1}(\theta_{1}',\theta_{2})}=-\mathrm{e}^{imL\sinh\theta_{1}'}S(\theta_{1}'-\theta_{2})=-S(\theta_{2}'-\theta_{1}')S(\theta_{1}'-\theta_{2})\end{equation}
and the appropriate residue takes the form

\begin{eqnarray}
 &  & \mathop{\mathrm{Res}}_{{Q_{2}=2\pi I_{2}\atop \theta_{1}=\theta_{1}'}}\frac{1}{2}\frac{F_{4}^{\mathcal{O}_{2}}(\theta_{2}+i\pi,\theta_{1}+i\pi,\theta_{1}',\theta_{2}')}{\left(\mathrm{e}^{iQ_{1}(\theta_{1},\theta_{2})}+1\right)\left(\mathrm{e}^{iQ_{2}(\theta_{1},\theta_{2})}+1\right)}K_{t_{1},t_{2}}(\theta_{1},\theta_{2},\theta_{1}',\theta_{2}')=\nonumber \\
 &  & \frac{1}{2}i^{2}\frac{i(1-S(\theta_{2}-\theta_{1}')S(\theta_{1}'-\theta_{2}'))F_{2}^{\mathcal{O}_{2}}(\theta_{2}+i\pi,\theta_{2}')}{(1-S(\theta_{2}'-\theta_{1}')S(\theta_{1}'-\theta_{2}))(-i)(mL\cosh\theta_{2}+\varphi(\theta_{2}-\theta_{1}'))}K_{t_{1},t_{2}}(\theta_{1}',\theta_{2},\theta_{1}',\theta_{2}')\nonumber \\
 & = & -\frac{1}{2}\frac{F_{2}^{\mathcal{O}_{2}}(\theta_{2}+i\pi,\theta_{2}')}{(mL\cosh\theta_{2}+\varphi(\theta_{2}-\theta_{1}'))}K_{t_{1},t_{2}}(\theta_{2},\theta_{1}',\theta_{2}',\theta_{1}')\end{eqnarray}
When we sum over $I_{2}$ we must exclude $I_{2}=J_{2}$ i.e. the
term $\theta_{1}=\theta_{1}'$ and $\theta_{2}=\theta_{2}'$ (this
singularity was taken into account in $C_{22}^{(3)\mathrm{diag}}$). However, its contribution
to the $I_{2}$ sum is \begin{equation}
-\frac{1}{2}\frac{F_{2s}^{\mathcal{O}_{2}}}{(mL\cosh\theta_{2}'+\varphi(\theta_{2}'-\theta_{1}'))}K_{t_{1},t_{2}}(\theta_{2}',\theta_{1}',\theta_{2}',\theta_{1}')\end{equation}
which vanishes when $L\rightarrow\infty$. 

One can calculate the contribution of all other singularities in a
similar way. Taking $L\rightarrow\infty$ the final result is
\begin{eqnarray}
D_{22}^{(3)} & = & \frac{1}{2}\iint_{C_{++}}\frac{d\theta_{1}}{2\pi}\frac{d\theta_{2}}{2\pi}
\iint\frac{d\theta_{1}'}{2\pi}\frac{d\theta_{2}'}{2\pi}
F_{4}^{\mathcal{O}_{2}}(\theta_{2}+i\pi,\theta_{1}+i\pi,\theta_{1}',\theta_{2}')
F_{2}^{\mathcal{O}_{1}}(\theta_{1},\theta_{2})F_{2}^{\mathcal{O}_{3}}(\theta_{2}',\theta_{1}')
\nonumber\\
 &  & \times
\mathrm{e}^{-m(t_{1}-t_{2})(\cosh\theta_{1}+\cosh\theta_{2})}
\mathrm{e}^{-mt_{2}(\cosh\theta_{1}'+\cosh\theta_{2}')}
\nonumber\\
 & + & \int\frac{d\theta_{2}}{2\pi}\iint\frac{d\theta_{1}'}{2\pi}\frac{d\theta_{2}'}{2\pi}F_{2}^{\mathcal{O}_{1}}(\theta_{2},\theta_{1}')F_{2}^{\mathcal{O}_{2}}(\theta_{2}+i\pi,\theta_{2}')F_{2}^{\mathcal{O}_{3}}(\theta_{2}',\theta_{1}')
\nonumber\\
&&\times \mathrm{e}^{-m(t_{1}-t_{2})\cosh\theta_{2}}\mathrm{e}^{-mt_{1}\cosh\theta_{1}'}\mathrm{e}^{-mt_{2}\cosh\theta_{2}'}
\nonumber\\
&-&\frac{1}{2}\iint\frac{d\theta_{1}}{2\pi}\frac{d\theta_{2}}{2\pi}F_{2}^{\mathcal{O}_{1}}(\theta_{1},\theta_{2})\left\langle \mathcal{O}_{2}\right\rangle F_{2}^{\mathcal{O}_{3}}(\theta_{2},\theta_{1})\mathrm{e}^{-mt_{1}(\cosh\theta_{1}+\cosh\theta_{2})}
\label{3ptD22final}\end{eqnarray}
where the $(\theta_1, \theta_2)$ contour $C_{++}$ is specified in (\ref{Cppcontourdef})

\section{Evaluating the thermal correlator}\label{lowT_2p}

\label{section4}

Now we show how evaluate the series (\ref{eq:2ptwithD}). Since according to (\ref{eq:zerotemppart}) the contributions $D_{0N}$ and $D_{M0}$ 
are identical to terms contributing to the zero-temperature two-point function, the first 
nontrivial temperature correction is given by $D_{11}$, which is
evaluated in the next subsection. The contributions $D_{12}$, $D_{1n}$
for arbitrary $n>2$, and $D_{22}$ are calculated in subsections \ref{sec:D12},
\ref{sec:D1n} and \ref{sec:D22} respectively. The final expressions (which constitute the
main results of this work) are given by equations
\eqref{final_D11}, \eqref{D12}, \eqref{D1n} and \eqref{D22final}, respectively.

\subsection{The $D_{11}$ correction}

According to (\ref{firstfewDtildes}) and \eqref{dnm-uj}
 
\begin{eqnarray}
D_{11}=\mathop{\mathrm{lim}}_{L\to\infty}\tilde{D}_{11}\nonumber\\
\tilde{D}_{11}=C_{11}-Z_1 C_{00}\label{D11def}
\end{eqnarray}
where
\begin{equation}
C_{00}=\left\langle \mathcal{O}_1 \right\rangle \left\langle \mathcal{O}_2 \right\rangle
\end{equation}
and
\begin{equation}
\label{I11}
  C_{11}=\sum_{I,J} 
\bra{\{I\}}  \mathcal{O}_1(0)\ket{\{J\}}_L\
\bra{\{J\}}\mathcal{O}_2(0)\ket{\{I\}}_L \
e^{i(p_1-p_2)x}e^{-E_1(R-t)}e^{-E_2t}
\end{equation}
where $E_1$, $E_2$ and $p_1$, $p_2$ are the finite size energies and momenta 
of the one-particle states. Using the Bethe-Yang quantization conditions 
(\ref{eq:betheyang}) we have 
\begin{equation}
 mL\sinh\theta=2\pi I\quad ,\quad mL\sinh\theta'=2\pi J
\end{equation}
and
\begin{eqnarray*}
&&E_1=m \cosh\theta \quad ,\quad p_1=m \sinh\theta
\\
&&E_2=m \cosh\theta' \quad ,\quad p_2=m \sinh\theta'
\end{eqnarray*}
According to (\ref{eq:genffrelation}) and (\ref{diagff}), 
the two-particle matrix elements are given by
\begin{eqnarray}
  \bra{\{I\}}  \mathcal{O}_1(0)\ket{\{J\}}_L&=&
\frac{F_2^{\mathcal{O}_1}(\theta+i\pi,\theta')}
{\sqrt{\rho_1(\theta)\rho_1(\theta')}}+\delta_{IJ}
\vev{\mathcal{O}_1}\\
  \bra{\{J\}}  \mathcal{O}_2(0)\ket{\{I\}}_L&=&
\frac{F_2^{\mathcal{O}_2}(\theta'+i\pi,\theta)}
{\sqrt{\rho_1(\theta)\rho_1(\theta')}}+\delta_{IJ}
\vev{\mathcal{O}_2}
\end{eqnarray}
Substituting the above formulas into \eqref{I11} one obtains
\begin{eqnarray}
\label{I11b}
\nonumber
  C_{11}=\sum_{I,J} 
\frac{F_2^{\mathcal{O}_1}(\theta+i\pi,\theta')F_2^{\mathcal{O}_2}(\theta'+i\pi,\theta)}
{\rho_1(\theta)\rho_1(\theta')}e^{i(p_1-p_2)x}e^{-E_1(R-t)}e^{-E_2t}\\
\nonumber
+\vev{\mathcal{O}_1} \sum_J 
\frac{F_2^{\mathcal{O}_2}(\theta'+i\pi,\theta')}{\rho_1(\theta')}
e^{-E_2R}+
\vev{\mathcal{O}_2} \sum_J 
\frac{F_2^{\mathcal{O}_1}(\theta+i\pi,\theta)}{\rho_1(\theta)}
e^{-E_1R}\\
+\sum_{I}\vev{\mathcal{O}_1}\vev{\mathcal{O}_2} e^{-E_1R}
\end{eqnarray}
The last term in \eqref{I11b} is $O(L)$, but it is canceled in $\tilde{D}_11$ by the term 
$Z_1C_{00}$. All the other terms have a finite limit as $L\to\infty$ which can be written 
in the form 
\begin{eqnarray}
D_{11}&=&\int \frac{d\theta}{2\pi}\int \frac{d\theta'}{2\pi}
F_2^{\mathcal{O}_1}(\theta+i\pi,\theta')F_2^{\mathcal{O}_2}(\theta'+i\pi,\theta)
e^{i(\sinh\theta-\sinh\theta')mx}
e^{-m(R-t)\cosh\theta-mt\cosh\theta'}
\nonumber\\
&+&(\vev{\mathcal{O}_1}F_{2s}^{\mathcal{O}_2}+\vev{\mathcal{O}_2}F_{2s}^{\mathcal{O}_1})
\int\frac{d\theta}{2\pi} e^{-mR\cosh\theta}
\label{final_D11}\end{eqnarray}
Note that according to (\ref{eq:kinematicalaxiom}) the two-particle form factor does not 
have kinematical singularities. The result \eqref{final_D11} was first
obtained in \cite{Pozsgay:2009pv}.

\subsection{More than just a warm-up: $D_{12}$}

\label{sec:D12}

According to (\ref{firstfewDtildes}) and \eqref{dnm-uj}
\begin{equation*}
  D_{12}=\lim_{L\to\infty}\big(C_{12}-Z_1 C_{01}\big)
\end{equation*}
where
\begin{equation}
\label{C12kiind}
\begin{split}
C_{12}&=\sum_{I}\sum_{J_1J_2}
\bra{\{I\}}\mathcal{O}_1(0)\ket{\{J_1,J_2\}}_L \times
\bra{\{J_1,J_2\}}\mathcal{O}_2(0)\ket{\{I\}}_L
\mathrm{e}^{i(P_1-P_2)x}\mathrm{e}^{-E_1(R-t)}\mathrm{e}^{-E_2t}\\
&=\sum_{I} \sum_{J_1J_2}
\frac{F_3^{\mathcal{O}_1}(\theta_1+i\pi,\theta'_1,\theta'_2)
F_3^{\mathcal{O}_2}(\theta_1+i\pi,\theta'_2,\theta'_1)}
{\rho_1(\theta_1)\rho_2(\theta'_1,\theta'_2)}
K_{t,x}^{(R)}(\theta_{1},\theta_{1}',\theta_{2}')
\end{split}
\end{equation}
\begin{eqnarray}
K_{t,x}^{(R)}(\theta_{1},\theta_{1}',\theta_{2}') & = &
\mathrm{e}^{imx(\sinh\theta_{1}-\sinh\theta_{1}'-\sinh\theta_{2}')}
\mathrm{e}^{-m(R-t)\cosh\theta_{1}}\mathrm{e}^{-mt(\cosh\theta_{1}'+\cosh\theta_{2}')}
\end{eqnarray}
and
\begin{equation*}
  Z_1 C_{01}=\left(\sum_I \mathrm{e}^{-ER}\right) 
 \left(\int \frac{d\theta}{2\pi}
F_1^{\mathcal{O}_1}F_1^{\mathcal{O}_2}
\mathrm{e}^{-im x \sih\theta -m t \coh\theta}\right)
\end{equation*}

\subsubsection{First summation: one-particle states}

\label{C21-1p}

We first perform the summation over $I$. The quantization condition
reads
\begin{equation*}
 Q_1(\theta_1)= mL\sinh\theta_1=2\pi I\qquad\qquad
 \rho_1=\frac{\partial Q_1}{\partial \theta_1}=mL\cosh\theta_1
\end{equation*}
with $I\in\mathbb{N}$. Therefore it is possible to convert the
summation into a sum over contour integrals
\begin{equation}
  \label{C21_k1}
  \sum_{J_1J_2} \sum_I \oint \frac{d\theta_1}{2\pi}
\frac{F_3^{\mathcal{O}_1}(\theta_1+i\pi,\theta'_1,\theta'_2)
F_3^{\mathcal{O}_2}(\theta_1+i\pi,\theta'_2,\theta'_1)}
{\rho_2(\theta'_1,\theta'_2)}
K_{t,x}^{(R)}(\theta_{1},\theta_{1}',\theta_{2}')
\frac{1}{\mathrm{e}^{iQ_1(\theta_1)}-1}
\end{equation}
In order to open up the contours one has to calculate the surplus
singularities of the integrand, which appear
at $\theta_1\to\theta'_1$ and at $\theta_1\to\theta'_2$. 
Each of the form factors have first order poles, therefore the
singularity is a second order pole. In the following
we calculate the residue at $\theta_1\to\theta'_1$; the case
$\theta_1\to\theta'_2$ will be given by a change of variables.

The residue of the form factors for $\theta_1\to\theta'_1$ read
\begin{equation}
  \label{eq:C12_res1}
  F_3^{\mathcal{O}_1}(\theta_1+i\pi,\theta'_1,\theta'_2)=
i\Big(1-S(\theta'_1-\theta'_2)\Big)\frac{F_1^{\mathcal{O}_1}}{\theta_1-\theta'_1}
+\dots
\end{equation}
Let us introduce the connected part of the three-particle form factor as
\begin{equation}
  \label{eq:maradek1}
  F_{3sc}^{\mathcal{O}_1}(\theta'_1|\theta'_1,\theta'_2)=\lim_{\theta_1\to\theta'_1} \Big(
F_3^{\mathcal{O}_1}(\theta_1+i\pi,\theta'_1,\theta'_2)-
i\Big(1-S(\theta'_1-\theta'_2)\Big)\frac{F_1^{\mathcal{O}_1}}{\theta_1-\theta'_1}\Big)
\end{equation}
The connected form factor defined above still has a pole at
$\theta'_1=\theta'_2$. In fact, the singularity structure of the
original form factor near $\theta_1=\theta'_1=\theta'_2$ is given by
\begin{equation}
  F_3^{\mathcal{O}_1}(\theta_1+i\pi,\theta'_1,\theta'_2)=
2i F_1^{\mathcal{O}_1} \left(\frac{1}{\theta_1-\theta'_1}-\frac{1}{\theta_1-\theta'_2}\right)
\end{equation}
and after subtracting the first pole there remains the second one
leading to
\begin{equation}
  F_{3sc}^{\mathcal{O}_1}(\theta'_1|\theta'_1,\theta'_2)=-2i F_1^{\mathcal{O}_1}
\frac{1}{\theta'_1-\theta'_2}+\dots
\end{equation}
Also, it can be proven that
\begin{equation}
   F_{3sc}^{\mathcal{O}_1}(\theta'_1|\theta'_1,\theta'_2)=
S(\theta'_1-\theta'_2)
 F_{3sc}^{\mathcal{O}_1}(\theta'_2|\theta'_2,\theta'_1)
\end{equation}
In the case of the crossed form factor one has
\begin{equation}
  \begin{split}
\label{F3sc-crossed}
 F_3^{\mathcal{O}_2}(\theta_1+i\pi,\theta'_2,\theta'_1)=
-i\Big(1-S(\theta'_2-\theta'_1)\Big)\frac{F_1^{\mathcal{O}_2}}{\theta_1-\theta'_1}
+S(\theta'_2-\theta'_1) F_{3sc}^{\mathcal{O}_2}(\theta'_1|\theta'_1,\theta'_2)+\dots
  \end{split}
\end{equation}

With these notations the residue of \eqref{C21_k1} at $\theta_1=\theta'_1$ is expressed as
\begin{equation}
  \begin{split}
 & K_{t,x}^{(R)}(\theta_{1}',\theta_{1}',\theta_{2}') 
\Big\{
\Big((S(\theta'_1-\theta'_2)-1)(im\coh\theta'_1x-m\sih\theta'_1(R-t))+imL\coh\theta'_1\Big)
F_1^{\mathcal{O}_1}F_1^{\mathcal{O}_2}\\
&\hspace{4cm}+i
F_{3sc}^{\mathcal{O}_1}(\theta'_1|\theta'_1,\theta'_2)F_1^{\mathcal{O}_2}+i
F_{3sc}^{\mathcal{O}_2}(\theta'_1|\theta'_1,\theta'_2)F_1^{\mathcal{O}_1}
\Big\}
  \end{split}
\end{equation}
There is a similar contribution at
$\theta_1=\theta'_2$, with the role of $\theta'_1$ and $\theta'_2$
exchanged. After integrating over $\theta'_1,\theta'_2$ one could make a
change of variables to obtain the same contribution
twice. However, one has to keep both residues separately because of
the poles  of the quantities $F_{3sc}$. Making the change of variables only in the regular terms
one obtains the two contributions
\begin{equation}
\label{koztesp1}
  \begin{split}
  \text{dsing}_{FF}=&2 K_{t,x}^{(R)}(\theta_{1}',\theta_{1}',\theta_{2}')  
\Big((S(\theta'_1-\theta'_2)-1)(im\coh\theta'_1x-m\sih\theta'_1(R-t))+imL\coh\theta'_1\Big)
F_1^{\mathcal{O}_1}F_1^{\mathcal{O}_2}\\
\text{ssing}_{FF}=&iK_{t,x}^{(R)}(\theta_{1}',\theta_{1}',\theta_{2}')  
\left(
F_{3sc}^{\mathcal{O}_1}(\theta'_1|\theta'_1,\theta'_2)F_1^{\mathcal{O}_2}+
F_{3sc}^{\mathcal{O}_2}(\theta'_1|\theta'_1,\theta'_2)F_1^{\mathcal{O}_1}
\right)+\\
&i
K_{t,x}^{(R)}(\theta_{2}',\theta_{2}',\theta_{1}') 
\left(
F_{3sc}^{\mathcal{O}_1}(\theta'_2|\theta'_2,\theta'_1)F_1^{\mathcal{O}_2}+
F_{3sc}^{\mathcal{O}_2}(\theta'_2|\theta'_2,\theta'_1)F_1^{\mathcal{O}_1}
\right)
  \end{split}
\end{equation}
Now is is possible to perform the summations over $\theta'_1,\theta'_2$.
The $\mathcal{O}(L)$ term of \eqref{koztesp1} can be transformed in the $L\to\infty$ limit into
\begin{equation}
\begin{split}
 & mLF_1^{\mathcal{O}_1}F_1^{\mathcal{O}_2}
  \left(\int \frac{d\theta'_2}{2\pi}\mathrm{e}^{-im\sih\theta'_2x-m\coh\theta'_2 t}\right)
\left(\int \frac{d\theta'_1}{2\pi} \coh\theta'_1
  \mathrm{e}^{-m\coh\theta'_1R}\right)-\\
&-F_1^{\mathcal{O}_1}F_1^{\mathcal{O}_2} \int \frac{d\theta'_2}{2\pi} 
 \mathrm{e}^{-m\coh\theta'_2 (R+t)-im\sih\theta'_2x}
\end{split}
\end{equation}
The first term gets exactly canceled by $Z_1C_{01}$ leaving only the finite
contribution
\begin{equation}
  -F_1^{\mathcal{O}_1}F_1^{\mathcal{O}_2} \int \frac{d\theta'_2}{2\pi} 
 \mathrm{e}^{-m\coh\theta'_2 (R+t)-im\sih\theta'_2x}
\end{equation}
The remaining terms of \eqref{koztesp1} are regular, therefore it is allowed
to replace the summation over $\theta'_1,\theta'_2$ with the appropriate integral.
The final result is
\begin{equation}
\begin{split}
D_{12}=& 
\frac{1}{2} \int_{C_{+}}\frac{d\theta_1}{2\pi}
\int \frac{d\theta'_1}{2\pi}\frac{d\theta'_2}{2\pi}
F_3^{\mathcal{O}_1}(\theta_1+i\pi,\theta'_1,\theta'_2)
F_3^{\mathcal{O}_2}(\theta_1+i\pi,\theta'_2,\theta'_1)
K_{t,x}^{(R)}(\theta_{1},\theta_{1}',\theta_{2}')\\
&+\int    \frac{d\theta'_1}{2\pi}\frac{d\theta'_2}{2\pi}
K_{t,x}^{(R)}(\theta_{1}',\theta_{1}',\theta_{2}') 
(S(\theta'_1-\theta'_2)-1)(m\coh\theta'_1x+im\sih\theta'_1(R-t))
F_1^{\mathcal{O}_1}F_1^{\mathcal{O}_2}\\
&+\frac{1}{2} \int  \frac{d\theta'_1}{2\pi}\frac{d\theta'_2}{2\pi}
\Big\{K_{t,x}^{(R)}(\theta_{1}',\theta_{1}',\theta_{2}') 
\Big(F_{3sc}^{\mathcal{O}_1}(\theta'_1|\theta'_1,\theta'_2)F_1^{\mathcal{O}_2}+
F_{3sc}^{\mathcal{O}_2}(\theta'_1|\theta'_1,\theta'_2)F_1^{\mathcal{O}_1}\Big)
+(\theta'_1\leftrightarrow\theta'_2)\Big\}\\
&-F_1^{\mathcal{O}_1}F_1^{\mathcal{O}_2} \int \frac{d\theta'_2}{2\pi} 
 \mathrm{e}^{-m\coh\theta'_2 (R+t)-im\sih\theta'_2x}
\label{D12}
\end{split}
\end{equation}

\subsubsection{Performing the two-particle summation first}\label{subsubsec:D12doublesum}

We can express $C_{12}$ as
\begin{eqnarray} 
\sum_{I}\frac{1}{\rho_{1}(\theta_{1})}\sum_{J_{1}J_{2}}
\oint\oint_{C_{J_{1}J_{2}}}\frac{d\theta'_1}{2\pi}\frac{d\theta'_2}{2\pi}
\frac{F_{3}^{\mathcal{O}_{1}}(\theta_{1}+i\pi,\theta'_1,\theta'_2)
F_{3}^{\mathcal{O}_{2}}(\theta'_2+i\pi,\theta'_1+i\pi,\theta_1)}
{\left(\mathrm{e}^{iQ_{1'}(\theta'_1,\theta'_2)}+1\right)\left(\mathrm{e}^{iQ_{2'}(\theta'_1,\theta'_2)}+1\right)}
K_{t,x}^{(R)}(\theta_{1},\theta'_1,\theta'_2)\nonumber\\
\label{C12-2a}
\end{eqnarray}
\begin{eqnarray*}
K_{t,x}^{(R)}(\theta_{1},\theta'_1,\theta'_2) & = &
\mathrm{e}^{imx(\sinh\theta_{1}-\sinh\theta'_1-\sinh\theta'_2)}\mathrm{e}^{-m(R-t)\cosh\theta_{1}}\mathrm{e}^{-mt(\cosh\theta'_1+\cosh\theta'_2)}
\end{eqnarray*}
Now we open the multi-contour to surround the real axes in $\theta'_1$ and $\theta'_2$; 
however, we encounter some {}``surplus''
singularities:
\begin{itemize}
\item QF poles, where the singularity in one of the variables arise from a Q-denominator
while in the other from a form factor:
\begin{eqnarray*}
 &  & \theta'_1=\theta_{1}\quad,\quad\mathrm{e}^{iQ_{2'}(\theta'_1,\theta'_2)}+1=0\\
 &  &
 \theta'_2=\theta_{1}\quad,\quad\mathrm{e}^{iQ_{1'}(\theta'_1,\theta'_2)}+1=0
\end{eqnarray*}
\item FF poles, where the singularity in both variables comes from the form factors:
\begin{equation*}
\theta'_1=\theta_{1}\quad,\quad\theta'_2=\theta_{1}
\end{equation*}
\end{itemize}
(note that positions where there is only a singularity in one of the variables do not 
contribute, as the contour in the other variable can be shrunk to a point). In the following 
we calculate the contributions of these singularities. 

\bigskip
\textbf{The FF singularity}
\bigskip

We can write
\begin{equation}
\begin{split}
&F_{3}^{\mathcal{O}_{1}}(\theta_{1}+i\pi,\theta'_1,\theta'_2)=\frac{1}{\theta_{1}-\theta'_1}i\left(1-S(\theta'_1-\theta'_2)\right)F_{1}^{\mathcal{O}_{1}}+\frac{1}{\theta_{1}-\theta'_2}i\left(S(\theta'_2-\theta'_1)-1\right)F_{1}^{\mathcal{O}_{1}}
\\
&+F_{3cc}^{\mathcal{O}_{1}}(\theta_{1}|\theta'_1,\theta'_2) 
\end{split}
\end{equation}
where $F_{3cc}^{\mathcal{O}_{1}}$ is the regular part of the form
factor around the singularity. The pole contribution is then 
\begin{eqnarray}
\nonumber
&  & \frac{1}{2}\oint_{\theta_{1}}
\frac{d\theta'_1}{2\pi}\oint_{\theta_{1}}\frac{d\theta'_2}{2\pi}\frac{K_{t,x}^{(R)}(\theta_{1},\theta'_1,\theta'_2)}{\left(\mathrm{e}^{iQ_{1'}(\theta'_1,\theta'_2)}+1\right)\left(\mathrm{e}^{iQ_{2'}(\theta'_1,\theta'_2)}+1\right)}
S(\theta'_1-\theta'_2)
\\
 & &\times \nonumber
 \left(\frac{1}{\theta_{1}-\theta'_1}i\left(1-S(\theta'_1-\theta'_2)\right)F_{1}^{\mathcal{O}_{1}}+\frac{1}{\theta_{1}-\theta'_2}i\left(S(\theta'_2-\theta'_1)-1\right)F_{1}^{\mathcal{O}_{1}}+F_{3cc}^{\mathcal{O}_{1}}(\theta_{1}|\theta'_1,\theta'_2)\right)\\
\nonumber
 & &\times 
 \left(\frac{1}{\theta_{1}-\theta'_1}i\left(1-S(\theta'_1-\theta'_2)\right)F_{1}^{\mathcal{O}_{2}}+\frac{1}{\theta_{1}-\theta'_2}i\left(S(\theta'_2-\theta'_1)-1\right)F_{1}^{\mathcal{O}_{2}}+F_{3cc}^{\mathcal{O}_{2}}(\theta_{1}|\theta'_1,\theta'_2)\right)
\end{eqnarray}
$F_{3cc}$ does not contribute since then either the $\theta'_1$
or the $\theta'_2$ integration contour can be contracted to a point.
For similar reasons, the only terms that could give a nonzero contribution
are the {}``cross-terms''
\begin{eqnarray*}
&  & \nonumber
\frac{1}{2}\oint_{\theta_{1}}\frac{d\theta'_1}{2\pi}\oint_{\theta_{1}}\frac{d\theta'_2}{2\pi}\frac{K_{t,x}^{(R)}(\theta_{1},\theta'_1,\theta'_2)}
{\left(\mathrm{e}^{iQ_{1'}(\theta'_1,\theta'_2)}+1\right)\left(\mathrm{e}^{iQ_{2'}(\theta'_1,\theta'_2)}+1\right)}2S(\theta'_1-\theta'_2)\\ 
\nonumber
 & \times &
 \left(\frac{1}{\theta_{1}-\theta'_1}i\left(1-S(\theta'_1-\theta'_2)\right)F_{1}^{\mathcal{O}_{1}}
\times\frac{1}{\theta_{1}-\theta'_2}i\left(S(\theta'_2-\theta'_1)-1\right)F_{1}^{\mathcal{O}_{2}}\right)
\end{eqnarray*}
We need the residue at $\theta'_1=\theta'_2=\theta_{1}$. Then 
\begin{equation*}
S(\theta'_1-\theta'_2)=S(0)=-1
\end{equation*}
and 
\begin{equation*}
\mathrm{e}^{iQ_{1'}(\theta_{1},\theta_{1})}=\mathrm{e}^{iQ_{2'}(\theta_{1},\theta_{1})}
=\mbox{e}^{imL\sinh\theta_{1}}=1
\end{equation*}
therefore the contribution of the pole is given by
\begin{equation*}
-F_{1}^{\mathcal{O}_{1}}F_{1}^{\mathcal{O}_{2}}\mathrm{e}^{-imx\sinh\theta_{1}-m(R+t)\cosh\theta_{1}}
\end{equation*}
After performing the $\theta_{1}$ sum converted to an integral we
find
\begin{equation}
S_{FF}=-\int\frac{d\theta}{2\pi}F_{1}^{\mathcal{O}_{1}}F_{1}^{\mathcal{O}_{2}}\mathrm{e}^{-imx\sinh\theta_{1}-m(R+t)\cosh\theta_{1}}
\end{equation}
which correctly reproduces the last term of $D_{12}$ \eqref{D12}.

\bigskip
{\bf The QF pole at $\theta'_1=\theta_{1}$}
\bigskip

The singular contribution is
\begin{equation}
\frac{1}{2}\oint_{\theta_{1}}\frac{d\theta'_1}{2\pi}\oint_{C_{J_{2}I}}\frac{d\theta'_2}{2\pi}\frac{F_{3}^{\mathcal{O}_{1}}(\theta_{1}+i\pi,\theta'_1,\theta'_2)F_{3}^{\mathcal{O}_{2}}(\theta_{1}+i\pi,\theta'_2,\theta'_1)}{\left(\mathrm{e}^{iQ_{1'}(\theta'_1,\theta'_2)}+1\right)\left(\mathrm{e}^{iQ_{2'}(\theta'_1,\theta'_2)}+1\right)}K_{t,x}^{(R)}(\theta_{1},\theta'_1,\theta'_2)\end{equation}
where $C_{J_{2}I}$ surrounds the $\theta'_2$ solution of \begin{equation}
Q_{2'}(\theta_{1},\theta'_2)=mL\sinh\theta'_2+\delta(\theta'_2-\theta_{1})=2\pi J_{2}\end{equation}
where $\theta_{1}$ is the solution of \begin{equation}
Q_{1}(\theta_{1})=mL\sinh\theta_{1}=2\pi I\end{equation}
The behaviours of the form factors are given by \eqref{eq:maradek1} and \eqref{F3sc-crossed}.
We can separate the integrand into two terms according to the order of the $\theta'_1=\theta_{1}$
singularity. The first order term has the form 
\begin{eqnarray*}
S_{QF}^1& =-\frac{1}{2} & \oint_{\theta_{1}}\frac{d\theta'_1}{2\pi}\oint_{C_{J_{2}I}}\frac{d\theta'_2}{2\pi}\frac{K_{t,x}^{(R)}(\theta_{1},\theta'_1,\theta'_2)}{\left(\mathrm{e}^{iQ_{1'}(\theta'_1,\theta'_2)}+1\right)\left(\mathrm{e}^{iQ_{2'}(\theta'_1,\theta'_2)}+1\right)}\frac{i}{\theta'_1-\theta_{1}}\left(S(\theta'_2-\theta'_1)-1\right)\\
 & \times &
 \left(F_{1}^{\mathcal{O}_{2}}F_{3c}^{\mathcal{O}_{1}}(\theta_{1}|\theta'_1,\theta'_2)+F_{1}^{\mathcal{O}_{1}}F_{3c}^{\mathcal{O}_{2}}(\theta_{1}|\theta'_1,\theta'_2)\right)
\end{eqnarray*}
This can be easily evaluated:
\begin{equation*}
-\frac{1}{2}\sum_{\theta'_2}\frac{K_{t,x}^{(R)}(\theta_{1},\theta_{1},\theta'_2)}{\left(1-S(\theta_{1}-\theta'_2)\right)\bar{\rho}_{3}(\theta'_2|\theta'_1)}\left(S(\theta'_2-\theta_{1})-1\right)\left(F_{1}^{\mathcal{O}_{2}}F_{3c}^{\mathcal{O}_{1}}(\theta_{1}|\theta_{1},\theta'_2)+F_{1}^{\mathcal{O}_{1}}F_{3c}^{\mathcal{O}_{2}}(\theta_{1}|\theta_{1},\theta'_2)\right)
\end{equation*}
where 
\begin{equation*}
\bar{\rho}_{3}(\theta'_2|\theta_{1})=\frac{\partial}{\partial\theta'_2}Q_{2'}(\theta_{1},\theta'_2)=mL\cosh\theta'_2+\varphi(\theta'_2-\theta_{1})
\end{equation*}
is the density of $\theta'_2$ solutions for a given $\theta_{1}$.
This gives 
\begin{equation}
S_{QF}^1=-\frac{1}{2}\sum_{\theta'_2}\frac{K_{t,x}^{(R)}(\theta_{1},\theta_{1},\theta'_2)}
{\bar{\rho}_{3}(\theta'_2|\theta_{1})}S(\theta'_2-\theta_{1})
\left(F_{1}^{\mathcal{O}_{2}}F_{3c}^{\mathcal{O}_{1}}(\theta_{1}|\theta_{1},\theta'_2)
+F_{1}^{\mathcal{O}_{1}}F_{3c}^{\mathcal{O}_{2}}(\theta_{1}|\theta_{1},\theta'_2)\right)
\end{equation}
Together with a similar term $S_{QF}^2$ obtained by swapping
$\theta'_1\leftrightarrow\theta'_2$ this gives the
contribution
\begin{eqnarray}
&&\frac{1}{2}\int\frac{d\theta'_1}{2\pi}\int\frac{d\theta'_2}{2\pi}
\Big[K_{t,x}^{(R)}(\theta'_1,\theta'_1,\theta'_2)S(\theta'_2-\theta'_1)
\left(F_{1}^{\mathcal{O}_{1}}F_{3c}^{\mathcal{O}_{2}}(\theta'_1|\theta'_1,\theta'_2)+
F_{1}^{\mathcal{O}_{2}}F_{3c}^{\mathcal{O}_{1}}(\theta'_1|\theta'_1,\theta'_2)\right)
\nonumber\\
&&+(\theta'_1\leftrightarrow\theta'_2)\Big]
\end{eqnarray}
The second order term reads
\begin{eqnarray}
D_{QF}^1
& = &  -\frac{1}{2}\oint_{\theta_{1}}\frac{d\theta'_1}{2\pi}\oint_{C_{J_{2}I}}\frac{d\theta'_2}{2\pi}\frac{K_{t,x}^{(R)}(\theta_{1},\theta'_1,\theta'_2)}{\left(\mathrm{e}^{iQ_{1'}(\theta'_1,\theta'_2)}+1\right)\left(\mathrm{e}^{iQ_{2'}(\theta'_1,\theta'_2)}+1\right)}
\nonumber\\
&&\times
\frac{\left(1-S(\theta'_1-\theta'_2)\right)\left(S(\theta'_2-\theta'_1)-1\right)}{\left(\theta_{1}-\theta'_1\right)^{2}}
F_{1}^{\mathcal{O}_{1}}F_{1}^{\mathcal{O}_{2}}
\end{eqnarray}
The contribution of the double pole can be evaluated by taking the
derivative with respect to $\theta'_1$. The result reads
\begin{eqnarray*}
D_{QF}^1 & = & -\frac{1}{2}i\oint_{C_{J_{2}I}}\frac{d\theta'_2}{2\pi}
\frac{F_{1}^{\mathcal{O}_{1}}F_{1}^{\mathcal{O}_{2}}K_{t,x}^{(R)}(\theta_{1},\theta_{1},\theta'_2)}
{\left(\mathrm{e}^{iQ_{2'}(\theta_{1},\theta'_2)}+1\right)}
\Bigg[(-imx\cosh\theta_{1}-mt\sinh\theta_{1})\left(S(\theta'_2-\theta_{1})-1\right)\\
 &  & +imL\cosh\theta_{1}-i\varphi(\theta_{1}-\theta'_2)S(\theta'_2-\theta_{1})\Bigg]\\
 &  &
 +\frac{1}{2}\oint_{C_{J_{2}I}}\frac{d\theta'_2}{2\pi}
\frac{F_{1}^{\mathcal{O}_{1}}F_{1}^{\mathcal{O}_{2}}K_{t,x}^{(R)}(\theta_{1},\theta_{1},\theta'_2)}
{\left(\mathrm{e}^{iQ_{2'}(\theta_{1},\theta'_2)}+1\right)^{2}}\left(S(\theta'_2-\theta_{1})-1\right)\varphi(\theta'_2-\theta_{1})
\end{eqnarray*}
The last term with the double pole yields zero for $L\to\infty$ since it is proportional
to $L^{-2}$ and becomes $L^{-1}$ after including the density. Explicitly it evaluates
to
\begin{equation*}
\frac{1}{2}\oint_{C_{J_{2}I}}\frac{d\theta'_2}{2\pi}
\frac{F_{1}^{\mathcal{O}_{1}}F_{1}^{\mathcal{O}_{2}}K_{t,x}^{(R)}(\theta_{1},\theta_{1},\theta'_2)}
{-Q_{2'}'(\theta_{1},\theta_{3*})^{2}(\theta'_2-\theta_{3*})^{2}}
\left(S(\theta'_2-\theta_{1})-1\right)\varphi(\theta'_2-\theta_{1})(-S(\theta'_2-\theta_{1}))
\end{equation*}
where $\theta_{3*}$ is the location of the solution of \begin{equation}
Q_{2'}(\theta_{1},\theta'_2)=2\pi J_{2}\end{equation}
but \begin{equation}
Q_{2'}'(\theta_{1},\theta_{3*})^{2}=(mL\cosh\theta_{3*}+\varphi(\theta_{3*}-\theta_{1}))^{2}=O(L^{-2})\end{equation}
Even after multiplying this by the density $Q_{2'}'(\theta_{1},\theta_{3*})$
when converting the summation over $\theta_{3*}$ to integral a suppression
$O(L^{-1})$ remains, resulting in zero large volume limit.

Putting in the $\theta_{1}$ summation and a factor $2$ to account
for the contribution obtained by exchanging $\theta'_1$ with $\theta'_2$,
plus a minus sign since this is to be subtracted in the end, and adding
the $-Z_{1}C_{01}$ term gives 
\begin{eqnarray*}
 &  & -\int\frac{d\theta_{1}}{2\pi}\int\frac{d\theta'_2}{2\pi}F_{1}^{\mathcal{O}_{1}}F_{1}^{\mathcal{O}_{2}}K_{t,x}^{(R)}(\theta_{1},\theta_{1},\theta'_2)\Bigg[(mx\cosh\theta_{1}-imt\sinh\theta_{1})\left(S(\theta'_2-\theta_{1})-1\right)\\
 &  & -mL\cosh\theta_{1}+\varphi(\theta_{1}-\theta'_2)S(\theta'_2-\theta_{1})\Bigg]\\
 &  &
 -mL\int\frac{d\theta}{2\pi}\cosh\theta\mathrm{e}^{-mR\cosh\theta}
F_{1}^{\mathcal{O}_{1}}F_{1}^{\mathcal{O}_{2}}\int\frac{d\theta'}{2\pi}\mathrm{e}^{-imx\sinh\theta'-mt\cosh\theta'}
\end{eqnarray*}
The $O(L)$ term cancels as expected,  and after a partial integration one obtains 
\begin{equation*}
D_{QF}=-\int\frac{d\theta_{1}}{2\pi}\int\frac{d\theta'_2}{2\pi}F_{1}^{\mathcal{O}_{1}}F_{1}^{\mathcal{O}_{2}}
K_{t,x}^{(R)}(\theta_{1},\theta_{1},\theta'_2)(mx\cosh\theta_{1}+im(R-t)\sinh\theta_{1})
\left(S(\theta'_2-\theta_{1})-1\right)
\end{equation*}

\bigskip
\textbf{End result}
\bigskip

Putting together everything
\begin{equation}
\label{D12masodik}
\begin{split}
    D_{12}=&
\frac{1}{2}\int\frac{d\theta_{1}}{2\pi}
\int\int_{C_{++}}\frac{d\theta'_1}{2\pi}\frac{d\theta'_2}{2\pi}
F_{3}^{\mathcal{O}_{1}}(\theta_{1}+i\pi,\theta'_1,\theta'_2)
F_{3}^{\mathcal{O}_{2}}(\theta_{1}+i\pi,\theta'_2,\theta'_1)
K_{t,x}^{(R)}(\theta_{1},\theta'_1,\theta'_2)\\
&+\frac{1}{2}\int\frac{d\theta'_1}{2\pi}\int\frac{d\theta'_2}{2\pi}
\Big[K_{t,x}^{(R)}(\theta'_1,\theta'_1,\theta'_2)S(\theta'_2-\theta'_1)
\left(F_{1}^{\mathcal{O}_{1}}F_{3c}^{\mathcal{O}_{2}}(\theta'_1|\theta'_1,\theta'_2)+
F_{1}^{\mathcal{O}_{2}}F_{3c}^{\mathcal{O}_{1}}(\theta'_1|\theta'_1,\theta'_2)\right)
\\
&+(\theta'_1\leftrightarrow\theta'_2)\Big]\\
&-\int\frac{d\theta_{1}}{2\pi}\int\frac{d\theta'_2}{2\pi}F_{1}^{\mathcal{O}_{1}}F_{1}^{\mathcal{O}_{2}}
K_{t,x}^{(R)}(\theta_{1},\theta_{1},\theta'_2)(mx\cosh\theta_{1}+im(R-t)\sinh\theta_{1})
\left(S(\theta'_2-\theta_{1})-1\right)\\
&-\int\frac{d\theta}{2\pi}\mathrm{e}^{-m(R+t)\cosh\theta-imx\sinh\theta}
F_{1}^{\mathcal{O}_{1}}F_{1}^{\mathcal{O}_{2}}
\end{split}
\end{equation}
It is a straightforward, although somewhat tedious exercise to show
that the above expression can be transformed in the form \eqref{D12}.
First of all observe, that shifting all three variables the first term
of \eqref{D12masodik} can be 
written as
\begin{equation*}
  \frac{1}{2}\int_{C_{-}}\frac{d\theta_{1}}{2\pi}
\int\frac{d\theta'_1}{2\pi}
\int\frac{d\theta'_2}{2\pi}
F_{3}^{\mathcal{O}_{1}}(\theta_{1}+i\pi,\theta'_1,\theta'_2)
F_{3}^{\mathcal{O}_{2}}(\theta_{1}+i\pi,\theta'_2,\theta'_1)
K_{t,x}^{(R)}(\theta_{1},\theta'_1,\theta'_2)
\end{equation*}
This differs from the corresponding term in \eqref{D12} in the
contour for $\theta_1$, which in this case runs below the real
axis. Shifting this contour to run above the real axis one picks up
the poles of the integrand, which can 
be evaluated using standard techniques. It can be shown that the
resulting contributions are
\begin{equation*}
  \begin{split}
&  -\int\frac{d\theta'_1}{2\pi}\int\frac{d\theta'_2}{2\pi}
(-2+S(\theta'_1-\theta'_2)+S(\theta'_2-\theta'_1))(mx\cosh\theta'_1+im(R-t)\sinh\theta'_1)
K_{t,x}^{(R)}(\theta'_1,\theta'_1,\theta'_2)\\
&+\frac{1}{2}\int\frac{d\theta'_1}{2\pi}\int\frac{d\theta'_2}{2\pi}
\Big[\left(S(\theta'_2-\theta'_1)-1\right)\left(F_{1}^{\mathcal{O}_{1}}F_{3c}^{\mathcal{O}_{2}}(\theta'_1|\theta'_1,\theta'_2)+F_{1}^{\mathcal{O}_{2}}F_{3c}^{\mathcal{O}_{1}}(\theta'_1|\theta'_1,\theta'_2)\right)
K_{t,x}^{(R)}(\theta'_1,\theta'_1,\theta'_2)
\\
& +(\theta'_1\leftrightarrow\theta'_2)\Big]
  \end{split}
\end{equation*}
Adding these terms to the second and third lines of \eqref{D12masodik}
one recovers \eqref{D12}.

We wish to note, that if instead of \eqref{C12-2a} we had started with
a similar formula including the factors 
\begin{equation*}
  \frac{1}
{\left(\mathrm{e}^{-iQ_{1'}(\theta'_1,\theta'_2)}+1\right)\left(\mathrm{e}^{-iQ_{2'}(\theta'_1,\theta'_2)}+1\right)}
\end{equation*}
we would have arrived immediately at the result \eqref{D12}. However,
the calculation presented above is a non-trivial cross-check of our methods.

\subsection{The contribution $D_{1n}$ for $n>2$}

\label{sec:D1n}

Based on the previous subsection it is now straightforward to evaluate
the contribution $D_{1n}$ for arbitrary $n$. It is given by
\begin{equation*}
  D_{1n}=\lim_{L\to\infty}\big(C_{1n}-Z_1 C_{0,n-1}\big)
\end{equation*}
where
\begin{eqnarray}
C_{1n}&=&\frac{1}{n!}\sum_{I}\sum_{J_1\dots J_n}
\bra{\{I\}}\mathcal{O}_1(0)\ket{\{J_1,\dots,J_n\}}_L 
\bra{\{J_1,\dots,J_n\}}\mathcal{O}_2(0)\ket{\{I\}}_L
\nonumber\\
&&\times
\mathrm{e}^{i(P_1-P_2)x}\mathrm{e}^{-E_1(R-t)}\mathrm{e}^{-E_2t}
\nonumber\\
&=&\frac{1}{n!}\sum_{I} \sum_{J_1\dots J_n}
\frac{F_{n+1}^{\mathcal{O}_1}(\theta+i\pi,\theta'_1,\dots,\theta_{n})
F_{n+1}^{\mathcal{O}_2}(\theta+i\pi,\theta'_n,\dots,\theta'_1)}
{\rho_1(\theta)\rho_n(\theta'_1,\dots,\theta'_n)}
\nonumber\\
&&\times
\mathrm{e}^{imx(\sinh\theta-mx\sum_j
  \sinh\theta_j)-m(R-t)\cosh\theta-mt\sum_j\cosh\theta_j}
\label{C1nkiind}
\end{eqnarray}
There are additional disconnected terms in the case of $n$ being odd, 
according to the rule explained in subsection \ref{finvolFF}. This happens in the
presence of zero-momentum particles, which requires $I=0$ and the set
$\{J_1,\dots,J_n\}$ to be parity symmetric; the disconnected term is
given by formula \eqref{eq:oddoddlyrule}. It is easy to show using
the constrained density of states, that all contributions associated
to these disconnected terms
scales with negative powers of $L$, therefore we neglect them in the following.

We consider \eqref{C1nkiind} and we first perform the summation over $I$. The quantization condition
is
\begin{equation}
 Q_1(\theta)= mL\sinh\theta=2\pi I\qquad\qquad \rho_1=Q_1'
\label{tempQ1}\end{equation}
with $I\in\mathbb{N}$. The converting the sum over $I$  to contour integrals we get 
\begin{eqnarray}
&& \sum_{J_1\dots J_n}\sum_{I}\oint_{C_I} \frac{d\theta}{2\pi}
\frac{F_{n+1}^{\mathcal{O}_1}(\theta+i\pi,\theta'_1,\dots,\theta_{n})
F_{n+1}^{\mathcal{O}_2}(\theta+i\pi,\theta'_n,\dots,\theta'_1)}
{\rho_n(\theta'_1,\dots,\theta'_n)}\times
\nonumber\\
&&\times
\mathrm{e}^{imx(\sinh\theta-mx\sum_j
  \sinh\theta_j)-m(R-t)\cosh\theta-mt\sum_j\cosh\theta_j}
\frac{1}{\mathrm{e}^{iQ_1(\theta)}-1}
  \label{C1n_k1}\end{eqnarray}
where the contour $C_I$ surrounds the solution of (\ref{tempQ1}). When opening the contour 
to surround the real axis in $\theta$, we get the following contribution in the $L\to\infty$ limit:
\begin{equation}
\begin{split}
& \frac{1}{n!} \int_{C_+}
\frac{d\theta}{2\pi}\int  \frac{d\theta'_1}{2\pi}\dots \frac{d\theta'_n}{2\pi}
F_{n+1}^{\mathcal{O}_1}(\theta+i\pi,\theta'_1,\dots,\theta_{n})
F_{n+1}^{\mathcal{O}_2}(\theta+i\pi,\theta'_n,\dots,\theta'_1)\times\\
& \hspace{4cm} \times \mathrm{e}^{imx(\sinh\theta-\sum_j
  \sinh\theta'_j)x-m(R-t)\cosh\theta-mt\sum_j\cosh\theta'_j}
\end{split}
\label{D1nnaive}\end{equation}
where the contour ${C_+}$ is defined as
\begin{equation}
\int_{C_+}\frac{d\theta}{2\pi}f(\theta)=\int_{\mathbb{R}}\frac{d\theta}{2\pi}f(\theta+i\epsilon)
\end{equation}
However, there are additional poles of the integrand for $\theta=\theta'_j$ for
$j=1\dots n$, whose contribution must be subtracted. 

First we calculate the residue at $\theta\to\theta'_1$.
The behaviour of the form factors is given by the kinematical residue 
equation (\ref{eq:kinematicalaxiom}). 
Let us introduce the (partially) connected part of the form factor as
\begin{equation}
\label{eq:maradek1b}
\begin{split}
  F_{n+1,sc}^{\mathcal{O}_1}(\theta'_1|\theta'_1,\dots,\theta'_n)=\lim_{\theta\to\theta'_1} \Bigg[
 & F_{n+1}^{\mathcal{O}_1}(\theta+i\pi,\theta'_1,\dots,\theta'_n)
\\
&-i\Big(1-\prod_{j=2}^n S(\theta-\theta_j)\Big)
\frac{F_{n-1}^{\mathcal{O}_1}(\theta'_1,\dots,\theta'_n)}{\theta-\theta'_1}
\Bigg]
\end{split}
\end{equation}
The form factor $F_{sc}$ defined above is only ``partially'' connected since only one 
of the singularities is subtracted and so it still has poles at
$\theta'_1=\theta_j$ for $j=2\dots n$. In fact, the singularity structure of the
original form factor near $\theta=\theta'_1=\theta'_1$ is given by
\begin{equation}
  F_{n+1}^{\mathcal{O}_1}(\theta+i\pi,\theta'_1,\theta'_1,\dots)=
i(1+\prod_{k=3}^n S(\theta'_1-\theta_k) )
 F_{n-1}^{\mathcal{O}_1}(\theta_2,\dots,\theta'_n)
 \left(\frac{1}{\theta-\theta'_1}-\frac{1}{\theta-\theta_2}\right)
\end{equation}
and after subtracting the first pole there remains the second one
leading to
\begin{equation}
  F_{n+1,sc}^{\mathcal{O}_1}(\theta'_1|\theta_2,\dots,\theta'_n)=
-i(1+\prod_{k=3}^n S(\theta'_1-\theta_k) )
 F_{n-1}^{\mathcal{O}_1}(\theta_2,\dots,\theta'_n)
\frac{1}{\theta'_1-\theta_2}+\dots
\end{equation}
The connected part satisfies the exchange equation
\begin{equation}
  \label{conn_exchange}
  F_{n+1,sc}^{\mathcal{O}_1}(\theta'_1|\theta'_1,\dots,\theta'_j,\theta'_k\dots,\theta'_n)
=S(\theta'_j-\theta'_k)
 F_{n+1,sc}^{\mathcal{O}_1}(\theta'_1|\theta'_1,\dots,\theta'_k,\theta'_j\dots,\theta'_n)
\end{equation}
In the case of the crossed form factor one has
\begin{equation*}
  \begin{split}
&     F_{n+1}^{\mathcal{O}_2}(\theta+i\pi,\theta'_n,\dots,\theta'_1)=\\
&\hspace{1cm}-i\Big(1-\prod_{j=2}^n S(\theta_j-\theta)\Big)
\frac{F_{n-1}^{\mathcal{O}_2}(\theta'_n,\dots,\theta_2)}{\theta-\theta'_1}
+\Big(\prod_{j=2}^n S(\theta_j-\theta'_1)\Big)
F_{n+1,sc}^{\mathcal{O}_2}(\theta'_1|\theta'_n,\dots,\theta_2)+\dots
  \end{split}
\end{equation*}
The residue of the integrand at $\theta=\theta'_1$ is then expressed as
\begin{equation}
  \begin{split}
 &\mathrm{e}^{-imx\sum_{j=2}^n
  \sinh\theta'_j -mR\cosh\theta'_1-mt\sum_{j=2}^n \cosh\theta'_j }
\Big\{F_{n-1}^{\mathcal{O}_1}(\theta'_2,\dots,\theta'_n)F_{n-1}^{\mathcal{O}_2}(\theta'_n,\dots,\theta'_2)
\\
&\times\Big[   \big(   \prod_{j=2}^n S(\theta'_1-\theta'_j)-1\big) 
 (imx\cosh\theta'_1-m(R-t)\sinh\theta'_1)+imL\cosh\theta'_1\Big]
\\&+i
 F_{n+1,c}^{\mathcal{O}_1}(\theta'_1|\theta'_1,\dots,\theta'_n)F_{n-1}^{\mathcal{O}_2}(\theta'_n,\dots,\theta'_2)+
i F_{n+1,c}^{\mathcal{O}_2}(\theta'_1|\theta'_n,\dots,\theta'_1)F_{n-1}^{\mathcal{O}_1}(\theta'_2,\dots,\theta'_n)
\Big\}
  \end{split}
\end{equation}
There are similar contributions at
$\theta=\theta_j$ for some $j\ge 2$, with the role of $\theta_j$ and $\theta'_1$
exchanged. After integrating over all the $\theta_j$ one could make a
change of variables to obtain the same contribution
$n$ times. However, one has to keep these residues separately because of
the poles of the connected form factors. Making the change of variables only in the regular terms
one obtains
\begin{equation}
\label{koztesp2}
  \begin{split}
 &n \mathrm{e}^{-imx\sum_{j=2}^n
  \sinh\theta'_j -mR\cosh\theta'_1-mt\sum_{j=2}^n \cosh\theta'_j }
F_{n-1}^{\mathcal{O}_1}(\theta'_2,\dots,\theta'_n)F_{n-1}^{\mathcal{O}_2}(\theta'_n,\dots,\theta'_2)
\\
&\times\Big[   \big(   \prod_{j=2}^n S(\theta'_1-\theta'_j)-1\big) 
 (imx\cosh\theta'_1-m(R-t)\sinh\theta'_1)+imL\cosh\theta'_1\Big]
  \end{split}
\end{equation}
The $O(L)$ term of \eqref{koztesp2} term can be transformed in the $L\to\infty$ limit into
\begin{equation}
\begin{split}
 & \frac{1}{(n-1)!}mL\int \frac{d\theta'_1}{2\pi} \cosh\theta'_1
  \mathrm{e}^{-mR\cosh\theta'_1}
\\
&\times  \left(\int \frac{d\theta'_2}{2\pi}\dots \frac{d\theta'_n}{2\pi} 
F_{n-1}^{\mathcal{O}_1}(\theta'_2,\dots,\theta'_n)F_{n-1}^{\mathcal{O}_2}(\theta'_n,\dots,\theta'_2)
 \mathrm{e}^{-imx\sum_{j=2}^n \sinh\theta'_{j} -mt\sum_{j=2}^n \cosh\theta'_j }
\right)\\
&-\frac{1}{(n-1)!}
\int \frac{d\theta'_2}{2\pi}\dots \frac{d\theta'_n}{2\pi} 
F_{n-1}^{\mathcal{O}_1}(\theta'_2,\dots,\theta'_n)F_{n-1}^{\mathcal{O}_2}(\theta'_n,\dots,\theta'_2)
\Big(\sum_{j=2}^n \mathrm{e}^{-mR\cosh\theta'_j}\Big)
\\
&\times \mathrm{e}^{-imx\sum_{j=2}^n \sinh\theta'_{j} -mt\sum_{j=2}^n \cosh\theta'_j }
\end{split}
\label{haladunk}
\end{equation}
The subtraction of the last term takes into account the exclusion principle 
$\theta'_1\ne \theta'_j$ for $j=2\dots n$, which is already present at the level of quantum numbers.
The first term in \eqref{haladunk} gets exactly canceled by $Z_1C_{0,n-1}$ leaving only the second one
which is finite as $L\to\infty$.

The $O(L^0)$ terms of \eqref{koztesp2} are regular, therefore it is allowed
to replace the summation over the rapidities with the appropriate integral.

Putting everything together, the net result is
\begin{equation}
\label{D1n}
  \begin{split}
D_{1n}=&
 \frac{1}{n!} \int_{C_+}
\frac{d\theta}{2\pi}\int  \frac{d\theta'_1}{2\pi}\dots \frac{d\theta'_n}{2\pi}
F_{n+1}^{\mathcal{O}_1}(\theta+i\pi,\theta'_1,\dots,\theta_{n})
F_{n+1}^{\mathcal{O}_2}(\theta+i\pi,\theta'_n,\dots,\theta'_1)
\\
&\times \mathrm{e}^{imx(\sinh\theta-\sum_j
  \sinh\theta'_j)-m(R-t)\cosh\theta-mt\sum_j\cosh\theta'_j }
\\
+&\frac{1}{(n-1)!}\int \frac{d\theta'_1}{2\pi}\dots \frac{d\theta'_n}{2\pi} 
\mathrm{e}^{-imx\sum_{j=2}^n  \sinh\theta'_j -mR\cosh\theta'_1-mt\sum_{j=2}^n \cosh\theta'_j }
 \big(   \prod_{j=2}^n S(\theta'_1-\theta'_j)-1\big)
\\
&\times  (mx\cosh\theta'_1+im(R-t)\sinh\theta'_1)
F_{n-1}^{\mathcal{O}_1}(\theta'_2,\dots,\theta'_n)F_{n-1}^{\mathcal{O}_2}(\theta'_n,\dots,\theta'_2)
\\
+&\frac{1}{n!} 
\int \frac{d\theta'_1}{2\pi}\dots \frac{d\theta'_n}{2\pi} 
\Bigg\{
\Big[
\mathrm{e}^{-imx\sum_{j=2}^n  \sinh\theta'_j -mR\cosh\theta'_1-mt\sum_{j=2}^n
  \cosh\theta'_j }
\\
&\times
\big(
F_{n+1,sc}^{\mathcal{O}_1}(\theta'_1|\theta'_1,\dots,\theta'_n)F_{n-1}^{\mathcal{O}_2}(\theta'_n,\dots,\theta'_2)+
 F_{n+1,sc}^{\mathcal{O}_2}(\theta'_1|\theta'_n,\dots,\theta'_1)F_{n-1}^{\mathcal{O}_1}(\theta'_2,\dots,\theta'_n)
\big)
\Big]
\\
&+\Big[\theta'_1\leftrightarrow\theta'_j \text{ for } j=2..n\Big]\Bigg\}
\\
&-\frac{1}{(n-1)!}
\int \frac{d\theta'_2}{2\pi}\dots \frac{d\theta'_n}{2\pi} 
F_{n-1}^{\mathcal{O}_1}(\theta'_2,\dots,\theta'_n)F_{n-1}^{\mathcal{O}_2}(\theta'_n,\dots,\theta'_2)
\Big(\sum_{j=2}^n \mathrm{e}^{-mR\cosh\theta'_j}\Big)
\\
&\times \mathrm{e}^{-imx\sum_{j=2}^n \sinh\theta'_{j} -mt\sum_{j=2}^n \cosh\theta'_j }
  \end{split}
\end{equation}

\subsection{Life is not that simple: $D_{22}$}
\label{sec:D22}

Now we turn to the evaluation of the 2-particle -- 2-particle contribution to the 
thermal correlator. The novel feature of this contribution is that the diagonal terms 
must be separated from the non-diagonal ones, since the four-particle form factors (in contrast 
to the two-particle one that appears in $D_{11}$) have nonzero residues for the kinematical poles 
according to (\ref{eq:kinematicalaxiom}). When evaluated at the diagonal, these singularities are 
eliminated but result in an ambiguity of the diagonal matrix element, which was discussed in 
much detail in \cite{Pozsgay:2007gx}. Once this complication is attended to, the evaluation proceeds 
similarly to that of $D_{12}$. 

According to the general formalism outlined in section \ref{subsec:2ptorganization} we can write
\begin{equation}
 D_{22}=\mathop{\mathrm{lim}}_{L\to\infty}\tilde{D}_{22}
\end{equation}
where
\begin{equation}
\tilde{D}_{22}=C_{22}-Z_{1}C_{11}+(Z_{1}^{2}-Z_{2})C_{00}
\end{equation}
Using
\begin{equation}
\tilde{D}_{11}=C_{11}-Z_{1}C_{00}
\end{equation}
gives
\begin{equation}
\tilde{D}_{22}=C_{22}-Z_{1}\tilde{D}_{11}-Z_{2}C_{00}
\end{equation}
The new contribution is
\begin{eqnarray}
C_{22} & = & \sum_{I_{1}>I_{2}}\sum_{J_{1}>J_{2}}\left\langle
  \{I_{1},I_{2}\}\left|\mathcal{O}_{1}\right|\{J_{1},J_{2}\}\right\rangle
_{L}\left\langle
  \{J_{1},J_{2}\}\left|\mathcal{O}_{2}\right|\{I_{1},I_{2}\}\right\rangle
_{L}\nonumber\\
&\times& K_{t,x}^{(R)}(\theta_{1},\theta_{2};\theta_{1}',\theta_{2}')
\end{eqnarray}
where 
\begin{eqnarray}
K_{t,x}^{(R)}(\theta_{1},\theta_{2};\theta_{1}',\theta_{2}') & = & 
\mathrm{e}^{imx(\sinh\theta_{1}+\sinh\theta_{2}-\sinh\theta_{1}'-\sinh\theta_{2}')}
\mathrm{e}^{-m(R-t)(\cosh\theta_{1}+\cosh\theta_{2})}
\nonumber\\
&\times&\mathrm{e}^{-mt(\cosh\theta_{1}'+\cosh\theta_{2}')}
\end{eqnarray}
We can separate the sum into diagonal and non-diagonal part:
\begin{equation}
\sum_{I_{1}>I_{2}}\sum_{J_{1}>J_{2}}=\sum_{I_{1}>I_{2}}
(\mbox{terms with }\{J_{1},J_{2}\}=\{I_{1},I_{2}\})+
\sum_{I_{1}>I_{2}}\sum_{J_{1}>J_{2}}{}'
\end{equation}
where the prime means that $\{J_{1},J_{2}\}\neq\{I_{1},I_{2}\}$.

\subsubsection{Evaluating $Z_2$}

First of all, we need the two-point contribution to the partition function. This is easy to obtain:
\begin{eqnarray}
Z_{2} & = & \sum_{I_{1}<I_{2}}\mathrm{e}^{-mR\left(\cosh\theta_{1}+\cosh\theta_{2}\right)}
\left(\frac{1}{2}\sum_{I_{1},I_{2}}-\frac{1}{2}\sum_{I_{1}=I_{2}}\right)
\mathrm{e}^{-mR\left(\cosh\theta_{1}+\cosh\theta_{2}\right)}\\
\end{eqnarray}
To convert the sums to integrals, we need the two-particle density of states
\begin{equation}
\rho_2(\theta_1,\theta_2)=
m^{2}L^{2}\cosh\theta_{1}\cosh\theta_{2}+mL(\cosh\theta_{1}+\cosh\theta_{2})\varphi(\theta_{1}-\theta_{2}) 
\end{equation}
and also the density of states on the diagonal $I_1=I_2$ which can be obtained as the derivative of the 
degenerate ($\theta_1=\theta_2$) Bethe-Yang quantization condition as follows\footnote{
Note that due to (\ref{deltadef}) $\delta(0)=0$.}
\begin{eqnarray}
Q_d(\theta_1)&=&mL\sinh\theta_1=2\pi I_1 \nonumber\\
Q_d'(\theta_1)&=&mL\cosh\theta_1=\rho_1(\theta_1)
\end{eqnarray}
i.e. it coincides with the one-particle density $\rho_1$. 
The result is 
\begin{eqnarray}
Z_{2} & = & \frac{1}{2}\int\frac{d\theta_{1}}{2\pi}\int\frac{d\theta_{2}}{2\pi}
\rho_2(\theta_1,\theta_2)\mathrm{e}^{-mR\left(\cosh\theta_{1}+\cosh\theta_{2}\right)}
-\frac{1}{2}\int\frac{d\theta}{2\pi}\rho_1(\theta)\mbox{e}^{-2mR\cosh\theta}
\label{Z2}\end{eqnarray}

\subsubsection{The diagonal sum}
Using (\ref{diagff}), the diagonal matrix element is
\begin{equation}
\left\langle \{I_{1},I_{2}\}\left|\mathcal{O}\right|\{I_{1},I_{2}\}\right\rangle_L = 
\frac{F_{4s}^{\mathcal{O}}(\theta_{1},\theta_{2})
 +\rho_{1}(\theta_{1})F_{2c}^{\mathcal{O}}+\rho_{1}(\theta_{2})F_{2c}^{\mathcal{O}}
+\rho_{2}(\theta_{1},\theta_{2})\left\langle \mathcal{O}\right\rangle}
{\rho_{2}(\theta_{1},\theta_{2})}
\end{equation}
Substituting these into the diagonal sum
\begin{equation}
\sum_{I_{1}>I_{2}}
\left\langle\{I_{1},I_{2}\}\left|\mathcal{O}_{1}\right|\{I_{1},I_{2}\}\right\rangle_{L}
\left\langle\{I_{1},I_{2}\}\left|\mathcal{O}_{2}\right|\{I_{1},I_{2}\}\right\rangle_{L}
\mathrm{e}^{-mR(\cosh\theta_{1}+\cosh\theta_{2})}
\end{equation}
we have terms that can be ordered by the number of $F_{4s}$ factors they contain. 
The term which contains two $F_{4s}$ factors can be written as
\begin{eqnarray}
 &  & \sum_{I_{1}>I_{2}}\frac{\mathrm{e}^{-mR(\cosh\theta_{1}+\cosh\theta_{2})}}{\rho_{2}(\theta_{1},\theta_{2})^{2}}
F_{4s}^{\mathcal{O}_{1}}(\theta_{1},\theta_{2})
F_{4s}^{\mathcal{O}_{2}}(\theta_{1},\theta_{2})=
\nonumber\\
 &  & \left(\frac{1}{2}\sum_{I_{1},I_{2}}-\sum_{I_{1}=I_{2}}\right)
\frac{\mathrm{e}^{-mR(\cosh\theta_{1}+\cosh\theta_{2})}}{\rho_{2}(\theta_{1},\theta_{2})^{2}}
F_{4s}^{\mathcal{O}_{1}}(\theta_{1},\theta_{2})F_{4s}^{\mathcal{O}_{2}}(\theta_{1},\theta_{2})=
\nonumber\\
&  & 
\frac{1}{2}\int\frac{d\theta_{1}}{2\pi}\int\frac{d\theta_{2}}{2\pi}\frac{\mathrm{e}^{-mR(\cosh\theta_{1}+\cosh\theta_{2})}}{\rho_{2}(\theta_{1},\theta_{2})}F_{4s}^{\mathcal{O}_{1}}(\theta_{1},\theta_{2})F_{4s}^{\mathcal{O}_{2}}(\theta_{1},\theta_{2})
\nonumber\\ 
& & -\int\frac{d\theta_{1}}{2\pi}\frac{\mathrm{e}^{-2mR(\cosh\theta_{1})}mL\cosh\theta_1}{\rho_{2}(\theta_{1},\theta_{1})^2}
F_{4s}^{\mathcal{O}_{1}}(\theta_{1},\theta_{1})F_{4s}^{\mathcal{O}_{2}}(\theta_{1},\theta_{1})
\end{eqnarray}
where we added and subtracted the diagonal $\theta_1=\theta_2$. We used that the density of diagonal ($\theta_1=\theta_2$) 
two-particle states is given by $mL\cosh\theta_1$. The first term is $O(L^{-2})$ while the second is $O(L^{-3})$ and so they vanish as $L\rightarrow\infty.$

There are two terms containing a single $F_{4s}$. One of them is
\begin{equation}
\frac{1}{2}\sum_{I_{1},I_{2}}\frac{\mathrm{e}^{-mR(\cosh\theta_{1}+\cosh\theta_{2})}}{\rho_{2}(\theta_{1},\theta_{2})^{2}}
F_{4s}^{\mathcal{O}_{1}}(\theta_{1},\theta_{2})
\left(\rho_{1}(\theta_{1})F_{2c}^{\mathcal{O}_{2}}
+\rho_{1}(\theta_{2})F_{2c}^{\mathcal{O}_{2}}
+\rho_{2}(\theta_{1},\theta_{2})\left\langle \mathcal{O}_{2}\right\rangle \right)
\end{equation}
and the other can be obtained by interchanging $\mathcal{O}_1$ and $\mathcal{O}_2$. In writing the above formula 
we already included the $I_1=I_2$ diagonal, using again that it is suppressed by an $L^{-1}$ factor.
For $L\rightarrow\infty$ we get
\begin{equation}
\frac{1}{2}\int\frac{d\theta_{1}}{2\pi}\frac{d\theta_{2}}{2\pi}\mathrm{e}^{-mR(\cosh\theta_{1}+\cosh\theta_{2})}
F_{4s}^{\mathcal{O}_{1}}(\theta_{1},\theta_{2})\left\langle \mathcal{O}_{2}\right\rangle \end{equation}
Similarly, the other term yields
\begin{equation}
\frac{1}{2}\int\frac{d\theta_{1}}{2\pi}\frac{d\theta_{2}}{2\pi}\mathrm{e}^{-mR(\cosh\theta_{1}+\cosh\theta_{2})}
F_{4s}^{\mathcal{O}_{2}}(\theta_{1},\theta_{2})\left\langle \mathcal{O}_{1}\right\rangle \end{equation}

The terms without $F_{4s}$ give
\begin{eqnarray}
&&\sum_{I_{1}>I_{2}}
\frac{\mathrm{e}^{-mR(\cosh\theta_{1}+\cosh\theta_{2})}}{\rho_{2}(\theta_{1},\theta_{2})^{2}}
\left(\rho_{1}(\theta_{1})F_{2c}^{\mathcal{O}_{1}}
+\rho_{1}(\theta_{2})F_{2c}^{\mathcal{O}_{1}}
+\rho_{2}(\theta_{1},\theta_{2})\left\langle \mathcal{O}_{1}\right\rangle \right)
\nonumber\\
&&\quad\times\left(\rho_{1}(\theta_{1})F_{2c}^{\mathcal{O}_{2}}
+\rho_{1}(\theta_{2})F_{2c}^{\mathcal{O}_{2}}
+\rho_{2}(\theta_{1},\theta_{2})\left\langle \mathcal{O}_{2}\right\rangle \right)
\end{eqnarray}
We can replace
\begin{equation}
\sum_{I_{1}>I_{2}}\rightarrow\frac{1}{2}\sum_{I_{1},I_{2}}-\frac{1}{2}\sum_{I_{1}=I_{2}}
\end{equation}
and after converting the sums to integrals we obtain
\begin{eqnarray}
 &  & \frac{1}{2}\int\frac{d\theta_{1}}{2\pi}\int\frac{d\theta_{2}}{2\pi}
\frac{\mathrm{e}^{-mR(\cosh\theta_{1}+\cosh\theta_{2})}}{\rho_{2}(\theta_{1},\theta_{2})}
\left(\rho_{1}(\theta_{1})F_{2c}^{\mathcal{O}_{1}}+\rho_{1}(\theta_{2})F_{2c}^{\mathcal{O}_{1}}+\rho_{2}(\theta_{1},\theta_{2})\left\langle \mathcal{O}_{1}\right\rangle \right)
\nonumber\\
 &\times& 
\left(\rho_{1}(\theta_{1})F_{2c}^{\mathcal{O}_{2}}+\rho_{1}(\theta_{2})F_{2c}^{\mathcal{O}_{2}}+\rho_{2}(\theta_{1},\theta_{2})\left\langle \mathcal{O}_{2}\right\rangle \right)
 - \mbox{diagonal term}\nonumber\\
 & = & \int\frac{d\theta_{1}}{2\pi}\int\frac{d\theta_{2}}{2\pi}
\Bigg[
\frac{1}{2}\mathrm{e}^{-mR(\cosh\theta_{1}+\cosh\theta_{2})}
F_{2c}^{\mathcal{O}_{1}}F_{2c}^{\mathcal{O}_{2}}
\frac{\left(\cosh\theta_{1}+\cosh\theta_{2}\right)^{2}}{\cosh\theta_{1}\cosh\theta_{2}}\nonumber\\
 & + & mL\cosh\theta_{1}\mathrm{e}^{-mR(\cosh\theta_{1}+\cosh\theta_{2})}
\left(\left\langle \mathcal{O}_{1}\right\rangle F_{2c}^{\mathcal{O}_{2}}
+\left\langle \mathcal{O}_{2}\right\rangle F_{2c}^{\mathcal{O}_{1}}\right)\nonumber\\
 & + & \frac{1}{2}\mathrm{e}^{-mR(\cosh\theta_{1}+\cosh\theta_{2})}
\left(m^{2}L^{2}\cosh\theta_{1}\cosh\theta_{2}+mL(\cosh\theta_{1}
+\cosh\theta_{2})\varphi(\theta_{1}-\theta_{2})\right)
\left\langle \mathcal{O}_{1}\right\rangle \left\langle \mathcal{O}_{2}\right\rangle 
\Bigg]\nonumber\\
 & - & \frac{1}{2}\int\frac{d\theta}{2\pi}
\mathrm{e}^{-2mR\cosh\theta}
\left[mL\cosh\theta\left\langle \mathcal{O}_{1}\right\rangle \left\langle \mathcal{O}_{2}\right\rangle 
+2\left(\left\langle \mathcal{O}_{1}\right\rangle F_{2c}^{\mathcal{O}_{2}}
+\left\langle \mathcal{O}_{2}\right\rangle F_{2c}^{\mathcal{O}_{1}}\right)\right]\end{eqnarray}
where we dropped terms that vanish as $L\to\infty$. This has terms which diverge in the limit; 
however, we must now add the ``counter terms''
\begin{eqnarray}
-Z_{2}C_{00} &=& \frac{1}{2}\int\frac{d\theta}{2\pi}
mL\cosh\theta\mbox{e}^{-2mR\cosh\theta}
\left\langle \mathcal{O}_{1}\right\rangle \left\langle \mathcal{O}_{2}\right\rangle \\
&& -\frac{1}{2}\int\frac{d\theta_{1}}{2\pi}\int\frac{d\theta_{2}}
{2\pi}\big[m^{2}L^{2}\cosh\theta_{1}\cosh\theta_{2}
\nonumber\\
&&+mL(\cosh\theta_{1}+\cosh\theta_{2})
\varphi(\theta_{1}-\theta_{2})\big]\mathrm{e}^{-mR\left(\cosh\theta_{1}+\cosh\theta_{2}\right)}
\left\langle \mathcal{O}_{1}\right\rangle \left\langle \mathcal{O}_{2}\right\rangle 
\nonumber
\end{eqnarray}
and
\begin{eqnarray}
-Z_{1}D_{11} & = & -Z_{1}\int\frac{d\theta_{1}}{2\pi}\int\frac{d\theta_{2}}{2\pi}
F_{2}^{\mathcal{O}_{1}}(\theta_{1}+i\pi,\theta_{2})F_{2}^{\mathcal{O}_{2}}(\theta_{1},\theta_{2}+i\pi)
\nonumber\\
 &  & \times
\mathrm{e}^{imx(\sinh\theta_{1}-\sinh\theta_{2})}\mathrm{e}^{-m(R-t)\cosh\theta_{1}}
\mathrm{e}^{-mt\cosh\theta_{2}}
\nonumber\\
 &  & -Z_{1}\int\frac{d\theta}{2\pi}\left(F_{2c}^{\mathcal{O}_{1}}\left\langle \mathcal{O}_{2}\right\rangle +F_{2c}^{\mathcal{O}_{2}}\left\langle \mathcal{O}_{1}\right\rangle \right)\mbox{e}^{-mR\cosh\theta}\end{eqnarray}
These cancel all the divergences leaving us with the final result for the diagonal contribution:
\begin{eqnarray}
\mbox{D}_{22}^{\mathrm{(diag)}} & = & 
\frac{1}{2}\int\frac{d\theta_{1}}{2\pi}\frac{d\theta_{2}}{2\pi}
\mathrm{e}^{-mR(\cosh\theta_{1}+\cosh\theta_{2})}
\left(F_{4s}^{\mathcal{O}_{1}}(\theta_{1},\theta_{2})\left\langle \mathcal{O}_{2}\right\rangle 
+F_{4s}^{\mathcal{O}_{2}}(\theta_{1},\theta_{2})\left\langle \mathcal{O}_{1}\right\rangle \right)
\nonumber\\
 & + & \frac{1}{2}\int\frac{d\theta_{1}}{2\pi}\int\frac{d\theta_{2}}{2\pi}
\mathrm{e}^{-mR(\cosh\theta_{1}+\cosh\theta_{2})}
F_{2c}^{\mathcal{O}_{1}}F_{2c}^{\mathcal{O}_{2}}
\frac{\left(\cosh\theta_{1}+\cosh\theta_{2}\right)^{2}}{\cosh\theta_{1}\cosh\theta_{2}}\nonumber\\
 & - & 
\int\frac{d\theta}{2\pi}\mathrm{e}^{-2mR\cosh\theta}\left(\left\langle \mathcal{O}_{1}\right\rangle F_{2c}^{\mathcal{O}_{2}}
+\left\langle \mathcal{O}_{2}\right\rangle F_{2c}^{\mathcal{O}_{1}}\right)
\label{D22diagresult}\end{eqnarray}
However, the first piece of $-Z_{1}D_{11}$: 
\begin{equation}
-Z_{1}\int\frac{d\theta_{1}}{2\pi}\int\frac{d\theta_{2}}{2\pi}F_{2}^{\mathcal{O}_{1}}(\theta_{1}+i\pi,\theta_{2})F_{2}^{\mathcal{O}_{2}}(\theta_{1},\theta_{2}+i\pi)\mathrm{e}^{imx(\sinh\theta_{1}-\sinh\theta_{2})}\mathrm{e}^{-m(R-t)\cosh\theta_{1}}\mathrm{e}^{-mt\cosh\theta_{2}}
\label{Z1D11remainder}\end{equation}
is not canceled by the diagonal part. We now turn to the evaluation of the non-diagonal contribution, 
which does eliminate this last divergence, as explicitly demonstrated in appendix \ref{D22QF}. 

\subsubsection{Evaluating the non-diagonal part}
Now we must evaluate
\begin{eqnarray}
C_{22}^{\mathrm{(nondiag)}} & = & \sum_{I_{1}>I_{2}}\sum_{J_{1}>J_{2}}{}'\left\langle
  \{I_{1},I_{2}\}\left|\mathcal{O}_{1}\right|\{J_{1},J_{2}\}\right\rangle
_{L}\left\langle
  \{J_{1},J_{2}\}\left|\mathcal{O}_{2}\right|\{I_{1},I_{2}\}\right\rangle
_{L}
\nonumber\\
&&\times K_{t,x}^{(R)}(\theta_{1},\theta_{2};\theta_{1}',\theta_{2}')
\end{eqnarray}
where
\begin{eqnarray}
K_{t,x}^{(R)}(\theta_{1},\theta_{2};\theta_{1}',\theta_{2}') & = & 
\mathrm{e}^{imx(\sinh\theta_{1}+\sinh\theta_{2}-\sinh\theta_{1}'-\sinh\theta_{2}')}
\mathrm{e}^{-m(R-t)(\cosh\theta_{1}+\cosh\theta_{2})}
\nonumber\\
&&\times\mathrm{e}^{-mt(\cosh\theta_{1}'+\cosh\theta_{2}')}
\end{eqnarray}
Using (\ref{eq:genffrelation}), the matrix elements are of the form
\begin{eqnarray}
\left\langle \{I_{1},I_{2}\}\left|\mathcal{O}_{1}\right|\{J_{1},J_{2}\}\right\rangle _{L} & = & 
\frac{F_{4}^{\mathcal{O}_{1}}(\theta_{2}+i\pi,\theta_{1}+i\pi,\theta_{1}',\theta_{2}')}
{\rho_{2}(\theta_{1},\theta_{2})^{1/2}\rho_{2}(\theta_{1}',\theta_{2}')^{1/2}}
\nonumber\\
\left\langle \{J_{1},J_{2}\}\left|\mathcal{O}_{2}\right|\{I_{1},I_{2}\}\right\rangle _{L} & = & 
\frac{F_{4}^{\mathcal{O}_{2}}(\theta_{2}'+i\pi,\theta_{1}'+i\pi,\theta_{1},\theta_{2})}
{\rho_{2}(\theta_{1},\theta_{2})^{1/2}\rho_{2}(\theta_{1}',\theta_{2}')^{1/2}}
\end{eqnarray}
The quantization conditions read
\begin{eqnarray}
Q_{1}(\theta_{1},\theta_{2}) & = & mL\sinh\theta_{1}+\delta(\theta_{1}-\theta_{2})=2\pi I_{1}\nonumber\\
Q_{2}(\theta_{1},\theta_{2}) & = & mL\sinh\theta_{2}+\delta(\theta_{2}-\theta_{1})=2\pi I_{2}
\label{D22Qs}\end{eqnarray}
and\begin{eqnarray}
Q_{1'}(\theta_{1}',\theta_{2}') & = & mL\sinh\theta_{1}'+\delta(\theta_{1}'-\theta_{2}')=2\pi J_{1}\nonumber\\
Q_{2'}(\theta_{1}',\theta_{2}') & = & mL\sinh\theta_{2}'+\delta(\theta_{2}'-\theta_{1}')=2\pi J_{2}
\label{D22Qps}\end{eqnarray}
Now we can write
\begin{eqnarray}
 &  & \left\langle \{I_{1},I_{2}\}\left|\mathcal{O}_{1}\right|\{J_{1},J_{2}\}\right\rangle _{L}\left\langle \{J_{1},J_{2}\}\left|\mathcal{O}_{2}\right|\{I_{1},I_{2}\}\right\rangle _{L}K_{t,x}^{(R)}(\theta_{1},\theta_{2};\theta_{1}',\theta_{2}')\\
 &  & =\oint\oint_{C_{J_{1}J_{2}}}\frac{d\theta_{1}'}{2\pi}\frac{d\theta_{2}'}{2\pi}
F_{4}^{\mathcal{O}_{1}}(\theta_{2}+i\pi,\theta_{1}+i\pi,\theta_{1}',\theta_{2}')
F_{4}^{\mathcal{O}_{2}}(\theta_{2}'+i\pi,\theta_{1}'+i\pi,\theta_{1},\theta_{2})
\nonumber\\
 &  & \times
\frac{
K_{t,x}^{(R)}(\theta_{1},\theta_{2};\theta_{1}',\theta_{2}')}
{\left(\mathrm{e}^{iQ_{1'}(\theta_{1}',\theta_{2}')}+1\right)\left(\mathrm{e}^{iQ_{2'}(\theta_{1}',\theta_{2}')}+1\right)}
\end{eqnarray}
where $C_{J_{1}J_{2}}$ is a multi-contour surrounding the solution of (\ref{D22Qps}). When we open the multi-contours to surround the real axes 
we encounter new singularities. These can be classified as follows:
\begin{enumerate}
\item $QF$-singularities: partly from the $Q$s, partly from the $F$s:
\begin{eqnarray}
\theta_{1}'=\theta_{1} & \quad\mbox{and}\quad & Q_{2'}(\theta_{1}',\theta_{2}')=2\pi J_{2}\\
\theta_{1}'=\theta_{2} & \quad\mbox{and}\quad & Q_{2'}(\theta_{1}',\theta_{2}')=2\pi J_{2}\\
\theta_{2}'=\theta_{1} & \quad\mbox{and}\quad & Q_{1'}(\theta_{1}',\theta_{2}')=2\pi J_{1}\\
\theta_{2}'=\theta_{2} & \quad\mbox{and}\quad & Q_{1'}(\theta_{1}',\theta_{2}')=2\pi J_{1}
\label{D22QFclass}\end{eqnarray}

\item $FF$-singularities: come from the $F$s\begin{eqnarray}
\theta_{1}' & = & \theta_{2}'=\theta_{1}\\
\theta_{1}' & = & \theta_{2}'=\theta_{2}
\label{D22FFclass}\end{eqnarray}

\item (Spurious) $QQ$-singularities: these result from 
\begin{eqnarray}
\theta_{1}'=\theta_{1} & \quad\mbox{and}\quad & \theta_{2}'=\theta_{2}\\
\theta_{1}'=\theta_{2} & \quad\mbox{and}\quad & \theta_{2}'=\theta_{1}
\label{D22QQclass}\end{eqnarray}
(it turns out that eventually these do not give any contributions in the $L\to\infty$ limit).
\end{enumerate}

Since the evaluation of these contributions is the same as for $D_{12}$, the details are relegated 
to appendix \ref{D22sing}. The upshot is that 
\begin{equation}
\sum_{J_{1}>J_{2}}{}'\oint\oint_{C_{J_{1}J_{2}}}\frac{d\theta_{1}'}{2\pi}\frac{d\theta_{2}'}{2\pi}=
\frac{1}{2}\oint\oint_{C}\frac{d\theta_{1}'}{2\pi}\frac{d\theta_{2}'}{2\pi}
-(FF\mbox{ terms})-(QF\mbox{ terms})
\end{equation}
where $C$ is the open multi-contour. Here we also used the fact that there are no singularities at $J_{1}=J_{2}$
because the form factors vanish.

\subsubsection{End result for $D_{22}$}
Putting together the results (\ref{D22QFfinal}), (\ref{D22SQFfinal}) and (\ref{D22FFfinal}) 
of appendix \ref{D22sing} with (\ref{D22diagresult}) one obtains
\begin{eqnarray}
 && D_{22}=\label{D22final} \\
 && \frac{1}{4}\iint\frac{d\theta_{1}}{2\pi}\frac{d\theta_{2}}{2\pi}\iint_{C_{++}}\frac{d\theta_{1}'}{2\pi}\frac{d\theta_{2}'}{2\pi}
F_{4}^{\mathcal{O}_{1}}(\theta_{2}+i\pi,\theta_{1}+i\pi,\theta_{1}',\theta_{2}')
F_{4}^{\mathcal{O}_{2}}(\theta_{2}'+i\pi,\theta_{1}'+i\pi,\theta_{1},\theta_{2})
\nonumber\\
 &&\times
K_{t,x}^{(R)}(\theta_{1},\theta_{2};\theta_{1}',\theta_{2}')
\nonumber\\
 &+& \frac{1}{2}\iint\frac{d\theta_{1}}{2\pi}\frac{d\theta_{2}}{2\pi}
\mathrm{e}^{-mR(\cosh\theta_{1}+\cosh\theta_{2})}
\left(F_{4s}^{\mathcal{O}_{1}}(\theta_{1},\theta_{2})\left\langle \mathcal{O}_{2}\right\rangle 
+F_{4s}^{\mathcal{O}_{2}}(\theta_{1},\theta_{2})\left\langle \mathcal{O}_{1}\right\rangle \right)
\nonumber\\
 &+& \frac{1}{2}\iint\frac{d\theta_{1}}{2\pi}\frac{d\theta_{2}}{2\pi}
\mathrm{e}^{-mR(\cosh\theta_{1}+\cosh\theta_{2})}F_{2c}^{\mathcal{O}_{1}}F_{2c}^{\mathcal{O}_{2}}
\frac{\left(\cosh\theta_{1}+\cosh\theta_{2}\right)^{2}}{\cosh\theta_{1}\cosh\theta_{2}}
\nonumber\\
 &-& \int\frac{d\theta}{2\pi}\mathrm{e}^{-2mR\cosh\theta}
\left(\left\langle \mathcal{O}_{1}\right\rangle F_{2c}^{\mathcal{O}_{2}}
+\left\langle \mathcal{O}_{2}\right\rangle F_{2c}^{\mathcal{O}_{1}}\right)
\nonumber\\
 &+& \iint\frac{d\theta_{1}}{2\pi}\frac{d\theta_{2}}{2\pi}
\int\frac{d\theta_{2}'}{2\pi}F_{2}^{O_{1}}(\theta_{2}+i\pi,\theta_{2}')F_{2}^{O_{2}}(\theta_{2}'+i\pi,\theta_{2})
\nonumber\\
 && \times
\mathrm{e}^{imx(\sinh\theta_{2}-\sinh\theta_{2}')}\mathrm{e}^{-mR\cosh\theta_{1}}
\mathrm{e}^{-m(R-t)\cosh\theta_{2}}\mathrm{e}^{-mt\cosh\theta_{2}'}
\nonumber\\
 && \times
 \left((mx\cosh\theta_{1}-imt\sinh\theta_{1})(1-S(\theta_{2}'-\theta_{1})S(\theta_{1}-\theta_{2}))
+\varphi(\theta_{1}-\theta_{2}')S(\theta_{2}'-\theta_{1})S(\theta_{1}-\theta_{2})\right)
\nonumber\\
 &+& \frac{1}{2}\iint\frac{d\theta_{1}}{2\pi}\frac{d\theta_{2}}{2\pi}\int\frac{d\theta_{2}'}{2\pi}
\Big[\mathrm{e}^{imx(\sinh\theta_{2}-\sinh\theta_{2}')}
\mathrm{e}^{-mR\cosh\theta_{1}}\mathrm{e}^{-m(R-t)\cosh\theta_{2}}\mathrm{e}^{-mt\cosh\theta_{2}'}
\nonumber\\
 && \times\left(F_{2}^{O_{1}}(\theta_{2}+i\pi,\theta_{2}')
F_{4sc}^{\mathcal{O}_{2}}(\theta_{2}',\theta_{1}|\theta_{1},\theta_{2})
+F_{2}^{O_{2}}(\theta_{2}'+i\pi,\theta_{2})
F_{4sc}^{\mathcal{O}_{1}}(\theta_{2}',\theta_{1}|\theta_{1},\theta_{2})\right)
\nonumber\\
 && +\left(\theta_{1}\leftrightarrow\theta_{2}\right)\Big]
\nonumber\\
 &-& \iint\frac{d\theta_{1}}{2\pi}\frac{d\theta_{2}}{2\pi}\Big[
\mathrm{e}^{imx(\sinh\theta_{2}-\sinh\theta_{1})}\mathrm{e}^{-m(R-t)(\cosh\theta_{1}+\cosh\theta_{2})}
\mathrm{e}^{-2mt\cosh\theta_{1}}
\nonumber\\
 && \times
S(\theta_{1}-\theta_{2})
F_{2}^{O_{1}}(\theta_{2}+i\pi,\theta_{1})F_{2}^{O_{2}}(\theta_{1}+i\pi,\theta_{2})
\nonumber\\
 && +\left(\theta_{1}\leftrightarrow\theta_{2}\right)\Big]
\nonumber\\
&-&\iint\frac{d\theta_{1}}{2\pi}\frac{d\theta_{2}}{2\pi}
F_{2}^{O_{1}}(\theta_{1}+i\pi,\theta_{2})F_{2}^{O_{2}}(\theta_{2}+i\pi,\theta_{1})
\mathrm{e}^{imx(\sinh\theta_{1}-\sinh\theta_{2})-m(2R-t)\cosh\theta_{1}-mt\cosh\theta_{2}}
\nonumber
\end{eqnarray}
where the function $F_{4sc}$ is defined in (\ref{F4scdef}), and $C_{++}$ denotes the 
integration contour specified in (\ref{Cppcontourdef}).
All the other integrals are taken over real values of their variables.

\subsection{Discussion of the proposal of LeClair and Mussardo}

\label{LM-discussion}

In \cite{Leclair:1999ys} LeClair and Mussardo introduced a 
regularization scheme for finite temperature 
correlation functions. The two main assumptions of the
proposal are that the spectral expansion should be built using the
zero-temperature form factors, and that the only effect of finite
temperature is an appropriate modification (dressing) of the
statistical weight functions and the one-particle energies and
momenta. In the case of one-point function the proposed formula was
proven to be correct up to the third order in \cite{Pozsgay:2007gx}; 
an all-orders proof is also possible \cite{pozsi-LM}. However, 
the two-point function seems to be more problematic 
\cite{Saleur:1999hq,CastroAlvaredo:2002ud}. 

In the following we compare our results to the proposal of
\cite{Leclair:1999ys}. For the two-point functions their formula reads
\begin{eqnarray}
\vev{\mathcal{O}(x,t)\mathcal{O}(0,0)}^R&=&
\big(\vev{\mathcal{O}}^R\big)^2+
\sum_{N=1}^\infty \frac{1}{N!} \sum_{\sigma_i=\pm 1}
\int \frac{d\theta_1}{2\pi}\dots \frac{d\theta_N}{2\pi}
\left[\prod_{j=1}^N f_{\sigma_j}(\theta_j) 
\mathrm{e}^{-\sigma_j(t\eps_j+ixk_j)}\right]\nonumber\\
&&\times \big|\bra{0}\mathcal{O}\ket{\theta_1\dots\theta_N}_{\sigma_1\dots\sigma_N}\big|^2
  \label{eq:2point_leclair_mussardo}
\end{eqnarray}
where $f_{\sigma_j}(\theta_j)=1/(1+e^{-\sigma_j \eps(\theta_j)})$,
$\eps_j=\eps(\theta_j)/R$ and $k_j=k(\theta_j)$ with 
$\eps(\theta)$ being the solution of the TBA equation
\begin{equation}
\epsilon(\theta)=mR\cosh\theta-\int\frac{d\theta'}{2\pi}\varphi(\theta-\theta')\log(1+\mathrm{e}^{-\epsilon(\theta')})\label{eq:TBA}
\end{equation}
and $k(\theta)$ is given by 
\begin{eqnarray}
k(\theta)&=&m\sinh\theta+\int d\theta' \delta(\theta-\theta')
  \rho_1(\theta')\nonumber\\
2\pi \rho_1(\theta)(1+\mathrm{e}^{\eps(\theta)})
&=&m\cosh\theta+\int d\theta' \varphi(\theta-\theta') \rho_1(\theta')
\end{eqnarray}
The form factors appearing in \eqref{eq:2point_leclair_mussardo} are
defined by
\begin{equation*}
  \bra{0}\mathcal{O}\ket{\theta_1\dots\theta_N}_{\sigma_1\dots\sigma_N}=
F^{\mathcal{O}}_N(\theta_1-i\pi\tilde{\sigma}_1,\dots,\theta_N-i\pi\tilde{\sigma}_N)\quad\quad
\tilde{\sigma}_j=(1-\sigma_j)/2 \in \{0,1\}
\end{equation*}
The interpretation of the series \eqref{eq:2point_leclair_mussardo} is
as follows: the excitations with $\sigma_j=+1$ or $\sigma_j=-1$
represent particles or holes over an infinite volume thermal
state. Therefore the statistical weight functions and the one-particle energies and
momenta are given by the dressed values as calculated in TBA. 

If the (\ref{eq:2point_leclair_mussardo}) series expression were correct, 
then a systematic double expansion in terms of
$\mathrm{e}^{-mt}$ and $\mathrm{e}^{-m(R-t)}$ should reproduce our results. 
Indeed, the first few terms indicate that this might be true. It was
already pointed out in \cite{Pozsgay:2009pv} that to the lowest order
in $\mathrm{e}^{-mR}$ the term $N=2$ reproduces our
$D_{02}+D_{11}+D_{20}$. Moreover, in the case of our $D_{12}$, the
last line of formula \eqref{D12} 
suggests the dressing
\begin{equation*}
  e^{-mt\cosh\theta +imx\sinh\theta } \quad\to\quad
  \frac{e^{-mt\cosh\theta +imx\sinh\theta }}{1+e^{-mR\cosh\theta }}+\ordo(e^{-2mR})
\end{equation*}
for the exponential factors in $D_{01}$. This pattern repeats itself
and similar contributions can be found in $D_{1n}$ \eqref{D1n}, which suggest a
dressing of the factors in $D_{0,n-1}$. However, the situation is more complicated as
we consider higher order terms. 

First of all observe, that the formula \eqref{eq:2point_leclair_mussardo} 
is not well-defined for $N\ge 3$. 
There appear second order poles whenever the rapidity of a particle
approaches the rapidity of a hole, and in the original work
\cite{Leclair:1999ys} it is not explained how to integrate over these
singularities. Note that it was the evaluation of these ill-defined
terms which required a lot of effort in our evaluation of the two-point function. 

Based on the form of our results \eqref{D12}, \eqref{D1n} and
\eqref{D22final} it seems unlikely, that any regularized form of
\eqref{eq:2point_leclair_mussardo} would be correct. However, at
present we
cannot make any definitive statement about this issue. The inspection
of higher order
terms ($D_{nm}$ with $n,m>2$) might decide whether there exists a neat
formula for the two-point function, possibly with a structure similar
to \eqref{eq:2point_leclair_mussardo} but with different dressing
prescriptions. This problem is left for future work.

\section{Second order form factor perturbation theory}

\label{section5}

As a further application of the framework presented here, 
we show how to derive the main results of the paper \cite{Takacs:2009fu} on second order 
form factor perturbation theory using the present formalism. 
We simplify the presentation by considering a theory with a single massive particle in 
its spectrum instead of the double sine-Gordon theory treated in \cite{Takacs:2009fu}. 
Consider modifying the Hamiltonian of an integrable model as follows: 
\begin{equation}
H_{\mathrm{nonintegrable}}=H_{\mathrm{integrable}}+\lambda\int dx\Psi(t,x)\end{equation}
where $\Psi$ denotes a local (Lorentz scalar) field which breaks integrability. 
Corrections that are first order in $\lambda$ were derived in \cite{Delfino:1996xp}, 
but when evaluating the second order one encounters the same difficulties with disconnected terms 
as in the case of the thermal two-point function. 
The principle of the solution to this problem is the same as for the thermal correlator: 
we perform perturbation theory in finite volume, express the quantities we are interested in 
and then take the limit $L\to\infty$. In the approach of \cite{Takacs:2009fu} it was necessary 
to compute some  part of the discrete sum over the finite volume quantum numbers explicitly; we show that 
this can be greatly simplified by applying the residue methods of the present work.

The general perturbation theory formula for second order corrections to energy levels is
\begin{equation}
\delta E_{i}=\sum_{k\neq i}
\frac{\left|\left\langle i\left|H_{1}\right|k\right\rangle \right|^{2}}
{E_{i}^{(0)}-E_{k}^{(0)}}
\quad ,\quad H_1=\lambda\int dx\Psi(t,x)
\end{equation}
therefore the correction to the vacuum level can be written as 
\begin{equation}
\delta E_{0}=-\lambda^{2}L^{2}\sum_{k\neq0}\frac{\left|\left\langle 0\left|:\exp i\frac{\beta}{2}\varphi(0,0):\right|k\right\rangle _{L}\right|^{2}}{E_{k}^{(0)}-E_{0}^{(0)}}
\label{vac2order}\end{equation}
The summation goes over all excited states in the spectrum (with zero total momentum selected for by 
translational invariance), which can be described using the Bethe-Yang picture of section \ref{finvolFF}. The leading contribution is given by the state containing a single stationary particle, and can be written as
\begin{equation}\delta E_{0}(L) = -\lambda^{2}L^{2}\frac{\left|\left\langle 0\left|:\exp i\frac{\beta}{2}\varphi(0,0):\right|\{0\}\right\rangle _{L}\right|^{2}}{m}
+O\left(\mathrm{e}^{-\mu L}\right)
\end{equation}
Using the relation (\ref{eq:genffrelation}) we obtain
\begin{equation}\delta E_{0}(L)=
-\lambda^{2}L^{2}\frac{\left|F^\Psi_{1}\right|^{2}}{\rho_{1}(0)m}+O\left(\mathrm{e}^{-\mu L}\right)
=-\lambda^{2}L\frac{\left|F^\Psi_{1}\right|^{2}}{m^{2}}+O\left(\mathrm{e}^{-\mu L}\right)
\label{E0corrfinal}\end{equation}
which results in the following shift of the bulk energy density
\begin{equation}
\delta\mathcal{E}=
-\lambda^{2}\frac{\left|F^\Psi_{1}\right|^{2}}{m^{2}}
\end{equation}
Next we are interested in the correction to the particle mass. This can be obtained by evaluating the correction 
to the first zero-momentum excited level in the finite volume system and then taking the limit
\begin{equation}
\delta m =  \lim_{L\rightarrow\infty}\delta E_{1}(L)-\delta E_{0}(L)
\label{massaslimit}\end{equation}
The correction to the first level can be written as
\begin{eqnarray}
\delta E_{1}(L) &=& 
\lambda^{2}L^{2}\frac{\left|\left\langle \{0\}\left|:\exp i\frac{\beta}{2}\varphi(0,0):\right|{0}\right\rangle _{L}\right|^{2}}{m}
+
\lambda^{2}L^{2}\sum_I\frac{\left|\left\langle \{0\}\left|:\exp i\frac{\beta}{2}\varphi(0,0):\right|\{I,-I\}\right\rangle _{L}\right|^{2}}{m-2m\cosh\theta}
\nonumber\\
&=& \lambda^{2}L^{2}\frac{\left|F_{1}\right|^{2}}{\rho_{1}(0)m}
-\lambda^{2}L^{2}\sum_{\theta}
\frac{F_{3}(i\pi,\theta,-\theta)F_{3}(\theta,-\theta,i\pi)}
{\rho_{1}(0)\rho_{2}(\theta,-\theta)(2m\cosh\theta-m)}
\end{eqnarray}
(where we omitted the states with three or more particles) where $\theta$ is the solution of (cf. subsection \ref{finvolFF}).
\begin{equation}
Q(\theta)=mL\sinh\theta+\delta(2\theta) = 2\pi I\qquad,\qquad I\in\mathbb{N}+\frac{1}{2}
\end{equation}
and
\begin{equation}
\rho_{2}(\theta,-\theta) =  mL\cosh\theta(mL\cosh\theta+2\varphi(2\theta))
\end{equation}
Extending the sum over $\theta$ to negative values and performing the residue trick 
we get 
\begin{eqnarray}
\delta E_{1}(L)&=&\lambda^{2}L\frac{\left|F_{1}\right|^{2}}{m^{2}}
\nonumber\\
&&+\frac{\lambda^{2}L}{2}\sum_{I\in\mathbb{Z}+\frac{1}{2}}\oint_{C_{I}}\frac{d\theta}{2\pi}\frac{1}{\mathrm{e}^{iQ(\theta)}+1}\frac{\tilde{\rho}_{2}(\theta)}{\rho_{2}(\theta,-\theta)}\frac{F_{3}(i\pi,\theta,-\theta)F_{3}(-\theta+i\pi,\theta+i\pi,0)}{m(2m\cosh\theta-m)}
\nonumber\\
&=&\lambda^{2}L\frac{\left|F_{1}\right|^{2}}{m^{2}}+\frac{\lambda^{2}}{2}\sum_{I\in\mathbb{Z}+\frac{1}{2}}\oint_{C_{I}}\frac{d\theta}{2\pi}\frac{1}{\mathrm{e}^{iQ(\theta)}+1}\frac{F_{3}(i\pi,\theta,-\theta)F_{3}(-\theta+i\pi,\theta+i\pi,0)}{m^{3}(2\cosh\theta-1)\cosh\theta}
\end{eqnarray}
where 
\begin{equation}
\tilde{\rho}_{2}(\theta)=mL\cosh\theta+2\varphi(2\theta)
\end{equation}
is nothing else than the density of two-particle states with zero total momentum. 
Using (\ref{eq:kinematicalaxiom}), the form factor has the following singularity at $\theta=0$
\begin{equation}
\left|F_3(i\pi,\theta,-\theta)\right|^{2}
\sim\frac{16\left|F_{1}\right|^{2}}{\theta^{2}}+O(\theta^{0})
\end{equation}
where we also used $S(0)=-1$. Subtracting and adding the singular term at the origin results in
\begin{eqnarray}
\delta E_{1}(L) & = & \lambda^{2}L\frac{\left|F_{1}\right|^{2}}{m^{2}}+\lambda^{2}\frac{1}{2}\sum_{I\in\mathbb{Z}+\frac{1}{2}}\oint_{C_{I}}\frac{d\theta}{2\pi}\frac{1}{\mathrm{e}^{iQ(\theta)}+1}\Bigg[\frac{F_{3}(i\pi,\theta,-\theta)F_{3}(-\theta+i\pi,\theta+i\pi,0)}{m^{3}(2\cosh\theta-1)\cosh\theta}\nonumber\\
 &  & -\frac{16\left|F_{1}\right|^{2}}{m^{3}\sinh^{2}\theta\cosh\theta}\Bigg]+\lambda^{2}\frac{1}{2}\sum_{I\in\mathbb{Z}+\frac{1}{2}}\oint_{C_{I}}\frac{d\theta}{2\pi}\frac{1}{\mathrm{e}^{iQ(\theta)}+1}\frac{16\left|F_{1}\right|^{2}}{m^{3}\sinh^{2}\theta\cosh\theta}
\end{eqnarray}
We can then open the contours to surround the real axis. In the first term, only the upper contour
contributes in the infinite volume limit, and one can also take $\epsilon\rightarrow0$. 
In the second term, we can open the contour, but we must also subtract the contribution 
of the double pole at the origin since that is not included in the original sum:
\begin{eqnarray}
 &  & \frac{\lambda^{2}}{2}\sum_{I\in\mathbb{Z}+\frac{1}{2}}\oint_{C_{I}}\frac{d\theta}{2\pi}\frac{1}{\mathrm{e}^{iQ(\theta)}+1}\frac{16\left|F_{1}\right|^{2}}{m^{3}\sinh^{2}\theta\cosh\theta}
\\
 &  & -\frac{\lambda^{2}}{2}\int_{-\infty}^{\infty}\frac{d\theta}{2\pi}\frac{16\left|F_{1}\right|^{2}}{m^{3}\sinh^{2}(\theta+i\epsilon)\cosh(\theta+i\epsilon)} 
 -\frac{\lambda^{2}}{2}\oint_{C_{0}}\frac{d\theta}{2\pi}\frac{1}{\mathrm{e}^{iQ(\theta)}+1}\frac{16\left|F_{1}\right|^{2}}{m^{3}\sinh^{2}\theta\cosh\theta}
\nonumber
\end{eqnarray}
The first integral is\begin{equation}
-\frac{\lambda^{2}}{2}\int_{-\infty}^{\infty}\frac{d\theta}{2\pi}\frac{16\left|F_{1}\right|^{2}}{m^{3}\sinh^{2}(\theta+i\epsilon)\cosh(\theta+i\epsilon)}=\lambda^{2}\frac{16\left|F_{1}\right|^{2}}{4m^{3}}\end{equation}
while the second integral is given by the residue theorem as
\begin{equation}
-\frac{\lambda^{2}}{2}i\left.\frac{\partial}{\partial\theta}\left(\frac{1}{\mathrm{e}^{iQ(\theta)}+1}\frac{16\left|F_{1}\right|^{2}}{m^{3}\cosh\theta}\right)\right|_{\theta=0}=-\frac{\lambda^{2}}{2}\frac{16\left|F_{1}\right|^{2}}{m^{3}}\frac{\mathrm{e}^{iQ(0)}Q'(0)}{(1+\mathrm{e}^{iQ(0)})^{2}}\end{equation}
Using
\begin{equation}
Q(0)=0\qquad,\qquad Q'(0)=mL+2\varphi(0)\end{equation}
we get 
\begin{eqnarray}
\delta E_{1}(L) & = & -\lambda^{2}L\frac{\left|F_{1}\right|^{2}}{m^{2}}-\lambda^{2}\int_{0}^{\infty}\frac{d\theta}{2\pi}\left(\frac{\left|F_{3}(i\pi,\theta,-\theta)\right|^{2}}{m^{3}(2\cosh\theta-1)\cosh\theta}-\frac{16\left|F_{1}\right|^{2}}{m^{3}\sinh^{2}\theta\cosh\theta}\right)
\nonumber\\
 & + & \lambda^{2}\frac{16\left|F_{1}\right|^{2}}{4m^{3}}(1-\varphi(0))\end{eqnarray}
Finally, using (\ref{massaslimit}) and (\ref{E0corrfinal}) we obtain the mass correction
\begin{equation}
\delta m = -\lambda^{2}\int_{0}^{\infty}\frac{d\theta}{2\pi}\left(\frac{\left|F_{3}(i\pi,\theta,-\theta)\right|^{2}}{m^{3}(2\cosh\theta-1)\cosh\theta}-\frac{16\left|F_{1}\right|^{2}}{m^{3}\sinh^{2}\theta\cosh\theta}\right)+\lambda^{2}\frac{16\left|F_{1}\right|^{2}}{4m^{3}}(1-\varphi(0))
\label{ffptmasscorr}\end{equation}
which agrees with the result in \cite{Takacs:2009fu}. 
Note that the leading bulk term drops out from the difference of the energy levels as it indeed should.

The contributions of higher-particle states to the spectral sum can be computed analogously. 
It is straightforward to verify that 
the $n$ particle state term always contains a double pole part analogous to 
the one treated above, which exactly cancels the $n-1$ particle contribution to the 
vacuum level in (\ref{vac2order}).

\section{Conclusions and outlook}

\label{section6}

First let us sum up what has been achieved in this paper. Using the idea of finite volume regularization 
and multi-dimensional residue techniques we have developed a systematic technique to evaluate the form 
factor expansion for the finite-temperature two-point function in integrable field theories. In fact, as 
the examples of the zero-temperature three-point function and of form factor perturbation theory show, the 
approach can be applied to many problems involving spectral sums with singularities coming from the 
presence of disconnected terms. Albeit it was expected on general physical grounds, it is an important 
fact that our calculation demonstrated that the resulting expressions for the correlators (and for the 
mass gap in the case of FFPT) are well-defined when removing the regulator by taking the infinite volume
limit.

For the three-point function it is apparent that the resulting formula is just the proper way of 
separating the disconnected pieces and can in fact be written down directly by inspection of the infinite 
volume expression. This is due to the disconnected pieces appearing linearly.  
However, the other two cases involve the disconnected terms squared, similarly to the case of 
one-point functions of bulk operators with boundaries \cite{Kormos:2010ae}. While it is 
well-known that the resulting terms, naively containing squares of Dirac $\delta$ functions can be regularized in 
a finite box, the correct result can only be obtained by carefully taking into account 
that the finite volume spectrum is different 
from that of a non-interacting system. This was pointed out for the one-point functions in our 
previous paper \cite{Pozsgay:2007gx}, and is manifested by the explicit dependence of the results 
(\ref{D12}), (\ref{D22final}) and (\ref{ffptmasscorr}) on the $S$ matrix (directly or via the derivative 
$\varphi$ of the phase-shift). 
The LeClair-Mussardo proposal for the one and two-point functions tries 
to capture this feature by a TBA dressing of the energy and momentum of the finite-temperature 
quasi-particles. Contrary to the one-point case \cite{Pozsgay:2007gx}, 
our result for the two-point function does not confirm their conjectured expression  
(which is, in any case, eventually ill-defined). Despite some partial indications of resummation, 
it is not obvious whether the interaction dependence can be summed up to yield some simple 
dressing prescription.

This leads us to one of the main open questions, namely, to investigate the possibility of such 
resummation and find out whether there is a way to introduce some sort of dressing prescription 
to simplify the expansion by systematically combining contributions. At this point this seems to 
require the evaluation of higher orders, which is in principle straightforward, but an extremely 
tedious task. Therefore another important (albeit technical) issue is to simplify the method of 
evaluating the contributions to the expansion.

It is also very important, especially in view of potential applications, to extend the method to 
non-diagonal theories. While this is in principle straightforward (for the basic ideas cf. 
\cite{Essler:2009zz} in the framework of $O(3)$ model), it would be desirable to have some 
efficient approach to characterizing the finite-volume form factors of non-diagonal models for 
general number of particles. Work in this direction is in progress. This is even more important, 
since at present the most we can show for testing the expansion for the thermal correlator is its 
internal consistency. Consistency is shown by two facts: (1) that terms divergent for large volumes 
cancel in the final result order-by-order and (2) that different orders of performing the summation 
lead to identical results, as demonstrated in for $D_{12}$. These are indeed very nontrivial tests of 
the calculation, but a physical application of the method would be much better.

Another important issue, especially in view of potential applications to non-relativistic systems
along the lines of \cite{Kormos:2009zz}, is the extension to include a nonzero chemical potential. 
At present it is not entirely clear how to do that, but rewriting the expansion through some partial 
resummation/dressing procedure could be helpful 
(in analogy to the way the dependence is introduced into the LeClair-Mussardo formula for the 
one-point function \cite{Leclair:1999ys}).

Finally we comment on the relation of our results to the recent work by Essler and Konik \cite{Essler:2009zz}. 
Their finite-volume calculation is essentially the evaluation of $D_{12}$ using the one-particle summation, with an explicit 
summation of the discrete part instead of a residue trick. However, this approach is very limited: it can only be applied to 
$D_{1n}$ since all other contributions require summation over states with two (or more) particles and there is no obvious way to 
perform the discrete sums directly. This is where 
the multi-dimensional residue method presented here is so powerful since it makes the evaluation of such sums a mechanical 
(albeit somewhat tedious) exercise. 

The infinite-volume regularization method of \cite{Essler:2009zz}, on the other hand, is plagued by (at least potential, but most likely 
actual) ambiguities: for states containing more than one particle, regularization by point-splitting in rapidity space is ambiguous 
(direction-dependent) at the locations in rapidity space where the form factors have either of the two types of disconnected contributions 
described in subsection \ref{finvolFF}. These ambiguities were analyzed in much detail 
in our previous paper \cite{Pozsgay:2007gx}. Therefore (at least at the present state of art) 
the only safe method to evaluate multi-particle 
contributions is by finite volume regularization, and the only systematic way to perform the summations is by using the 
multi-dimensional residue theorem, i.e. in the framework presented here. As mentioned above, however, it is an important goal 
to simplify the method of calculation, which could potentially lead to dispensing with these technical requirements in the end. 

\subsubsection*{Acknowledgments}
We are grateful to M\'arton Kormos for useful discussions. BP was supported 
by the Stichting voor Fundamenteel Onderzoek der
Materie (FOM) in the Netherlands, while GT was partially supported by the 
Hungarian OTKA grants K60040 and K75172.   

\appendix

\section{Multi-dimensional residue formula}

Suppose that we have functions $g(z),\, f_{1}(z),\dots,\, f_{n}(z)$
of $n$ complex variables $z=(z_{1},\dots,z_{n})$. Let us take a
multi-contour $C$ in $\mathbb{C}^{n}$ (i.e. a direct sum of elementary
multi-contours that are defined as direct products of $n$ one-dimensional
contours; without loss of generality -- due to the linearity of integration
-- we may suppose that it is a single (i.e. monomial) product contour
$C=C_{1}\times\dots\times C_{n}$). Let us suppose that the equations
\begin{equation}
f_{k}(z)=0\qquad k=1,\dots,n\end{equation}
have a single solution $z_{*}=(z_{*1},\dots,z_{*n})$ such that for
each $k$, $z_{k*}$ is inside $C_{k}$. Then we have the formula
\begin{equation}
\oint_{C}\frac{dz_{1}}{2\pi i}\dots\frac{dz_{n}}{2\pi i}\frac{g(z)}{f_{1}(z)\dots f_{n}(z)}=\frac{g(z_{*})}{\left.\det\left(\frac{\partial f_{k}}{\partial z_{l}}\right)\right|_{z=z_{*}}}
\label{multidimresidue}
\end{equation}
provided that the determinant does not vanish (this will always be
the case in our calculations). Note that if any of the $f_{k}$ is
nonzero everywhere inside its contour $C_{k}$ then the integral vanishes
since $C_{k}$ can be shrunk to a point. Therefore, all the usual
contour deformation arguments work as long as the contour deformations
take place away from the analytic variety defined by 
\begin{equation}
\det\left(\frac{\partial f_{k}}{\partial z_{l}}\right)=0
\end{equation}

\section{Finite volume FF and phase conventions}

One has to evaluate the products of finite volume form factors. It
follows from the crossing formula (\ref{eq:ffcrossing}) that 
\begin{equation}
\begin{split}
&  \bra{\{I_1\dots I_N\}}\mathcal{O}_1(0)\ket{\{J_1\dots J_M\}}_L 
\bra{\{J_1\dots J_M\}}\mathcal{O}_2(0)\ket{\{I_1\dots I_N\}}_L=\\
&=\frac{F_{N+M}^{\mathcal{O}_1}(\theta_1+i\pi,\dots,\theta_N+i\pi,\theta'_M,\dots,\theta'_1)
F_{N+M}^{\mathcal{O}_2}(\theta_1'+i\pi,\dots,\theta_M'+i\pi,\theta_N,\dots,\theta_1)}
{\rho_N(\theta_1,\dots,\theta_N)\rho_M(\theta_1',\dots,\theta_M')}
\end{split}
\label{phase-conv}
\end{equation}

In the following we show that in unitary models the crossing procedure
described by
\eqref{phase-conv} reproduces the usual complex conjugation of the
matrix element. 
In unitary models the phase of the form factor is given by (allow for
an extra sign ambiguity) 
\begin{equation}
  \label{eq:phase1}
  F_N(\theta_1,\dots,\theta_N)= \Big|
  F_N(\theta_1,\dots,\theta_N)\Big| \times 
\sqrt{\prod_{i<j} S(\theta_i-\theta_j)}
\qquad\qquad \theta_i\in\valos
\end{equation}
In particular $F_1$ is always real. The extension to include also
``bra'' vectors reads
\begin{equation}
  \label{altalanos_phase}
\begin{split}
&   F_{N+M}(\theta'_1+i\pi,\dots,\theta'_M+i\pi,\theta_1,\dots,\theta_N)=\\ 
& \hspace{1cm}  \Big| F_{N+M}(\theta'_1+i\pi,\dots,\theta'_M+i\pi,\theta_1,\dots,\theta_N)\Big| \times 
\sqrt{\prod_{i<j} S(\theta_i-\theta_j)}\times
\sqrt{\prod_{k<l} S(\theta_k'-\theta_l')}
\\
& \qquad\qquad \theta_i,\theta'_k\in\valos
\end{split}
\end{equation}
The complex conjugation property is then given by
\begin{equation}
\Big(   F_{N+M}(\theta'_1+i\pi,\dots,\theta'_M+i\pi,\theta_1,\dots,\theta_N)\Big)^*=
 F_{N+M}(\theta'_M+i\pi,\dots,\theta'_1+i\pi,\theta_N,\dots,\theta_1)
\end{equation}
This way one can avoid the operation of complex conjugation and one
can work with analytic functions in the complex plane.


\section{Evaluating the subtractions in the non-diagonal part of $D_{22}$}\label{D22sing}

We first substitute
\begin{equation}
\sum_{I_{1}>I_{2}}\rightarrow\frac{1}{2}\sum_{I_{1},I_{2}}-\frac{1}{2}\sum_{I_{1}=I_{2}}
\end{equation}

\subsection{Spurious QQ singularities}\label{D22QQ}
Let us consider the first family of such singularities, which is when 
\begin{equation}
\theta_{1}'=\theta_{1}\quad\mbox{and}\quad \theta_{2}'=\theta_{2}
\end{equation}
The evaluation of such terms is complicated by the fact that the diagonal limit of $F_{4}$ 
is undefined (i.e. direction dependent). Fortunately, after converting the sums to integrals 
a factor of $1/\rho_2$ remains therefore all such terms are of order $\ordo(L^{-2})$. Similar 
considerations apply to the other case
\begin{equation}
\theta_{1}'=\theta_{2}\quad\mbox{and}\quad \theta_{2}'=\theta_{1}
\end{equation}

\subsection{QF singularities}\label{D22QF}
We consider the case
\begin{equation}
\theta_{1}'=\theta_{1}\quad\mbox{and}\quad Q_{2'}(\theta_{1}',\theta_{2}')=2\pi J_{2}
\end{equation}
Just as in the case of $D_{12}$, the contribution can be split into a double and a single pole part.
\subsubsection{The double pole part}
The double pole part is given by
\begin{eqnarray}
\mbox{D}_{QF}^{11} & = & 
\sum_{I_1>I_2}\frac{1}{\rho_2(\theta_1,\theta_2)}
\oint_{C_{J_{2}I_{1}}}\frac{d\theta_{2}'}{2\pi}\oint_{\theta_{1}}\frac{d\theta_{1}'}{2\pi}\frac{K_{t,x}^{(R)}(\theta_{1},\theta_{2};\theta_{1}',\theta_{2}')}{\left(\mathrm{e}^{iQ_{1'}(\theta_{1}',\theta_{2}')}+1\right)\left(\mathrm{e}^{iQ_{2'}(\theta_{1}',\theta_{2}')}+1\right)}\nonumber \\
 & \times & \frac{i}{\theta_{1}-\theta_{1}'}(1-S(\theta_{2}-\theta_{1})S(\theta_{1}'-\theta_{2}'))F_{2}^{O_{1}}(\theta_{2}+i\pi,\theta_{2}')\nonumber \\
 & \times & \frac{i}{\theta_{1}'-\theta_{1}}(1-S(\theta_{2}'-\theta_{1}')S(\theta_{1}-\theta_{2}))F_{2}^{O_{2}}(\theta_{2}'+i\pi,\theta_{2})\end{eqnarray}
We have
\begin{equation}
\mathrm{e}^{iQ_{1'}(\theta_{1},\theta_{2}')}=-S(\theta_{1}-\theta_{2}')\mbox{e}^{imL\sinh\theta_{1}}
\end{equation}
On the other hand 
\begin{equation}
\mbox{e}^{imL\sinh\theta_{1}}S(\theta_{1}-\theta_{2})=1
\end{equation}
i.e
\begin{equation}
\mathrm{e}^{iQ_{1'}(\theta_{1},\theta_{2}')}=-S(\theta_{1}-\theta_{2}')S(\theta_{2}-\theta_{1})
\label{temp1}\end{equation}
Using $S(\theta)S(-\theta)=1$ we can rearrange the contribution as
\begin{eqnarray}
\mbox{D}_{QF}^{11} & = & \sum_{I_1>I_2}\frac{1}{\rho_2(\theta_1,\theta_2)} 
\oint_{C_{J_{2}I_{1}}}\frac{d\theta_{2}'}{2\pi}F_{2}^{O_{1}}(\theta_{2}+i\pi,\theta_{2}')
F_{2}^{O_{2}}(\theta_{2}'+i\pi,\theta_{2})
\nonumber\\
&&\times\oint_{\theta_{1}}\frac{d\theta_{1}'}{2\pi}\frac{K_{t,x}^{(R)}(\theta_{1},\theta_{2};\theta_{1}',\theta_{2}')}{\left(\mathrm{e}^{iQ_{1'}(\theta_{1}',\theta_{2}')}+1\right)\left(\mathrm{e}^{iQ_{2'}(\theta_{1}',\theta_{2}')}+1\right)}\nonumber \\
 && \times \frac{1}{(\theta_{1}-\theta_{1}')^{2}}\left((1-S(\theta_{2}-\theta_{1})S(\theta_{1}'-\theta_{2}'))+(1-S(\theta_{2}'-\theta_{1}')S(\theta_{1}-\theta_{2}))\right)\end{eqnarray}
We now use
\begin{equation}
\oint_{\theta_{1}}\frac{d\theta_{1}'}{2\pi}\frac{1}{(\theta_{1}-\theta_{1}')^{2}}f(\theta_{1})=if'(\theta_{1}'=\theta_{1})\end{equation}
and after manipulations similar to those in subsection \ref{subsubsec:D12doublesum} we arrive at
\begin{eqnarray}
\mbox{D}_{QF}^{11}  &=&  -\sum_{I_1>I_2}\frac{1}{\rho_2(\theta_1,\theta_2)} 
\oint_{C_{J_{2}I_{1}}}\frac{d\theta_{2}'}{2\pi}F_{2}^{O_{1}}(\theta_{2}+i\pi,\theta_{2}')
F_{2}^{O_{2}}(\theta_{2}'+i\pi,\theta_{2})
 \\
 & &\times
\frac{K_{t,x}^{(R)}(\theta_{1},\theta_{2};\theta_{1},\theta_{2}')}
{\left(\mathrm{e}^{iQ_{2'}(\theta_{1},\theta_{2}')}+1\right)}
 \Big((mx\cosh\theta_{1}-imt\sinh\theta_{1})(1-S(\theta_{2}'-\theta_{1})S(\theta_{1}-\theta_{2}))
\nonumber\\
& &+\varphi(\theta_{1}-\theta_{2}')S(\theta_{2}'-\theta_{1})S(\theta_{1}-\theta_{2})+mL\cosh\theta_{1}\Big)
\nonumber
\end{eqnarray}
The full contribution is then obtained by adding the three other double pole terms pertaining to 
the other QF-singularities in (\ref{D22QFclass}), 
which can also be obtained by suitably permuting the rapidity variables. 
Converting the sum over $\theta_1$, $\theta_2$ to integrals, taking care to subtract
the diagonal $\theta_1=\theta_2$ we obtain for the full double pole contribution the expression:
\begin{eqnarray}
\mbox{D}_{QF} & = & \iint\frac{d\theta_{1}}{2\pi}\frac{d\theta_{2}}{2\pi}\int\frac{d\theta_{2}'}{2\pi}
F_{2}^{O_{1}}(\theta_{2}+i\pi,\theta_{2}')F_{2}^{O_{2}}(\theta_{2}'+i\pi,\theta_{2})\nonumber \\
 & \times &
\mathrm{e}^{imx(\sinh\theta_{2}-\sinh\theta_{2}')}\mathrm{e}^{-mR\cosh\theta_{1}}
\mathrm{e}^{-m(R-t)\cosh\theta_{2}}\mathrm{e}^{-mt\cosh\theta_{2}'}\nonumber \\
 & \times &
 \Big((mx\cosh\theta_{1}-imt\sinh\theta_{1})(1-S(\theta_{2}'-\theta_{1})S(\theta_{1}-\theta_{2}))\nonumber \\
 & &
+\varphi(\theta_{1}-\theta_{2}')S(\theta_{2}'-\theta_{1})S(\theta_{1}-\theta_{2})+mL\cosh\theta_{1}\Big)
\nonumber\\
&-&
\int\frac{d\theta_{1}}{2\pi}\int\frac{d\theta_{2}'}{2\pi}
F_{2}^{O_{1}}(\theta_{1}+i\pi,\theta_{2}')F_{2}^{O_{2}}(\theta_{2}'+i\pi,\theta_{1})
\nonumber \\
 & \times &
\mathrm{e}^{imx(\sinh\theta_{1}-\sinh\theta_{2}')}
\mathrm{e}^{-m(2R-t)\cosh\theta_{1}}\mathrm{e}^{-mt\cosh\theta_{2}'} 
\end{eqnarray}
Now we recall the leftover counter term from (\ref{Z1D11remainder})
\begin{equation}
-Z_{1}\int\frac{d\theta_{1}}{2\pi}\int\frac{d\theta_{2}}{2\pi}
F_{2}^{\mathcal{O}_{1}}(\theta_{1}+i\pi,\theta_{2})
F_{2}^{\mathcal{O}_{2}}(\theta_{1},\theta_{2}+i\pi)\mathrm{e}^{imx(\sinh\theta_{1}-\sinh\theta_{2})}
\mathrm{e}^{-m(R-t)\cosh\theta_{1}}\mathrm{e}^{-mt\cosh\theta_{2}}
\end{equation}
with 
\begin{eqnarray}
Z_{1} & = & mL\int\frac{d\theta}{2\pi}\cosh\theta\mathrm{e}^{-mR\cosh\theta}
\end{eqnarray}
and see that it exactly cancels the $O(L)$ part, leaving us with the finite expression
\begin{eqnarray}
\mbox{D}_{QF}^{\mathrm{finite}} & = & 
\int\frac{d\theta_{1}}{2\pi}\int\frac{d\theta_{2}}{2\pi}\int\frac{d\theta_{2}'}{2\pi}
F_{2}^{O_{1}}(\theta_{2}+i\pi,\theta_{2}')F_{2}^{O_{2}}(\theta_{2}'+i\pi,\theta_{2})
\nonumber \\
 & \times &
\mathrm{e}^{imx(\sinh\theta_{2}-\sinh\theta_{2}')}\mathrm{e}^{-mR\cosh\theta_{1}}
\mathrm{e}^{-m(R-t)\cosh\theta_{2}}\mathrm{e}^{-mt\cosh\theta_{2}'}\nonumber \\
 & \times &
 \Big((mx\cosh\theta_{1}-imt\sinh\theta_{1})(1-S(\theta_{2}'-\theta_{1})S(\theta_{1}-\theta_{2}))
\nonumber \\
 &  &
+\varphi(\theta_{1}-\theta_{2}')S(\theta_{2}'-\theta_{1})S(\theta_{1}-\theta_{2})\Big)
\nonumber\\
&-&
\int\frac{d\theta_{1}}{2\pi}\int\frac{d\theta_{2}'}{2\pi}
F_{2}^{O_{1}}(\theta_{1}+i\pi,\theta_{2}')F_{2}^{O_{2}}(\theta_{2}'+i\pi,\theta_{1})
\nonumber \\
 & \times &
\mathrm{e}^{imx(\sinh\theta_{1}-\sinh\theta_{2}')}
\mathrm{e}^{-m(2R-t)\cosh\theta_{1}}\mathrm{e}^{-mt\cosh\theta_{2}'} 
\label{D22QFfinal}
\end{eqnarray}

\subsubsection{Single pole contributions}
Once again, we consider the $\theta_{1}'=\theta_{1}$ case and introduce the notation: 
\begin{eqnarray}
F_{4}^{\mathcal{O}_{1}}(\theta_{2}+i\pi,\theta_{1}+i\pi,\theta_{1}',\theta_{2}') & = & \frac{i}{\theta_{1}-\theta_{1}'}(1-S(\theta_{2}-\theta_{1})S(\theta_{1}'-\theta_{2}'))F_{2}^{O_{1}}(\theta_{2}+i\pi,\theta_{2}')\nonumber \\
 & + & F_{4sc}^{\mathcal{O}_{1}}(\theta_{2},\theta_{1}|\theta_{1}',\theta_{2}')\label{F4scdef}
\end{eqnarray}
and similarly
\begin{eqnarray}
F_{4}^{\mathcal{O}_{2}}(\theta_{2}'+i\pi,\theta_{1}'+i\pi,\theta_{1},\theta_{2}) & = & \frac{i}{\theta_{1}'-\theta_{1}}(1-S(\theta_{2}'-\theta_{1}')S(\theta_{1}-\theta_{2}))F_{2}^{O_{2}}(\theta_{2}'+i\pi,\theta_{2})\nonumber \\
 & + & F_{4sc}^{\mathcal{O}_{2}}(\theta_{2}',\theta_{1}'|\theta_{1},\theta_{2})
\end{eqnarray}
The contribution has the following form
\begin{eqnarray}
\mbox{S}_{QF}^{11} & = & -\frac{1}{2}\oint_{\theta_{1}}\frac{d\theta_{1}'}{2\pi}\oint_{C_{J_{2}I_{1}}}\frac{d\theta_{2}'}{2\pi}\frac{K_{t,x}^{(R)}(\theta_{1},\theta_{2};\theta_{1}',\theta_{2}')}{\left(\mathrm{e}^{iQ_{1'}(\theta_{1}',\theta_{2}')}+1\right)\left(\mathrm{e}^{iQ_{2'}(\theta_{1}',\theta_{2}')}+1\right)}\nonumber \\
 & \times & \Big[\frac{i}{\theta_{1}-\theta_{1}'}(1-S(\theta_{2}-\theta_{1})S(\theta_{1}'-\theta_{2}'))F_{2}^{O_{1}}(\theta_{2}+i\pi,\theta_{2}')F_{4sc}^{\mathcal{O}_{2}}(\theta_{2}',\theta_{1}'|\theta_{1},\theta_{2})\nonumber \\
 &  & +\frac{i}{\theta_{1}'-\theta_{1}}(1-S(\theta_{2}'-\theta_{1}')S(\theta_{1}-\theta_{2}))F_{2}^{O_{2}}(\theta_{2}'+i\pi,\theta_{2})F_{4sc}^{\mathcal{O}_{1}}(\theta_{2},\theta_{1}|\theta_{1}',\theta_{2}')\Big]\end{eqnarray}
Recalling (\ref{temp1}) and evaluating the residue integrals
\begin{eqnarray}
\mbox{S}_{QF}^{11} & = & +\frac{1}{2}\oint_{C_{J_{2}I_{1}}}\frac{d\theta_{2}'}{2\pi}\frac{K_{t,x}^{(R)}(\theta_{1},\theta_{2};\theta_{1},\theta_{2}')}{\left(1-S(\theta_{1}-\theta_{2}')S(\theta_{1}-\theta_{2})\right)\left(\mathrm{e}^{iQ_{2'}(\theta_{1},\theta_{2}')}+1\right)}\nonumber \\
 & \times & \Big[(1-S(\theta_{2}-\theta_{1})S(\theta_{1}-\theta_{2}'))F_{2}^{O_{1}}(\theta_{2}+i\pi,\theta_{2}')F_{4sc}^{\mathcal{O}_{2}}(\theta_{2}',\theta_{1}|\theta_{1},\theta_{2})\nonumber \\
 &  & -(1-S(\theta_{2}'-\theta_{1})S(\theta_{1}-\theta_{2}))F_{2}^{O_{2}}(\theta_{2}'+i\pi,\theta_{2})F_{4sc}^{\mathcal{O}_{1}}(\theta_{2},\theta_{1}|\theta_{1},\theta_{2}')\Big]\end{eqnarray}
Note that this vanishes when $\theta_{1}=\theta_{2}$ due to the form
factors vanishing, so when putting in the $\theta_{1},\theta_{2}$
summation we can include the diagonal $\theta_{1}=\theta_{2}$. Converting the summation 
to integrals we obtain
\begin{eqnarray}
\mbox{S}_{QF}^{11} & = & \frac{1}{4}\iint\frac{d\theta_{1}}{2\pi}\frac{d\theta_{2}}{2\pi}\int\frac{d\theta_{2}'}{2\pi}\mathrm{e}^{imx(\sinh\theta_{2}-\sinh\theta_{2}')}\mathrm{e}^{-mR\cosh\theta_{1}}\mathrm{e}^{-m(R-t)\cosh\theta_{2}}\mathrm{e}^{-mt\cosh\theta_{2}'}\nonumber \\
 &  &
 \times\Big(F_{2}^{O_{1}}(\theta_{2}+i\pi,\theta_{2}')
F_{4sc}^{\mathcal{O}_{2}}(\theta_{2}',\theta_{1}|\theta_{1},\theta_{2})
\nonumber\\
&&+S(\theta_{2}'-\theta_{1})S(\theta_{1}-\theta_{2})F_{2}^{O_{2}}(\theta_{2}'+i\pi,\theta_{2})
F_{4sc}^{\mathcal{O}_{1}}(\theta_{2},\theta_{1}|\theta_{1},\theta_{2}')\Big)
\end{eqnarray}
or, using the definition of $F_{4sc}$ and the form factor equation (\ref{eq:exchangeaxiom})
\begin{eqnarray}
\mbox{S}_{QF}^{11} & = & \frac{1}{4}\iint\frac{d\theta_{1}}{2\pi}\frac{d\theta_{2}}{2\pi}
\int\frac{d\theta_{2}'}{2\pi}
\mathrm{e}^{imx(\sinh\theta_{2}-\sinh\theta_{2}')}\mathrm{e}^{-mR\cosh\theta_{1}}
\mathrm{e}^{-m(R-t)\cosh\theta_{2}}\mathrm{e}^{-mt\cosh\theta_{2}'}\nonumber \\
 &  &
 \times\Big(F_{2}^{O_{1}}(\theta_{2}+i\pi,\theta_{2}')
F_{4sc}^{\mathcal{O}_{2}}(\theta_{2}',\theta_{1}|\theta_{1},\theta_{2})
\nonumber\\
&&+F_{2}^{O_{2}}(\theta_{2}'+i\pi,\theta_{2})
F_{4sc}^{\mathcal{O}_{1}}(\theta_{2}',\theta_{1}|\theta_{1},\theta_{2})\Big)
\end{eqnarray}
The full contribution is then obtained by adding the three other single pole terms pertaining to 
the other QF-singularities in (\ref{D22QFclass}), which can also be obtained by suitably permuting 
the rapidity variables:
\begin{eqnarray}
\mbox{S}_{QF} & = & \frac{1}{2}\iint\frac{d\theta_{1}}{2\pi}\frac{d\theta_{2}}{2\pi}
\int\frac{d\theta_{2}'}{2\pi}
\Big[\mathrm{e}^{imx(\sinh\theta_{2}-\sinh\theta_{2}')}
\mathrm{e}^{-mR\cosh\theta_{1}}
\mathrm{e}^{-m(R-t)\cosh\theta_{2}}\mathrm{e}^{-mt\cosh\theta_{2}'}\nonumber \\
 &  & \times\left(F_{2}^{O_{1}}(\theta_{2}+i\pi,\theta_{2}')
F_{4sc}^{\mathcal{O}_{2}}(\theta_{2}',\theta_{1}|\theta_{1},\theta_{2})
+F_{2}^{O_{2}}(\theta_{2}'+i\pi,\theta_{2})
F_{4sc}^{\mathcal{O}_{1}}(\theta_{2}',\theta_{1}|\theta_{1},\theta_{2})\right)\nonumber \\
 &  & +\left(\theta_{1}\leftrightarrow\theta_{2}\right)\Big]
\label{D22SQFfinal}\end{eqnarray}

\subsection{FF singularities}\label{D22FF}

First let us consider the $\theta_{1}'=\theta_{2}'=\theta_{1}$ case. We need to separate the singular 
terms from the form factors, fro which we introduce a new function $F_{4dc}$ defined by 
\begin{eqnarray}
F_{4}^{\mathcal{O}_{1}}(\theta_{2}+i\pi,\theta_{1}+i\pi,\theta_{1}',\theta_{2}') & = & \frac{i}{\theta_{1}-\theta_{1}'}(1-S(\theta_{2}-\theta_{1})S(\theta_{1}'-\theta_{2}'))F_{2}^{O_{1}}(\theta_{2}+i\pi,\theta_{2}')\nonumber \\
 & + & \frac{i}{\theta_{1}-\theta_{2}'}(S(\theta_{1}'-\theta_{2}')-S(\theta_{2}-\theta_{1}))F_{2}^{O_{1}}(\theta_{2}+i\pi,\theta_{1}')\nonumber \\
 & + & F_{4dc}^{\mathcal{O}_{1}}(\theta_{2},\theta_{1}|\theta_{1}',\theta_{2}')
\label{F4dcdef}\end{eqnarray}
and similarly
\begin{eqnarray}
F_{4}^{\mathcal{O}_{2}}(\theta_{2}'+i\pi,\theta_{1}'+i\pi,\theta_{1},\theta_{2}) & = & \frac{i}{\theta_{1}'-\theta_{1}}(1-S(\theta_{2}'-\theta_{1}')S(\theta_{1}-\theta_{2}))F_{2}^{O_{2}}(\theta_{2}'+i\pi,\theta_{2})\nonumber \\
 & + & \frac{i}{\theta_{2}'-\theta_{1}}(S(\theta_{2}'-\theta_{1}')-S(\theta_{1}-\theta_{2}))F_{2}^{O_{2}}(\theta_{1}'+i\pi,\theta_{2})\nonumber \\
 & + & F_{4dc}^{\mathcal{O}_{2}}(\theta_{2}',\theta_{1}'|\theta_{1},\theta_{2})\end{eqnarray}
Only the cross terms can contribute, otherwise at least one of the
contour integrals can be shrunk to a point. We obtain

\begin{eqnarray}
\mbox{S}_{FF}^{1} & = &
-\sum_{I_1>I_2}\frac{1}{\rho_2(\theta_1,\theta_2)}
\frac{1}{2}\oint_{\theta_{1}}\frac{d\theta_{2}'}{2\pi}\oint_{\theta_{1}}\frac{d\theta_{1}'}{2\pi}
\frac{K_{t,x}^{(R)}(\theta_{1},\theta_{2};\theta_{1}',\theta_{2}')}
{\left(\mathrm{e}^{iQ_{1'}(\theta_{1}',\theta_{2}')}+1\right)
\left(\mathrm{e}^{iQ_{2'}(\theta_{1}',\theta_{2}')}+1\right)}
\nonumber\\
& &\times
\frac{1}{\theta_{1}-\theta_{1}'}\frac{1}{\theta_{1}-\theta_{2}'}
F_{2}^{\mathcal{O}_{1}}(\theta_{2}+i\pi,\theta_{2}')F_{2}^{\mathcal{O}_{2}}(\theta_{2}'+i\pi,\theta_{2})
\nonumber\\
& &\times
2(1+S(\theta_1-\theta_2)) (1+S(\theta_2-\theta_1))
\end{eqnarray}
We have
\begin{equation}
K_{t,x}^{(R)}(\theta_{1},\theta_{2};\theta_{1},\theta_{1})=
\mathrm{e}^{imx(\sinh\theta_{2}-\sinh\theta_{1})}
\mathrm{e}^{-m(R-t)(\cosh\theta_{1}+\cosh\theta_{2})}\mathrm{e}^{-2mt\cosh\theta_{1}}
\end{equation}
and
\begin{eqnarray}
\mathrm{e}^{iQ_{1'}(\theta_{1},\theta_{1})} & = & \mathrm{e}^{imL\sinh\theta_{1}}=S(\theta_{2}-\theta_{1})\\
\mathrm{e}^{iQ_{2'}(\theta_{1},\theta_{1})} & = & \mathrm{e}^{imL\sinh\theta_{1}}=S(\theta_{2}-\theta_{1})\end{eqnarray}
The contribution $\mbox{S}_{FF}^{2}$ from  $\theta_{1}'=\theta_{2}'=\theta_{2}$ can be obtained by interchanging 
$\theta_1$ and $\theta_2$. Putting in the $\theta_{1}$ and $\theta_{2}$ integrals we obtain
\begin{eqnarray}
\mbox{S}_{FF} & = &
-\int\frac{d\theta_{1}}{2\pi}\int\frac{d\theta_{2}}{2\pi}
\Big[\mathrm{e}^{imx(\sinh\theta_{2}-\sinh\theta_{1})}\mathrm{e}^{-m(R-t)(\cosh\theta_{1}+\cosh\theta_{2})}
\mathrm{e}^{-2mt\cosh\theta_{1}}\nonumber\\
&\times &S(\theta_{1}-\theta_{2})F_{2}^{\mathcal{O}_{1}}(\theta_{2}+i\pi,\theta_{1})
F_{2}^{\mathcal{O}_{2}}(\theta_{1}+i\pi,\theta_{2})
 +\left(\theta_{1}\leftrightarrow\theta_{2}\right)\Big]
\label{D22FFfinal}\end{eqnarray}
(in this case the diagonal subtraction is $\ordo(L^{-1})$, 
so it does not give a contribution in the infinite volume limit). 
\addcontentsline{toc}{section}{References}
\bibliography{finiteTcorr}
\bibliographystyle{utphys}

\end{document}